\documentclass[twocolumn,prb,amsmath,amssymb,floatfix]{revtex4}

\usepackage{graphics}
\usepackage{graphicx}
\usepackage{dcolumn}
\usepackage{bm}
\usepackage{subfigure}

\def\Tr{\mbox{Tr}}

\def\vp{\varphi}

\def\al{\alpha}

\def\eps{\epsilon}

\def\om{\omega}

\def\be{\begin{equation}}

\def\ee{\end{equation}}

\def\bea{\begin{eqnarray}}

\def\eea{\end{eqnarray}}

\def\bse{\begin{subequations}}

\def\ese{\end{subequations}}

\def\bc{\begin{center}}

\def\ec{\end{center}}

\def\ra{\rightarrow}

\def\nonum{\nonumber}

\def\la{\langle}

\def\ra{\rangle}

\def\up{\uparrow}

\def\dn{\downarrow}

\begin{document}

\title{Transport and magnetization dynamics \\ in a superconductor/single-molecule magnet/superconductor junction}

\author{S. Teber$^1$, C. Holmqvist$^2$ and M. Fogelstr\"om$^2$}
\affiliation{$^1$Laboratoire de Physique Th\'eorique et Hautes Energies, Universit\'e Pierre et Marie Curie, 4 place Jussieu, 75005, Paris, France.
\\ $^2$Department of Microtechnology and Nanoscience - MC2, Chalmers University of Technology, S-412 96 G\"oteborg, Sweden.}

\date{\today}

\begin{abstract}
We study dc-transport and magnetization dynamics in a junction of arbitrary transparency consisting of two spin-singlet superconducting leads 
connected via a single classical spin precessing at the frequency $\Omega$. 
The presence of the spin in the junction provides different transmission amplitudes for spin-up and spin-down quasiparticles 
as well as a time-dependent spin-flip transmission  term. 
For a phase biased junction, we show that a steady-state superconducting charge current flows through the junction and
that an out-of-equilibrium circularly polarized spin current, of frequency $\Omega$, is emitted in the leads.
Detailed understanding of the charge and spin currents is obtained in the entire parameter range.
In the adiabatic regime, $\hbar \Omega \ll 2\Delta$ where $\Delta$ is the superconducting gap, 
and for high transparencies of the junction, a strong suppression of the current takes place around $\vp \approx 0$ due to an abrupt change in the occupation
of the Andreev bound-states. 
At higher values of the phase and/or precession frequency, extended (quasi-particle like) states compete with the bound-states in order to carry the current.
Well below the superconducting transition, these results are shown to be weakly affected by the back-action of the 
spin current on the dynamics of the precessing spin.
Indeed, we show that the Gilbert damping due to the quasi-particle spin current is strongly suppressed at low-temperatures,
which goes along with a shift of the precession frequency due to the condensate.
The results obtained may be of interest for on-going experiments in the field of molecular spintronics.
\end{abstract}

\maketitle

%%%%%%%%%%%%%%%%%%%%%%%%%%%%%%%%%%%%%%%
\section{Introduction}
\label{Sec.Introduction}

Spintronics exploits the fact that an electron current consists of spinful carriers, 
with information stored in their spin state, which interact in a controlled way with their magnetic environment. 
Research in this field was pioneered in the 70s with the experiments of Tedrow and Meservey~\cite{TedrowMeservey71} on ferromagnet/superconductor tunnel junctions
as well as the experiments of Julli\`ere~\cite{Julliere75} on magnetic tunnel junctions. 
It fully emerged in the 80s with the observation of spin-polarized electron injection from a ferromagnetic metal to a normal metal~\cite{Johnson85}
followed by the discovery of the well known giant magnetoresistance (GMR) effect~\cite{Baibich88,Binasch89}.
Since then, most of the studies focused on stationary magnetic states and the control of the electrical current 
by tuning the state of the magnet as in, e.g. GMR, see Ref.~[\onlinecite{Wolf01}] for a review. 
The study of the magnetization dynamics in ferromagnet (F)/normal metal (N) as well as F/superconductor (S) 
hybrids is much more recent and basically involves
controlling the state of the magnet with the help of an applied electrical current. 
The main trigger was the experimental confirmation of the ideas of Slonczewski~\cite{Slonczewski96} and Berger~\cite{Berger96} that 
an electrical current may affect the state of the magnet via a spin transfer torque, see Ref.~[\onlinecite{Ralph08}] for a
pedagogical introduction. 
Indeed, a spin-polarized current of high enough density injected into a ferromagnet was shown to 
reverse the magnetization or generate a steady-state precessing magnetization in accordance with theoretical predictions.
Many recent experiments have confirmed these facts, see Ref.~[\onlinecite{TserkovnyakRMP05}] for a review. 
Conversely, a microwave driven precessing magnetization of a ferromagnetic layer under ferromagnetic resonance (FMR)
was shown to emit pure spin currents, i.e. without any associated net charge transfer, into the adjacent normal metal layers, see
Refs.~[\onlinecite{Tserkovnyak02}] and [\onlinecite{Waintal03}] for the theory.
Such a spin current was only indirectly measured as an enhancement of the Gilbert damping of the magnetization dynamics~\cite{Mizukami01}.
Recently, a similar experiment has been performed on S/F hybrids~\cite{Bell08}. 
Contrary to the case of N/F hybrids, it was shown that the Gilbert damping is reduced at temperatures well below
the superconducting transition temperature, see Ref.~[\onlinecite{Morten08}] for the theory. 

At the same time, there is a fast growing interest in controlling the spin orientation of single molecules or 
even a single or a few atoms in order to perform basic quantum operations. 
Recently, the control of the spin orientation of a single manganese atom in a semiconductor quantum dot could be achieved
with the help of optical techniques~\cite{LeGall09}. 
At the molecular level, the experimental challenge is already at the level of designing a molecular junction.
Single-molecule magnets (SMM) were recently contacted to metallic leads
allowing electron-transport measurement through them~\cite{Jo06,Heersche06} to probe their properties,
see Refs.~[\onlinecite{Sanvito06,Wernsdorfer08}] for recent reviews on SMMs.
The access resistance due to the normal-contacts may however be a source of limitation
which motivated the design of superconducting molecular junctions.  
This was first done with semiconducting nanowires~\cite{Doh05}, carbon-nanotubes~\cite{Cleuziou06,Jarillo06,Jorgensen06,Jorgensen07} and more recently
with a single C$_{60}$ fullerene molecule~\cite{Winkelmann09}.
Magnetically active metallofullerene molecules could also be contacted to superconducting leads and the proximity-induced superconductivity was studied
via low-temperature transport measurements~\cite{Kasumov05}.
On the theoretical side, equilibrium properties of such Josephson junctions were extensively studied since a long time, see Refs.~[\onlinecite{Appelbaum66,Anderson66,Kulik66,Bulaevskii77,Cuevas01,Benjamin07}]
without being extensive;
the effect of magnetization dynamics has been discussed more recently, see e.g.~[\onlinecite{Balatsky03,Balatsky04,Shumeiko08,Teber09,Holmqvist09}]. 

\begin{figure}
\begin{tabular}{cc}
        \includegraphics[width=5.cm,height=1.75cm]{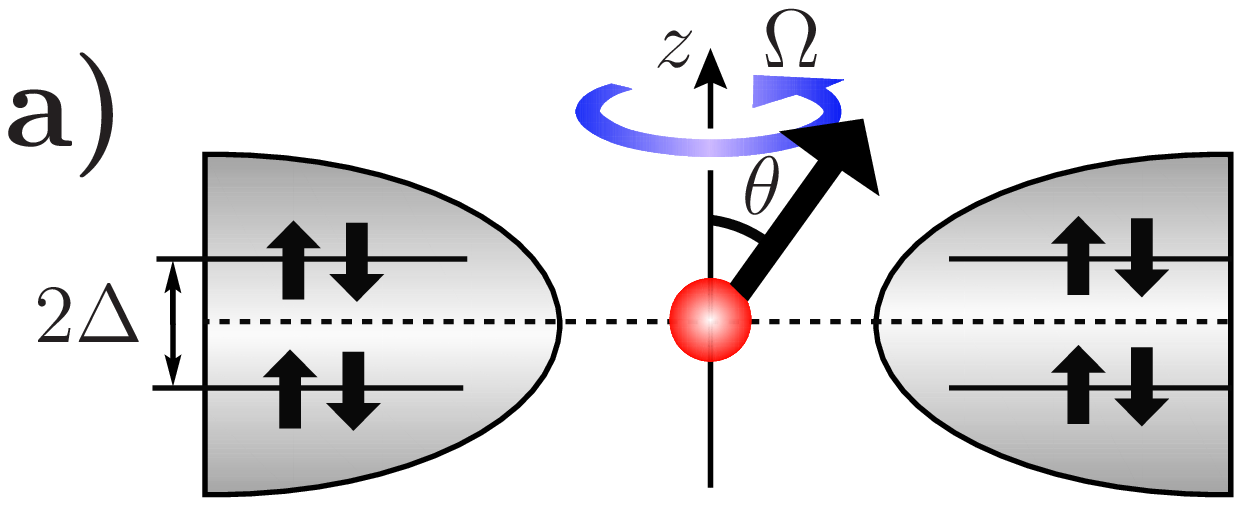}
\\
        \includegraphics[width=5.cm,height=1.75cm]{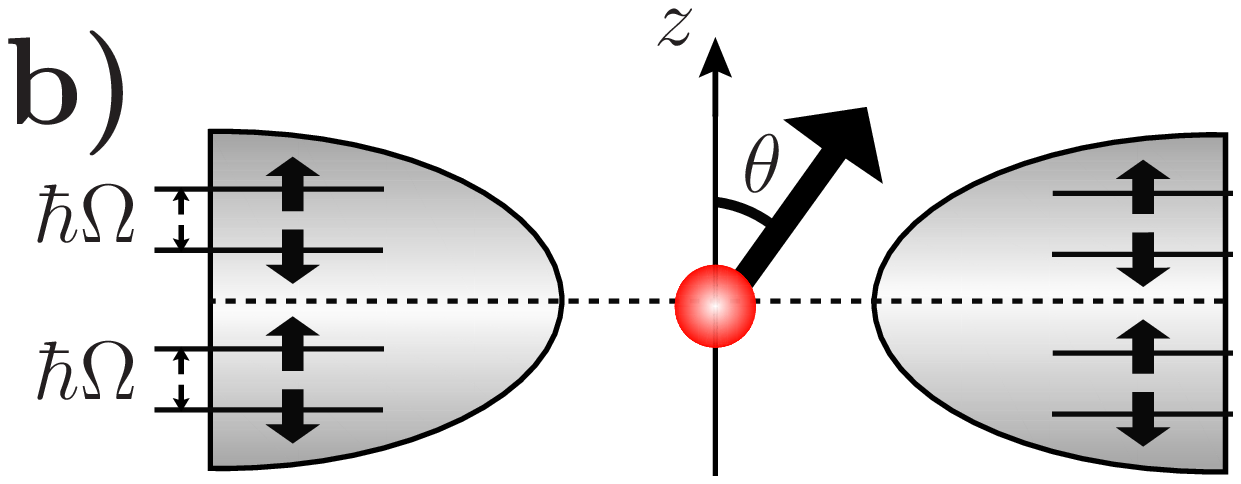}
\end{tabular}
\caption{ \label{Fig.Junction} (color online) Schematic representation of a localized spin (red dot) contacted to superconducting leads
representing a short S/SMM/S junction (S is for superconductor and SMM for single-molecule magnet).
a) In the laboratory frame, the spin is precessing and the leads are spin-degenerate.
b) In the rotating frame, the tilted spin is static and an effective magnetic field, $h_z = \Omega/2$, 
lifts the spin-degeneracy as well as modifies the occupation functions of the particles in the leads.
}
\end{figure}

Motivated by the recent experimental breakthroughs of molecular spintronics 
we study, in the present paper, a model, first proposed in Refs.~[\onlinecite{Balatsky03,Balatsky04}], 
describing a short (shorter than the superconducting coherence length) 
clean junction consisting of a single precessing spin, with precession frequency $\Omega$, 
connected to spin-singlet superconducting BCS leads, see Fig.~\ref{Fig.Junction}a. 
The precessing spin provides different transmission probabilities for spin-up and down quasi-particles.
It is assumed to be the one of, e.g. a molecular magnet. 
The latter have a total spin of quite large magnitude, e.g. $S=10$ for Fe$_{8}$ in its ground state. 
For simplicity, we will neglect its quantum fluctuations and consider that this spin is classical.
On the basis of such a model, our goal will be to compute the dc-transport properties of the junction. 
With respect to Refs.~[\onlinecite{Balatsky03,Balatsky04}]
we will consider a junction of arbitrary transparency and, besides the dc charge current, 
we will also compute the dc spin-current across the junction.
Even though such a set-up is still challenging to realize experimentally
the corresponding simplified model constitutes an interesting theoretical playground in order to study the 
combined effects of both superconductivity and magnetization dynamics at the molecular scale.

Magnetization dynamics drives the system out-of-equilibrium.
This can be most easily seen by going to the rotating frame, Fig.~\ref{Fig.Junction}b,
where the tilted molecular spin is static and an effective $z-$directed magnetic field, $h_z = \Omega/2$, acts on the leads. 
Contrary to a usual magnetic field, see e.g. Ref.~[\onlinecite{Melin00}] and references therein, 
we will show in the following that $h_z$ affects the occupation functions
of the leads translating the out-of-equilibrium nature of the problem.
In the case of normal leads no dc charge current can flow through the junction.
Nevertheless, the non-equilibrium spin accumulation, 
i.e. the difference in chemical potential, $\hbar \Omega$ at $T=0$, between spin-up and down quasi-particles, 
drives the emission of a spin current in the leads. 
In analogy with the case of the macro-magnet mentioned above, this spin-current will in turn
damp the precession of the molecular spin eventually leading to its complete alignment with the
external applied magnetic field.

Combining these out-of-equilibrium effects with superconductivity in the leads
brings new interesting features, even in the dc case~\cite{Note:Houzet}. 
Indeed, integrating the junction into a superconducting loop yields a superconducting 
phase difference between the leads: $\vp = -2\pi \Phi/\Phi_0$, where $\Phi$ is the total magnetic flux across the loop 
and $\Phi_0=h/2e$ is the magnetic flux quantum. 
Cooper pairs may then be transferred across this phase biased junction leading to the dc Josephson effect. 
Microscopically, this transfer relies on the existence of Andreev bound-states~\cite{Kulik69}. 
The phase dependence of the current is directly related to the phase dependence of the 
spectrum of these subgap states.   
In the case of a tunnel junction, without any magnetic impurity, the current phase relation (CPR) follows the well-known sinusoidal relation:
$I^c(\vp) = I_C \sin \vp$, where $I_C$ is the critical current depending on the normal resistance of the junction.
In more complex cases significant departures from this sinusoidal relation are known to take place, see Ref.~[\onlinecite{Golubov04}] for a review.
One well known origin for such a departure is the increase of the transparency, $\mathcal{T}$, of the junction leading to
phase-coherent transfer of multiple Cooper pairs across the junction. 
In the ballistic limit, $\mathcal{T}=1$, and at low temperatures ($T$): $I^c(\vp) \propto \sin (\vp/2)$.
Another known source of departure originates from preparing the system in an out-of-equilibrium state, e.g. such
as by a microwave irradiation of the junction, 
see Refs.~[\onlinecite{Fuechsle09,Chiodi09}] for recent experiments and Ref.~[\onlinecite{Virtanen10}] for the theory.
In the present problem, both of the these features are present.
We will show that they strongly affect the CPR in a way which is proper to the (time-dependent) magnetic interface we consider. 
Concerning the out-of-equilibrium spin current we will show, in accordance with the case of the macro-magnet considered above, 
that it is strongly reduced deep in the superconducting phase.
This will in turn suppress the damping of the precession providing a self-consistent check
of the robustness of the transport properties of the junction. 
The remaining superconducting part of the spin current will also be shown to shift the precession frequency. 

The paper is organized as follows. In Sec.~\ref{Sec.Model} we present the model\cite{Balatsky03,Balatsky04} of a Josephson junction with 
a precessing spin. 
In Sec.~\ref{Sec.Tunnel} we solve this model analytically, in the tunnel limit and zero temperature, to understand the basic physics of the problem.
In order to deal with a junction of arbitrary transparency we present, in Sec.~\ref{Sec.GeneralTransport}, a re-summation method~\cite{Teber09}
based on a usual Green's function approach combined with a unitary transformation to the rotated frame. 
This method provides a convenient interpretation of the out-of-equilibrium features of the system. 
It is also simple to implement numerically.
In Secs.~\ref{Sec.ResultsCharge} and \ref{Sec.ResultsSpin} the combined analytical and numerical results, 
for arbitrary transparency and finite temperatures,
are presented and discussed for the charge and spin currents, respectively.
Finally in Sec.~\ref{Sec.Conclusion} we summarize our results and conclude. 

In the following $\hbar=k_B=1$ everywhere unless specified.

%%%%%%%%%%%%%%%%%%%%%%%%%%%%%%%%%%%%%%
\section{The model}
\label{Sec.Model}

The model we shall consider is based on the following time-dependent tunnel junction Hamiltonian~\cite{Balatsky03,Balatsky04}:
\be
H(t) = H_L + H_R + H_T(t),
\label{Hamiltonian}
\ee
where the perturbation theory series for the time-dependent perturbation, $H_T$, will be summed to infinite order in what follows.
In Eq.~(\ref{Hamiltonian}), the first two terms correspond to the Hamiltonians of the right ($R$) and left ($L$) spin-singlet BCS superconducting leads:
\be
H_\al = \sum_{{\bf k},\sigma}\,\xi_{\bf k}\,c^\dagger_{\al,{\bf k},\sigma} c_{\al,{\bf k},\sigma} 
+ \sum_{\bf k}\,\big[\Delta_\al\,c^\dagger_{\al,{\bf k},\up}  c^\dagger_{\al,-{\bf k},\dn}  +\, h.\,c.\big],
\label{HBCS}
\ee
where $\xi_{\bf k}$ is the single particle spectrum, the (temperature-dependent) gap is given by: 
$\Delta_\al = \Delta(T)\,e^{i \chi_\al}$ and $\chi_\al$ is the superconducting phase in lead $\al=R,L$. 
The last term in Eq.~(\ref{Hamiltonian}) corresponds to the tunneling Hamiltonian between the leads:
\be
H_T(t) = \sum_{{\bf k},{\bf q},\sigma,\sigma'}\,\big[c^\dag_{R,{\bf k},\sigma} ~T_{\sigma,\sigma'}(t)~c_{L,{\bf q},\sigma'} +\,h.\,c.\,\big],
\label{HT}
\ee
with a time-dependent transmission amplitude reading:
\be
T (t)= T_0 {\bf 1}_2 + T_1 {\bf{S}}(t) . \vec{\sigma},
\label{TA0}
\ee
where ${\bf 1}_2$ is the unit $2\times2$ matrix and $\vec{\sigma} = (\sigma_x,\,\sigma_y,\,\sigma_z)$ is the vector of Pauli spin matrices. 
The time-dependence in Eq.~(\ref{TA0}) originates from the precessing motion of the classical spin localized in the junction.
The corresponding classical equation of motion reads: 
\be
\partial_t {\bf S} = -\gamma {\bf S} \times {\bf H}_{eff},
\label{LLE}
\ee
where $\gamma$ is the gyromagnetic ratio and ${\bf H}_{eff} = H_{eff} \hat{{\bf z}}$ is an effective $z-$directed magnetic field
including the applied field as well as other contributions such as crystal anisotropy and demagnetization fields. 
The solution to Eq.~(\ref{LLE}) reads:
\be
{\bf{S}}(t) = S\,(\sin \theta \cos \Omega t,\, \sin \theta \sin \Omega t,\, \cos \theta), 
\label{SpinVec}
\ee
where $\theta$ is the tilt angle of the spin with respect to the $z-$axis and $\Omega = \gamma H_{eff}$ the precession frequency around this axis. 

In spin space, the transmission amplitude matrix reads:
\be
T (t)= T_0 {\bf 1}_2 + T_\parallel \sigma_z + T_\bot \sigma_x e^{-i \sigma_z \Omega t},
\label{TAMatrix}
\ee
where $T_0$ is the direct transmission amplitude whereas 
$T_\parallel = T_1 S_z$  and $T_\bot = T_1 S_\bot$ are the spin-conserving and spin-flip transmission amplitudes, respectively. 
The latter two depend on the magnitude, $S$, of the spin localized in the junction as well as on its orientation, $\theta$, with respect to the quantization axis:
\be
T_\parallel = T_S \cos \theta, \quad T_\bot = T_S \sin \theta, \quad T_S=T_1\,S.
\label{TA1}
\ee

In the absence of precession, $\Omega=0$, and for normal metallic leads, the model in Eq.~(\ref{Hamiltonian}) was first proposed by Appelbaum~\cite{Appelbaum66}
in order to explain tunnel conductance anomalies due to magnetic impurities. Its microscopic justification was provided by Anderson~\cite{Anderson66}
who has shown, on the basis of the Schrieffer-Wolff transformation, that:
\be
T_1 = 2 V_{mix}^2\,\frac{U}{\eps_d(\eps_d + U)},
\ee
where $V_{mix}$ mixes the spin-states of the conduction electrons with the one of the localized spin in the junction, 
$\eps_d < 0$ is the energy of an electron on the localized state and $\eps_d+U$ that to add a second one
taking into account of the Coulomb charging energy $U$.
If either $\eps_d$ or $\eps_d + U$ are close to the Fermi level, tunneling through the magnetic impurity 
will dominate direct tunneling, i.e. $T_1 \gg T_0$. 
In the case of a superconducting junction, still with $\Omega=0$, the model was first considered 
by Kulik~\cite{Kulik66} and Bulaevskii et al.~\cite{Bulaevskii77}
for an ensemble of impurities in the dielectric layer between the superconductors. 
In the lowest order of perturbation theory, they have shown that, in the case where $T_S > T_0$, the junction is in
a $\pi-$state.
Recently, the model was reconsidered by Zhu and Balatsky~\cite{Balatsky03} for a single classical precessing spin in the junction. 
At the lowest order in perturbation theory, they have shown that the dc Josephson current
is not modulated in time by the precession of the spin. 
Latter studies have focused on the back-action of the current on the precessing spin showing a possible nutation~\cite{Balatsky04}. 

We reconsider here the model of a single classical precessing spin in a superconducting 
junction keeping in mind that the single spin may correspond to, e.g. an SMM such as Mn{$_{12}$} or Fe{$_8$}. 
The latter have large $S$ which should favor the tunneling through the spin: $T_S > T_1$. 
Moreover, at the molecular level, one may expect that the tilt angle $\theta$ may be varied in a
larger range than for a ferromagnet under FMR~\cite{Houzet08}. 
The cases where either $T_\parallel \ll T_\bot$ (large tilt angles)
or $T_\parallel \gg T_\bot$ (small tilt angles) will therefore be considered in the following.
 
As stated in the Introduction, our goal is to explore the transport properties of such a 
superconducting junction: charge as well as spin currents, for a single conducting channel, 
arbitrary transparency and temperatures below the superconducting critical temperature.

%%%%%%%%%%%%%%%%%%%%%%%%%%%%%%%%%%%%%%%%%%%%%%%%%%%%%%%%%%%%%%%%%%%%
\section{The tunnel limit}
\label{Sec.Tunnel}

In this section we consider the simple limit of a tunnel junction. The results obtained in this limit  
allow for a direct understanding of the basic influence of the precessing spin on the current flowing through the junction
as well as the basic back-action of the (spin) current on the magnetization dynamics.

%%%%%%%%%%%%%%%%%%%%%%%%%%%%%%%5
\subsection{Charge current}
\label{SubSec.ChargeCurr}

In the tunnel limit, the current at lead $\al=R,L$ may formally be separated into normal an anomalous contributions:
\begin{subequations}
\label{CCT}
\bea
&&I_{\al}^c(t) = I_{\al,G}^c(t) + I_{\al,F}^c(t),
\label{CCTG} \\
&&I_{\al,G}^c(t) = -e \int_{-\infty}^{t} {{dt'}}\,\Big[ \la [A_{\al}^c(t), A_{\al}^{c\,\dag}(t')] \ra +\,h.\,c. \Big],
\label{Gc} \\
&&I_{\al,F}^c(t) = -e \int_{-\infty}^{t} {{dt'}}\,\Big[ \la [A_{\al}^c(t), A_{\al}^c(t')] \ra +\,h.\,c. \Big],
\label{Fc}
\eea
\end{subequations}
where $I_{\al,G}^c$ is the normal contribution to the charge current and $I_{\al,F}^c$ the anomalous one.
In the following we will compute the current at the left lead ($\al=L$). The current at the 
right lead follows from charge conservation: $I_{R}^c=-I_{L}^c$, so that in the rest of this section we
will drop the lead index. In the interaction representation, the operator: $A_{L}^c \equiv A^c$, appearing in 
Eq.~(\ref{CCT}), is defined as~\cite{MahanBook}:
\be
A^c(t) = \sum_{{\bf k},{\bf p},\sigma,\sigma'}\,c^\dag_{R,{\bf k},\sigma}(t)\,T_{\sigma,\sigma'}(t)\,c_{L,{\bf p},\sigma'}(t),
\label{Ac}
\ee
and depends on the spin- and time-dependent tunneling amplitudes which were defined in Eq.~(\ref{TAMatrix}) and below.
For a spin-singlet superconducting junction the dc charge current is time-independent~\cite{Balatsky03}
and, in what follows, we shall focus on its precession frequency dependence.

In the absence of applied bias: $I_{G}^c=0$, and only the anomalous part contributes to the total dc current.
The latter reads:
\be
I_G^c = 4e \, \Big(\, (T_0^2-T_{\parallel}^2)\,\mbox{Re}\mathcal{D}^R(0) - T_\bot^2\,\mbox{Re}\mathcal{D}^R( \Omega)\, \Big )\, \sin \vp,
\label{CCT2}
\ee
where $\vp = \chi_R - \chi_L$, the total phase difference across the junction, was gauged
out from the gap so that: $\Delta_\al = \Delta$ is real.
The charge current of Eq.~(\ref{CCT2}) depends on the reactive part of the anomalous
2-particle propagator defined (in imaginary time) as:
\be
\mathcal{D}(i\om) = \sum_{{\bf{k}},{\bf{q}}} \frac{1}{\beta} \sum_{i\eps} \, \mathcal{F}^\dag_{R}({\bf{k}},i\eps) \mathcal{F}_{L}({\bf{q}},i\eps-i\om),
\label{Ddef}
\ee
from which the retarded and advanced functions are obtained by an analytic continuation: 
$i\om \rightarrow \om \pm i\eta$ ($\eta=0^+$), respectively.
The Matsubara superconducting Green's functions are defined in the usual way~\cite{MahanBook}:
$\mathcal{G}_{\al}$ is the normal component and $\mathcal{F}_{\al}=\mathcal{F}_{\al}^\dag$ ($\Delta \in \mathbb{R}$) are the anomalous components
of the Green's function in lead $\al$. 
In the following, we will mainly work with quasi-classical Green's functions that we simply define as:
$g_{\al} = \sum_{{\bf k}} \mathcal{G}_{\al,{\bf k}}$, $f_{\al} = f_{\al}^\dag = \sum_{{\bf{k}}} \mathcal{F}_{\al,{\bf k}}$.
For non-interacting leads they read: 
\be
g_{\al}^{(0)}(i\epsilon) = - \pi \nu_N \frac{i\eps}{\sqrt{\epsilon^2 + \Delta^2}},~ f_{\al}^{(0)}(i\epsilon) = \pi \nu_N \frac{\Delta}{\sqrt{\epsilon^2 + \Delta^2}},
\label{QCGF}
\ee
where the single particle spectrum of the electrons in the leads has been linearized around the Fermi surface,
$\nu_N$ is the normal state density of states and $\Delta \equiv \Delta(T)$ is temperature-dependent. 
With the help of Eq.~(\ref{QCGF}), Eq.~(\ref{Ddef}) reads:
\be
\mathcal{D}(i\om) =  \pi^2 \nu_N^2 \Delta^2 \frac{1}{\beta } \sum_{i \nu} \frac{1}{\sqrt{\big[\nu^2 + \Delta^2 \big] \big[(\nu+\omega)^2 + \Delta^2 \big]}},
\nonum
\ee
where the integrand~\cite{Note:Analogy1} is a pure branch-cut. This is a consequence of the singular BCS density of states at the gap edges.
These gap edge singularities will affect the response function, a signature of the fact that extended states
contribute to the current. As a result, the super-current will have a non-analytic dependence on the precession frequency.
This can readily be seen from the zero temperature ($T=0$) expression of the propagator
(the reactive part being even and the dissipative part odd with respect to $x$ we only consider $x>0$ in what follows):
\bse
\label{DGF}
\bea
\mbox{Re}\mathcal{D}^R(x) &&=
\left\{
          \begin{array}{ll}
                \pi \nu_N^2 \Delta \, K\big(x\big), & \quad x < 1, \\
                \pi \nu_N^2 \frac{\Delta}{x}\,K\big(\frac{1}{x}\big), & \quad x > 1, \\
          \end{array}
\right.
\label{ReD} \\
\mbox{Im}\mathcal{D}^R(x) &&= \frac{\pi \nu_N^2 \Delta}{x+1}\,K\Big(\frac{x - 1}{x + 1}\Big)\,\Theta \big(x-1 \big),
\label{ImD}
\eea
\ese
where $x = \omega / 2\Delta$, $K$ is the complete elliptic integral of the first kind
and $\Theta$ is the Heaviside function. The elliptic integral is logarithmically singular for $\om=2\Delta$ which, by Kramers-Kronig,
is related to the finite jump of $\mbox{Im}\mathcal{D}^R$ at this frequency.
With the help of Eqs.~(\ref{CCT2}) and (\ref{ReD}) the charge current reads:
\be
I^c =
\left\{
        \begin{array}{ll}
                2e \Delta \left( (t_0^2 - t_\parallel^2)  - t_\bot^2\frac{2}{\pi}\,K\big(\frac{\Omega}{2 \Delta }\big)\right) \sin \vp,\, & \Omega < 2 \Delta, \\
                2e \Delta \left( (t_0^2 - t_\parallel^2)  - t_\bot^2\frac{4\Delta}{\pi \Omega}\,K\big(\frac{\Omega}{2 \Delta }\big)\right) \sin \vp,\, & \Omega > 2 \Delta, \\
          \end{array}
\right.
\label{ChargeCurrentTunnel}
\ee
where reduced hopping amplitudes have been introduced: $t_i = T_i / W$ ($W = 1 / \pi \nu_N$ is the band-width).

As first noticed by Kulik~\cite{Kulik66} and Bulaevskii et al.~\cite{Bulaevskii77}, for a static spin,
the results of Eq.~(\ref{ChargeCurrentTunnel}) first show that the presence of the spin in the junction reduces the current.
As will be shown by a calculation to all orders in the tunnel amplitude, in Sec.~\ref{Sec.ResultsCharge}, a crossover to a $\pi-$junction indeed takes place
when the spin-conserving, $t_\parallel$, and/or spin-flip, $t_\bot$, tunnel amplitudes become of the order of the amplitude, $t_0$, for direct tunneling.

Moreover, at zero precession frequency, the current due to $t_0$ or $t_\parallel$ and $t_\bot$ 
corresponds to the usual tunnel limit of the Josephson current for a $0-$ or $\pi-$junction, respectively.
The situation becomes more interesting when the system is driven out-of-equilibrium by the external classical source
and the spin in the junction precesses. As can be seen from Eq.~(\ref{ChargeCurrentTunnel}) the
precession frequency dependent part of the super-current is entirely carried by the spin-flip term.
In order to single out this contribution we will therefore focus, in what follows, on: 
$\delta I^{c} = I^c(\Omega) - I^c(0)$.

\begin{figure}
\includegraphics[width=5.5cm,height=2.5cm]{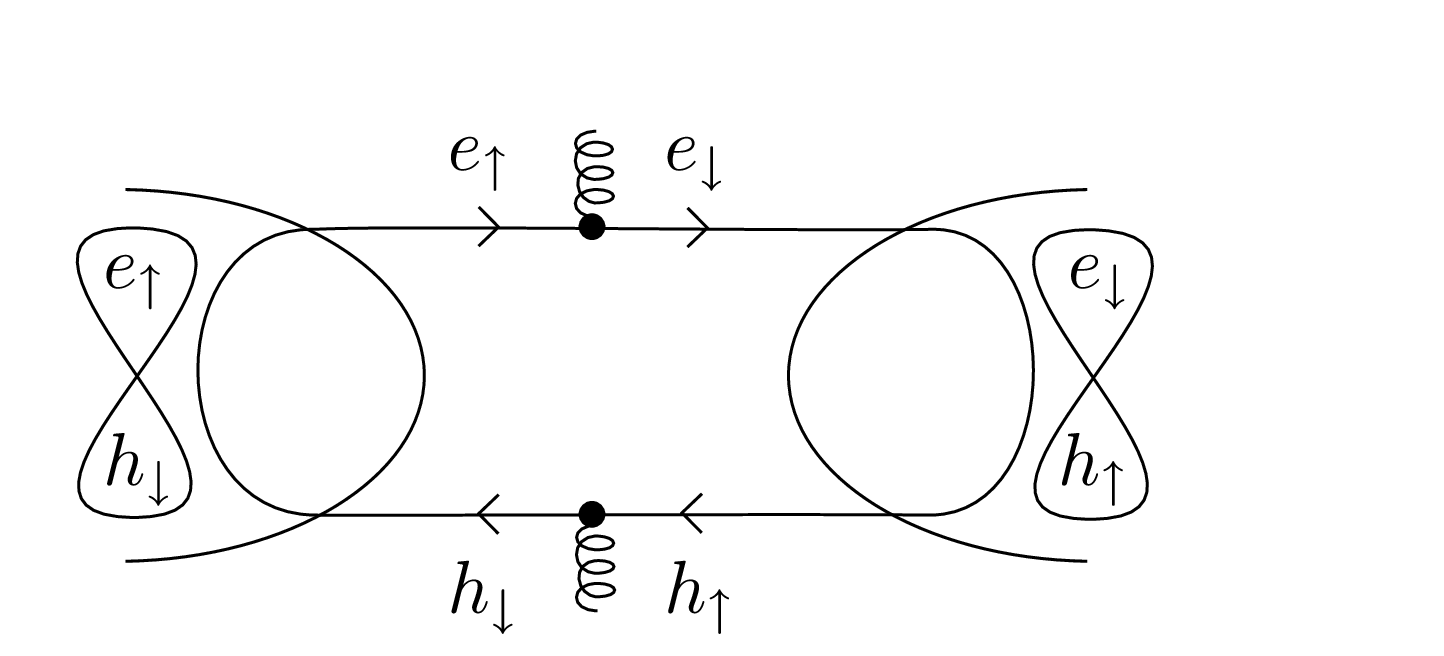}
\caption{ \label{Fig:ChargeCurrent} Lowest order diagram for the Josephson current 
showing the transfer of a Cooper pair from one lead to the other. 
The curvy lines correspond to the absorption/emission of a quantum of precession ($\hbar\Omega$)
while $e_\sigma$ and $h_\sigma$ denote particles and holes of spin $\sigma$, respectively.
This diagram shows that the Josephson current is of second order in $\Omega / 2\Delta$.}
\end{figure}

At low frequencies, $\Omega < 2 \Delta$, the charge current increases with increasing precession frequency (recall that $\Omega>0$).
In particular, in the adiabatic limit ($\Omega \ll 2 \Delta)$, the supercurrent arising from spin-flips reads:
\be
\delta I^c \approx -\frac{e t_\bot^2 \Delta}{2}\,\left(\frac{\Omega}{2 \Delta} \right)^2\,\sin \vp,~~\Omega \ll 2 \Delta,
\label{ChargeCurrentTunnelSmallOmega}
\ee
where the lowest order term is of second order in the precession frequency.
This comes from the two-particle nature of the Josephson current, see Fig.~\ref{Fig:ChargeCurrent}.
Indeed, in order to transfer a Cooper pair from one lead to another, an Andreev spin-down/up electron together with 
its retro-reflected spin-up/down hole will flip their spins absorbing/emitting a quantum of precession. 
The spin-up band is therefore separated from the
spin-down one by an energy interval equal to $\Omega$. As will be shown in more details in Sec.~\ref{SubSec.UT},
$\Omega/2$ corresponds to an effective $z-$directed magnetic field applied to the leads. This agrees with
interpreting $\Omega$ as a Zeeman splitting between the spin-bands, see Fig.~\ref{Fig.Junction}b.

On the other hand, at high frequencies, $\Omega \gg 2 \Delta$, the supercurrent arising from spin-flip processes decreases:
\be
\delta I^c \approx - 2et_\bot \Delta\,\frac{2 \Delta}{\Omega}\,\sin\vp,~~\Omega \gg 2 \Delta.
\label{ChargeCurrentTunnelLargeOmega}
\ee

Finally, in the intermediate range, a resonance appears at $\Omega = 2 \Delta$ where the spin-flip super-current diverges logarithmically:
\be
\delta I^c \approx \frac{2e t_\bot^2}{\pi }~\Delta~\log \bigg( |\frac{\Omega}{2 \Delta} - 1| \bigg)\,\sin\vp,~~\Omega \approx 2\Delta.
\label{ChargeCurrentTunnelResonance}
\ee
Following the discussion at the level of the 2-particle propagator, such a singularity translates the fact that, 
when the system is driven at the frequency $2\Delta$ by the external source, an infinite number
of extended states are available to carry the current~\cite{Note:Analogy2}. 

The regime in which Eqs.~(\ref{ChargeCurrentTunnelLargeOmega}) and (\ref{ChargeCurrentTunnelResonance}) 
are valid requires high precession frequencies, at least of the order of the amplitude of the superconducting gap.
It may be difficult to reach, by an order of magnitude, in practice and would lead to a heating up of the system
as witnessed by the fact that transport in this regime is dominated by extended states. 
In the following, we will therefore be mainly interested in the adiabatic regime where: $\Omega \ll 2 \Delta$,
and the electronic degrees of freedom (of time scale $\hbar /\Delta$) adjust instantaneously to the 
magnetic ones (of time scale $2\pi/\Omega$).

%%%%%%%%%%%%%%%%%%%%%%%%%
\subsection{Spin current}
\label{SubSec.SpinCurr}

The spin current at a given lead~\cite{Note:SpinCurrent} can be calculated along the same lines as the charge current (recall that $\hbar=1$ unless specified):
\bse
\label{SCT}
\bea
&&{\bf I}^s(t) = {\bf I}_{G}^s(t) + {\bf I}_{F}^s(t),
\label{SCTG} \\
&&{\bf I}_{G}^s(t) = \frac{1}{2}\,\int_{-\infty}^{t} {{dt'}}\,\Big[ \la [{\bf A}^s(t), A^{c\,\dag}(t')] \ra +\,h.\,c. \Big],
\label{NsC} \\
&&{\bf I}_{F}^s(t) = \frac{1}{2}\,\int_{-\infty}^{t} {{dt'}}\,\Big[ \la [{\bf A}^s(t), A^c(t')] \ra +\,h.\,c. \Big],
\label{AsC}
\eea
\ese
where ${\bf I}_{G}^s$ is the normal contribution to the spin current, ${\bf I}_{F}^s$ the anomalous one.
Contrary to the case of the charge current, there is no conservation law related to the spin of the
electrons in the leads. Rather than transferring spins from one lead to the other, the precessing single-molecule magnet
ejects spins in a symmetric way in the two leads (${\bf I}_R^s = {\bf I}_L^s$), see Ref.~\onlinecite{TserkovnyakRMP05} where this point was emphasized. 
Therefore, there is no global spin current flowing through the junction.
In Eq.~(\ref{SCT}) the operator $A^c$ is the one which was defined in Eq.~(\ref{Ac}) 
and, still in the interaction representation, the operator ${\bf A}^s$ is defined as:
\be
{\bf A}^s(t) = \sum_{{\bf k},{\bf p},\sigma,\sigma',\sigma''}\,c^\dag_{R,{\bf k},\sigma}(t)\,\vec{\sigma}_{\sigma,\sigma'}\,T_{\sigma',\sigma''}(t)\,c_{L,{\bf p},\sigma''}(t).
\label{As}
\ee

Both the normal and anomalous parts contribute to the total dc spin current which reads:
\bse
\label{SCT2}
\bea
I_z^s =&& 2T_\bot^2 \Big[ \mbox{Im}\mathcal{D}^R(\Omega) \cos \vp - \mbox{Im}\mathcal{Q}^R(\Omega) \Big],
\label{SCTz} \\
I_x^s =&&2T_\parallel T_\bot \Big[\mbox{Im}\mathcal{Q}^R(\Omega) \cos(\Omega t) -
\label{SCTx} \\
&& \qquad \qquad - \big( \mbox{Re}\mathcal{Q}^R(\Omega) - \mbox{Re}\mathcal{Q}^R(0) \big) \sin (\Omega t) \Big] +
\nonum \\
+&&2T_0 T_\bot \Big[ \mbox{Im}\mathcal{D}^R(\Omega) \sin(\Omega t) +
\nonum \\
&& \qquad \qquad + \big(\mbox{Re}\mathcal{D}^R(\Omega) - \mbox{Re}\mathcal{D}^R(0) \big) \cos (\Omega t) \Big] \sin \vp +
\nonum \\
+&&2T_\parallel T_\bot \Big[ \mbox{Im}\mathcal{D}^R(\Omega) \cos(\Omega t) -
\nonum \\
&& \qquad \qquad - \big( \mbox{Re}\mathcal{D}^R(\Omega) - \mbox{Re}\mathcal{D}^R(0) \big) \sin (\Omega t) \Big] \cos \vp,
\nonum
\eea
\ese
and the $y-$component is derived from the $x-$component with the help of the substitution:
$\Omega t \rightarrow \Omega t - \pi /2$, in $I_x^s$. This general result shows that the precession
of the spin in the junction transfers spins into the adjacent leads (see below for the difference between normal
and superconducting leads).
Moreover, Eq.~(\ref{SCTx}) shows that, contrary to the charge current, the spin current is circularly polarized in the $xy-$plane 
and rotates in time at the precession frequency of the spin localized in the junction.
Finally, the expression of the spin current involves not
only the reactive part of the anomalous propagator but also its dissipative part.
This is related to the fact that the spin current arises from breaking singlet pairs. Concomitantly,
the normal part also contributes to the spin current and is associated with $\mathcal{Q}$, the normal 2-particle propagator,
defined (in imaginary time) as:
\be
\mathcal{Q}(i\om) = \sum_{{\bf{k}},{\bf{q}}} \frac{1}{\beta} \sum_{i\eps}\,\mathcal{G}_{R}({\bf{k}},i\eps-i\om) \mathcal{G}_{L}({\bf{q}},i\eps),
\label{Qdef}
\ee
where $\mathcal{G}$ is the normal component of the superconducting Green's function and the 
expression of the related non-interacting quasiclassical function has been given in Eq.~(\ref{QCGF}).
With the help of this equation, and at zero temperature, a direct computation of Eq.~(\ref{Qdef}) yields ($x>0$):
\bse
\label{QGF}
\bea
\mbox{Re}\mathcal{Q}^R(x) &&= \mbox{Re}\mathcal{Q}^R(0) - \pi \nu_N^2 \Delta\,q(x),
\label{ReQ} \\
\nonum \\
\mbox{Im}\mathcal{Q}^R(x) &&= -\frac{2\pi \nu_N^2 \Delta}{1+x}\,\Big[ K\Big(\frac{x - 1}{x + 1}\Big) +
\label{ImQ} \\
&& \quad + \quad \big(x^2-1 \big) E\Big(\frac{x - 1}{x + 1}\Big)\,\Big]\,\Theta\big(x-1\big),
\nonum
\eea
\ese
where $x = \omega / 2\Delta$, $K$ and $E$ are the complete elliptic integrals of the first and second kinds, respectively,
and $\Theta$ is the Heaviside function. 
Notice that, in Eq.~(\ref{ReQ}), the combination $\mbox{Re}\mathcal{Q}^R(x)- \mbox{Re}\mathcal{Q}^R(0)$
eliminates an ultra-violet singularity in the propagator. The remaining function $q(x)$ is smooth and $q(x) \rightarrow 0$ in the limit $x\rightarrow \infty$ 
(i.e where $\Delta \rightarrow 0$), while $q(x) \rightarrow 3\pi x^2/8$ in the limit $x \rightarrow 0$. Because the imaginary part has a finite
jump at $x=1$, the real part diverges logarithmically at this point. The fact that the normal component enters the expression of the spin current
allows us to consider both cases of normal and superconducting leads.

%%%%%%%%%%%%%%%%%%%%%%%%%%%%%
\subsubsection{Normal leads}

\begin{figure}
\includegraphics[width=5.5cm,height=2.5cm]{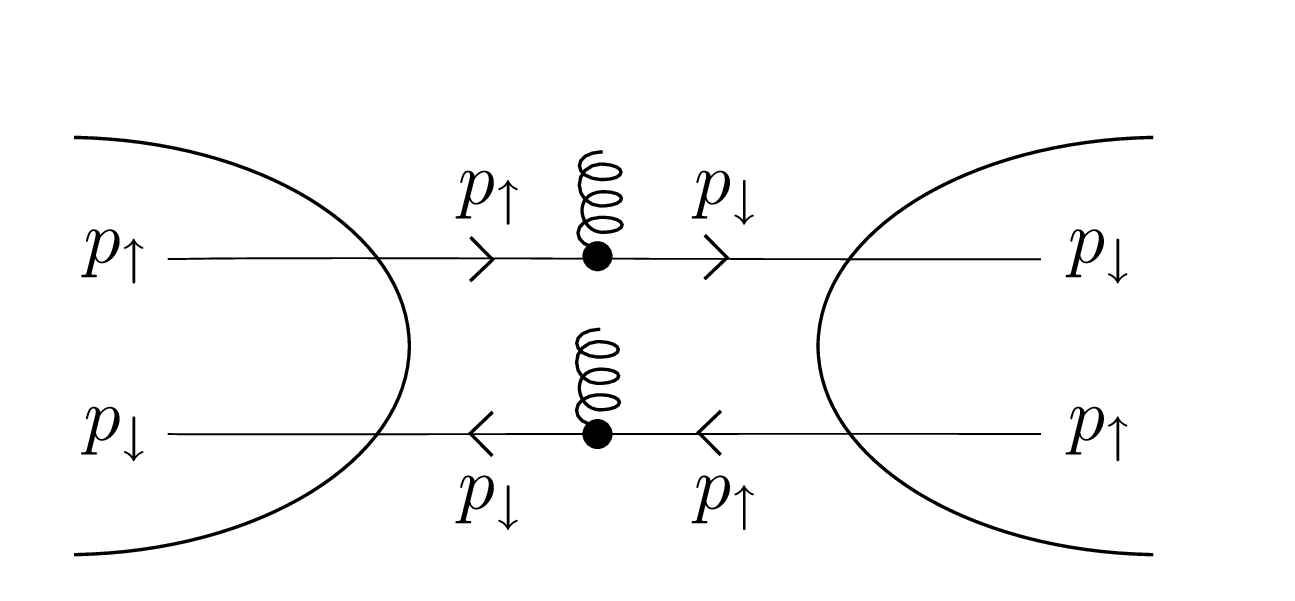}
\caption{ \label{Fig:NormalSpinCurrent} Lowest order diagram for the normal spin-current showing the transfer of quasi-particles of 
spin $\sigma$, denoted as $p_\sigma$. The curvy lines correspond to the emission/absorption of a quantum of precession.
This diagram shows that the normal spin current is of first order in $\Omega/2\Delta$ and that it is symmetric between the two leads.}
\end{figure}

Being particularly simple, the case of normal leads where $\Delta=0$ is worth examining.
In this case, the $T=0$ spin-current arises solely from quasi-particles and reads:
\bse
\label{SCTM}
\bea
I_{z}^s =&& \frac{2t_\bot^2}{\pi}\Omega,
\label{SCTzM} \\
I_{x}^s =&& -\frac{2t_\parallel t_\bot}{\pi} \Omega \cos(\Omega t)
\label{SCTxM} \\
I_{y}^s =&& -\frac{2t_\parallel t_\bot}{\pi} \Omega \sin(\Omega t).
\label{SCTyM}
\eea
\ese
This equation can be written in a more compact form as:
\be
{\bf I}^s = \frac{2 t_S^2}{\pi S^2}\,{\bf S} \times \partial_t {\bf S},
\label{SCTM_vec}
\ee
where Eq.~(\ref{LLE}), (\ref{SpinVec}) and (\ref{TA1}) have been used
and the reduced hopping amplitude through the spin has been introduced: $t_S = T_S / W$ ($W = 1 / \pi \nu_N$ is the band-width).
Eqs.~(\ref{SCTxM}) and (\ref{SCTyM}) explicitly show the circular polarization
of the spin current in the $xy-$plane. The $x-$ and $y-$components of the average number of spins:
\be
\la I_{j}^s \ra = \int_0^{\frac{2\pi}{\Omega}} \frac{dt}{2\pi}\,I_{j}^s(t),
\label{Average_SC}
\ee
where $j=x,\,y$, emitted in the leads is zero.
On the other hand, the $z-$component of the spin current is time independent and
has a linear dependence on the precession frequency. 
From Eq.~(\ref{Average_SC}), we see that $\la I_{z}^s \ra = 4t_S^2$. 
As schematically represented on Fig.~\ref{Fig:NormalSpinCurrent}, this transfer
of spin is symmetric between the two leads and we have:  $I^c_\up = - I^c_\dn$, 
which confirms the fact that there is a non-zero 
$z-$component of the spin current and no associated charge current.
These arguments show that the normal junction is a pure spin pump
where the two periodic parameters are the projections of the in-plane 
spin along the $x$ and $y$ axis, see Ref.~[\onlinecite{Wang03}] in the case of a quantum dot
as well as Ref.~[\onlinecite{TserkovnyakRMP05}] where similar 
results are reviewed in the case of an F/N junction.
The linearity of the spin current with respect to the precession frequency
is related to the single-particle nature of the tunneling quasi-particles.

%

%%%%%%%%%%%%%%%%%%%%%5
\subsubsection{Superconducting leads}

The N/SMM/N case considered in the previous paragraph may formally correspond to the S/SMM/S case
in the limit where: $\Omega \gg 2 \Delta \rightarrow 0$, and extended (quasi-particle like) states carry all the current. 
In the opposite regime of low-precession frequencies, $\Omega \ll 2\Delta$, 
the dissipative part of the propagator vanishes suggesting that the (spin-singlet) condensate plays a dominant role. 
Nevertheless, we still get a non-zero spin current arising from the reactive part of the normal as well as
anomalous propagators. 
With the help of Eqs.~(\ref{SCT2}) and (\ref{QGF}) the $T=0$ dc spin current reads:
\bse
\label{SCTSC}
\bea
I_z^s =&& 0,
\label{SCTzCS} \\
I_x^s =&& \frac{t_0 t_\bot \Delta}{4} \Big (\frac{\Omega}{2\Delta} \Big)^2 \cos(\Omega t) \sin \vp +
\label{SCTxSC} \\
&& + \frac{t_\parallel t_\bot \Delta}{4} \Big (\frac{\Omega}{2\Delta} \Big)^2 \Big ( 3 - \cos \vp \Big) \sin(\Omega t),
\nonum \\
I_y^s =&& \frac{t_0 t_\bot \Delta}{4} \Big (\frac{\Omega}{2\Delta} \Big)^2 \sin(\Omega t) \sin \vp -
\label{SCTySC} \\
&& - \frac{t_\parallel t_\bot \Delta}{4} \Big (\frac{\Omega}{2\Delta} \Big)^2 \Big ( 3 - \cos \vp \Big) \cos(\Omega t).
\eea
\ese
These equations show that, in analogy with the charge superconducting current, 
the spin-current depends on the superconducting phase difference and 
has a quadratic dependence on the precession frequency at low frequencies. 
This suggests that this spin-current is of superconducting nature, i.e. related to
the tunneling of pairs of particles. 
Its quadratic dependence on the precession frequency implies that it
is smaller than the quasi-particle spin-current by a factor of the order of $\Omega/\Delta$, i.e. at least one
order of magnitude in the adiabatic regime.
Actually, because of the singlet nature of the pairs
the average number of spins emitted in the leads: $\la I_{j}^s \ra = 0$ for $j=x,\,y,\,z$, as can be seen
with the help of Eqs.~(\ref{Average_SC}) and (\ref{SCTSC}).
Moreover, its phase dependence shows that the spin-current in the case of
a $\pi-$junction is twice larger than in the case of a $0-$junction.
In these two cases, the expression of the spin current slightly simplifies
and its time-dependence is seen to be out of phase with respect to the one of the 
quasi-particle spin current of Eq.~(\ref{SCTM}) by $90^\circ$.
It can be written in a more compact form as:
\be
{\bf I}^s(\vp) = \mathcal{N}_\vp \frac{t_S^2}{4S^2}\,\frac{\gamma {\bf S}.{\bf H}_{eff}}{2\Delta}\,\gamma {\bf S} \times {\bf H}_{eff},
\label{SCTSC_vec}
\ee
where $\mathcal{N}_0=1$ and $\mathcal{N}_\pi = 2$, for $\vp=0$ and $\vp=\pi$, respectively.

%%%%%%%%%%%%
\subsection{Gilbert damping}
\label{SubSec.DampingT}

In the last paragraph, we have shown that a precessing molecular spin transfers a spin current
to its environment (either metallic or superconducting leads), even if no voltage bias is applied.
Following the Introduction, the emission of a spin current by the precessing spin in the junction implies in turn that it is loosing angular momentum.
The corresponding spin transfer torque~\cite{Slonczewski96} is given by:
\be
\vec{\tau}^s = {\bf I}_L^s - {\bf I}_R^s = 2{\bf I}_L^s,
\label{STT}
\ee
and has to be added as a source of damping into the classical equations of motion of the precessing spin:
\be
\partial_t {\bf S} = -\gamma {\bf S} \times {\bf H}_{eff} + \vec{\tau}^s.
\label{LLGSE}
\ee
If the external source does not compensate for this loss, the damping of the precession will lead to a complete alignment of the spin with
the applied magnetic field.  

In the case of normal leads, we see from Eq.~(\ref{SCTM_vec}) that the spin transfer 
torque corresponds to a Gilbert term leading indeed to a damping of the precession. 
The corresponding Gilbert constant, at $T=0$, has the following expression:
\be
\al_S = \frac{4 t_S^2}{\pi},
\label{GibertC_TM}
\ee
and only depends on the transparency of the junction.
On the other hand, in the case of superconducting leads with $\vp=0$ or $\pi$, Eq.~(\ref{SCTSC_vec})
shows that the damping is suppressed and 
that the superconducting spin-current only leads to a shift of the precession frequency:
\be
\Omega' = \Omega\Big[1 - \frac{\zeta_S}{S} \Big],
\ee
where, at $T=0$, $\zeta_S$ reads:
\be
\zeta_S = \mathcal{N}_\vp \frac{t_S^2}{2}\,\frac{\Omega}{2\Delta}\,\cos \theta.
\label{FreqShift_TSC}
\ee
Eq.~(\ref{FreqShift_TSC}) shows that the shift of the precession frequency depends on the precession frequency 
itself as well as on the tilt angle of the precessing spin and the superconducting phase.

In agreement with the results of the previous paragraph, we arrive at the conclusion 
that a spin precessing between superconducting leads is only weakly perturbed by its environment
with respect to a spin precessing between normal metallic leads.
Uniting these two descriptions within a common framework, based on the two-fluid model, 
this dramatic reduction of the dissipation is natural since quasi-particles are frozen deep in the superconducting state
and that the main current-carriers are spin-singlets.

%%%%%%%%%%%%%%%%%%%%%%%%%%%%%%%%%%%%%%%%%%%%%%%%%%%%%%%%
\section{General transport equations}
\label{Sec.GeneralTransport}

In order to go beyond the tunnel limit, we present in this section a unitary transformation which relates the time-dependent problem 
to a time-independent one with non-colinear magnetization.
This will allow us to reduce the Dyson equations for the dressed Green's function to a set of algebraic equations. 
In particular, an explicit expression for the Keldysh Green's function can be obtained from which the 
charge and spin currents may easily be computed, at least numerically.

%%%%%%%%%%%%%%%%%%%%%%%%%%%%%%%%%%
\subsection{Unitary transformation}
\label{SubSec.UT}

In order to take into account of both spin-flip processes and superconductivity in a compact way, we consider a space
which is the tensor product of spin and Gorkov-Nambu spaces, denoted as spin$\otimes$Nambu in what follows. 
In this $4-$dimensional space, we introduce the following bispinor: 
\be
\hat{\psi}_{\al,{\bf k}}^\dag = (c^\dag_{\al,{\bf k},\up},c^\dag_{\al,{\bf k},\dn},c_{\al,-{\bf k},\up},c_{\al,-{\bf k},\dn}),
\label{SpinorSpinNambu}
\ee
where the hat denotes operators acting in spin$\otimes$Nambu space. 
For example, the spin and Nambu Pauli matrices, $\hat{\sigma}$ and $\hat{\tau}$, respectively, acting on
the bispinor of Eq.~(\ref{SpinorSpinNambu}) are defined as:
\bea
\hat{\sigma}_i = \left(
\begin{array}{cccc}
\sigma_i & 0 \\
0 & -\sigma_i
\end{array}
\right), \quad
\hat{\tau}_i = \left(
\begin{array}{cccc}
0 & \sigma_i \\
-\sigma_i & 0
\end{array}
\right),
\label{PauliSN}
\eea   
where $i=\,0,\,1,\,2,\,3$ and $\sigma_0={\bf 1}_2$ while the $1,\,2,\,3$ 
components of $\sigma_i$ correspond to the usual $x,\,y,\,z$ matrices, respectively.

With these conventions in hand, the action associated with the Hamiltonian in Eq.~(\ref{Hamiltonian}) 
and evaluated along an arbitrary (for the moment) $\mathcal{C}$ contour reads:
\be
S[\hat{\psi}, \hat{\psi}^\dag] = \int_{\mathcal{C}} {{dt}}~\Big[ \sum_{\al,{\bf k}} \hat{\psi}_{\al,{\bf k}}^\dag(t)\,i \partial_t \hat{\psi}_{\al,{\bf k}}(t) - H(t) \Big].
\label{Action}
\ee
In Eq.~(\ref{Action}) the Hamiltonian reads:
\bea
H(t) =&& \sum_{\al, {\bf k}} \hat{\psi}_{\al,{\bf k}}^\dag(t) \hat{H}_{\al,{\bf k}} \hat{\psi}_{\al,{\bf k}}(t) +
\nonum \\
&&+ \sum_{{\bf k},{\bf q}} \left( \hat{\psi}_{R,{\bf k}}^\dag(t) \hat{T}(t) \hat{\psi}_{L,{\bf q}}(t) +\,h.\,c.\, \right),
\nonum
\eea
where $\hat{H}_{\al,{\bf k}}$ is the matrix Hamiltonian of the BCS leads:
\be
\hat{H}_{\al, {\bf k}}  = \left(
\begin{array}{cccc}
\xi_{\bf k} & i \sigma_y \Delta_\al \\
-i\sigma_y \Delta_\al^* & -\xi_{\bf k} \end{array} 
\right), 
\label{BCSM}
\ee
and $\hat{T}(t)$ is the matrix tunnel amplitude:
\be
\hat{T}(t)  = \left(
\begin{array}{cccc}
T(t) & 0 \\
0 & -T^*(t) \end{array} 
\right).
\label{TASN}
\ee
where $T(t)=T_0 + T_\parallel \sigma_z + T_\bot \sigma_x e^{-i \sigma_z \Omega t}$, see Eq.~(\ref{TAMatrix}). 
Defining the time-dependent unitary matrix as:
\be
\hat{\mathcal{U}}(t)  = 
\left(
\begin{array}{cccc}
e^{i \sigma_z \frac{\Omega}{2}t} & 0 \\
0 & e^{-i \sigma_z \frac{\Omega}{2}t} 
\end{array}
\right)
=
e^{i \hat{\sigma}_3 \frac{\Omega}{2}t}, 
\label{UTM}
\ee
where the $4\times4$ Pauli spin matrix $\hat{\sigma}_3$ was defined in Eq.~(\ref{PauliSN}), the field operators transform as:
\be
\hat{\tilde{\psi}}_{\al,{\bf k}}^\dag(t) = \hat{\psi}_{\al,{\bf k}}^\dag (t)\,\hat{\mathcal{U}}(t), 
\qquad 
\hat{\tilde{\psi}}_{\al,{\bf k}}(t) = \hat{\mathcal{U}}^\dagger(t) \hat{\psi}_{\al,{\bf k}}(t).
\nonum
\ee
It is then straightforward to check that this unitary transformation leaves the BCS Hamiltonian, see Eq.~(\ref{BCSM}), invariant
while the transmission matrix, see Eq.~(\ref{TASN}), becomes time-independent:
\be
\hat{\tilde{T}} = \hat{\mathcal{U}}^\dagger(t) \hat{T}(t) \hat{\mathcal{U}}(t)  \equiv  \hat{T}(\Omega = 0).
\label{TunnelAmplitudeRotated}
\ee
The unitary-transformed action therefore corresponds to the one of a static tilted spin with an effective magnetic field, $h_z$, acting on the leads~\cite{Note:GT}:
\be
S[\hat{\tilde{\psi}}, \hat{\tilde{\psi}}^\dag] = \int_{\mathcal{C}} {{dt}}\,
\big[ \sum_{\al,{\bf k}} \hat{\tilde{\psi}}_{\al,{\bf k}}^\dag~(i \partial_t - h_z \hat{\sigma}_z) \hat{\tilde{\psi}}_{\al,{\bf k}} - \tilde{H} \big].
\label{ActionGT}
\ee
In Eq.~(\ref{ActionGT}) the transformed Hamiltonian is time-independent and equals the Hamiltonian of Eq.~(\ref{Hamiltonian}) with zero precession frequency. 
The latter has been absorbed in the effective magnetic field:
\be
h_z = \frac{\Omega}{2},
\label{hz}
\ee
acting on the superconducting leads.
From Eq.~(\ref{hz}), we recover the fact, already noticed in Sec.~\ref{Sec.Tunnel}, that the precession frequency corresponds to an effective Zeeman splitting energy. 

Gauging out the precession frequency in the transmission term is equivalent to the usual transformation of 
going from the laboratory frame, where the spin is precessing, to the rotating frame where the spin appears to be static, see Fig.~\ref{Fig.Junction}b. 
The transformation is related to the peculiar harmonic dependence of the spin-flip transmission amplitude on time, 
see Eq.~(\ref{TAMatrix}), and implies that the problem has a {\it steady-state} solution. 
It allows us to replace the time-dependent problem by a time-independent one at the expense of 
dealing with a static problem with non-colinear magnetization. 
Indeed, the effective magnetic field in the leads, $h_z$, is oriented along the $z-$axis whereas 
the spin in the junction is tilted in an arbitrary direction fixing the amplitudes $T_\parallel$ and $T_\bot$. 
However, the system is out-of-equilibrium no matter what frame one considers. 
In particular, in the rotating frame, the out-of-equilibrium nature of the problem is related to the non-trivial 
action of the effective magnetic field on the Green's functions of the leads, i.e. with respect to what a usual magnetic field would do. 
We will show this in the next paragraph and end this one with a simple relation between the Green's functions in both frames.

The Green's functions are matrices in spin$\otimes$Nambu space and are defined with the help of:
\be
\hat{g}_{\al,\beta}(t,t') = -i \la \mathcal{T}_\mathcal{C} \hat{\psi}_\al(t) \hat{\psi}_\beta^\dag(t') \ra,
\ee
where $\mathcal{T}_\mathcal{C}$ is the time-ordering operator, for the Grassmann fields, along the $\mathcal{C}-$contour
and $\al,\beta=R,L$ are lead indices (see Eq.~(\ref{MSGFUT}) with $h_z=0$ for the form of the matrix Green's function). 
The unitary transformation then allows the following correspondence between the Green's functions in both frames:
\be
\hat{\tilde{g}}_{\al,\beta}(t,t') = \hat{\mathcal{U}}^\dagger(t) \hat{g}_{\al,\beta}(t,t') \hat{\mathcal{U}}(t'). 
\label{UTGF}
\ee
% 

%%%%%%%%%%%
\subsection{Consequences for the leads}
\label{SubSec.Leads}

Before using the unitary transformation to compute the transport properties of the junction 
we start by focusing on its consequences at the level of the leads (Green's functions, gap equation and thermodynamic quantities).

\subsubsection{Green's functions}
 
To deal with the steady-state out-of-equilibrium situation we are considering we will use the standard Keldysh technique, 
see e.g. Refs.~[\onlinecite{KeldyshReviews}] for reviews, and compute the Green's functions on a closed time or Keldysh, $\mathcal{C}_K$, 
contour. This procedure leads to a doubling of the degrees of freedom by introducing $\psi_+$ fields on the upper, forward,
branch and $\psi_-$ fields on the lower, backward, branch. Therefore, besides the usual retarded, $\hat{g}^R$, and advanced, $\hat{g}^A$,
Green's functions defined in Sec.~\ref{Sec.Tunnel}, there will be an additional component:
\be
\hat{g}_{\al,\beta}^-(t,t') = -i \la \mathcal{T}_{\mathcal{C}_K} \psi_{+,\al}(t) \psi_{-,\beta}^\dag(t') \ra,
\label{LKGF}
\ee
which is the lesser Keldysh function. In what follows, we will work with $\hat{g}^-$, 
keeping in mind that it is related to the standard Keldysh Green's function, $\hat{g}^K$,
with the help of the following identity: $\hat{g}^K = \hat{g}^R - \hat{g}^A + 2 \hat{g}^-$.
%
%\be
%\hat{g}^K = \hat{g}^R - \hat{g}^A + 2 \hat{g}^-.
%\label{KeldyshVsMinus}
%\ee
%

The unitary transformation of Eq.~(\ref{UTGF}) will therefore not only affect the spectral Green's functions (retarded and advanced)
but also the Keldysh Green's function which contains information about the occupation of states and hence the out-of-equilibrium 
properties of the system. For non-interacting leads the lesser component reads:
\be
\hat{g}_{\al,\beta}^{(0)-}(t,t') = -\delta_{\al,\beta}\,n_F^\al \circ \left(\hat{g}_{\al}^{(0) R} - \hat{g}_\al^{(0) A} \right)(t,t'),
\label{freeLKGF}
\ee
where $n_F^\al$ is the Fermi occupation function in lead $\al$, the retarded and advanced functions were defined in Eq.~(\ref{QCGF})  
and the symbol $\circ$ implies time convolution.

In spin$\otimes$Nambu space, the result of the unitary transformation on the spectral  quasiclassical 
matrix Green's function reads:
\be
\hat{\tilde{g}}_{\al}(i\om) =  \left(
\begin{array}{cccc}
g_\al(i\om_+) & 0 & 0& -f_\al(i\om_+) \\
0 & g_\al(i\om_-) & f_\al(i\om_-) & 0 \\
0 & f_\al^\dag(i\om_-) & g_\al(i\om_-) & 0 \\
-f_\al^\dag(i\om_+) & 0 & 0 & g_\al(i\om_+) \end{array}
\right),
\label{MSGFUT}
\ee
where $i\om_{\pm}=i\om \pm h_z$. At the level of the advanced and retarded Green's functions,
Eq.~(\ref{MSGFUT}) shows that the effective magnetic field acts in a way similar to a usual magnetic field.

The unusual nature of the present magnetic field appears from the effect of the unitary transformation on the Keldysh component of the Green's function:
\be
\hat{\tilde{g}}_{\al,\beta}^{(0)-} (t,t') = -\delta_{\al,\beta}\,\hat{\tilde{n}}_F^\al \circ \left(\hat{\tilde{g}}_\al^{(0)R} - \hat{\tilde{g}}_\al^{(0)A} \right)(t,t'),
\label{KeldyshGT}
\ee
and equivalently for the function $\hat{\tilde{g}}^K$. 
Eq.~(\ref{KeldyshGT}) shows that, contrary to what happens with a usual magnetic field, 
the occupation function is also affected by the transformation. 
The latter becomes a matrix in spin$\otimes$Nambu space and reads:
\bea
\hat{\tilde{n}}_F^\al (t)
= 
n_F^\al(t)\,\hat{\mathcal{U}}^\dag(t) = n_F^\al(t)\, e^{-i \hat{\sigma}_z \frac{\Omega}{2} t}.
\label{nfRotated}
\eea
Though the system is in a steady state, as it has been shown in Sec.~\ref{SubSec.UT}, 
the very fact that the time-dependent unitary transformation modifies the occupation 
functions reflects the out-of-equilibrium nature of the problem. 
For example, we recover the fact that the spin-accumulation (the difference in chemical potential between
spin-up and spin-down particles):
\be
\vec{\mu}^s = \int_0^\infty {{d\eps}}~\Tr \left[ \hat{\vec{\sigma}} \hat{\tilde{n}}_F(\eps) \right],
\label{SpinAcc}
\ee
is non-zero only in the $z-$direction and, at $T=0$, has an amplitude: $\mu_z^s(T=0) = \Omega$, corresponding to 
the splitting between spin-up and spin-down bands, see Fig.~\ref{Fig.Junction}b. 

\subsubsection{Density of states}

In the rotated frame the BCS density of states (DOS) of a given lead subject to the effective magnetic field reads:
\be
\tilde{\rho}_{\up,\dn}(\eps) = \nu_N~\frac{|\eps \mp h_z|}{\sqrt{(\eps \mp h_z)^2 - \Delta^2}}.
\label{rho_TM}
\ee
Such a DOS is similar to the one determined by Tedrow and Meservey~\cite{TedrowMeservey71}
and has been schematically represented on Fig.~\ref{Fig.Junction}b. 
In particular an effective (precession-frequency dependent) gap of 
$2\Delta - 2 h_z$ between spin up and down quasi-particle bands appears. 
For $\Omega = 2 \Delta$ the spin up and down quasi-particle peaks overlap. 
This is another manifestation of the fact that the nonanalyticity of the current at $\Omega = 2 \Delta$, 
see Eq.~(\ref{ChargeCurrentTunnelResonance}), originates from extended states.

\subsubsection{Gap equation and thermodynamic properties}

Another proof of the unusual nature of the effective magnetic field is that it 
has no effect on the superconducting gap $\Delta$. This can be seen from the gap equation,
$\Delta_\al = V_0\,f_\al(\tau=0)$ where $V_0$ is the effective attractive coupling, 
which involves the equal time component of the anomalous part of the quasiclassical BCS Green's function. 
The gap equation is therefore invariant under the unitary transformation. 
The effective magnetic field we are considering does not affect superconductivity in the leads in a way similar to a usual magnetic field.

More generally, the thermodynamic properties of the leads involve energy integrations over products of the DOS and the occupation function. 
Changing variables of integration therefore eliminates the effective magnetic field and makes these quantities invariant under the unitary transformation.

%%%%%%%%%%%%%%%%%%%
\subsection{Transport equations}
\label{SubSec.TranspEq}

With the help of the unitary transformation of Sec.~\ref{SubSec.UT}, we derive in this paragraph general transport formulas valid for a junction
of arbitrary transparency.
To this end, we will follow the standard approach of computing the current from the Keldysh Green's function, see
Refs.~[\onlinecite{Caroli71,Cuevas96}] for some references on the subject. 

In the compact spin$\otimes$Nambu space the charge and spin current operators are defined as:
\bea
&&\hat{I}^c(t) = - i e \sum_{{\bf k},{\bf q}}~\hat{\psi}_{R,{\bf k}}^\dag(t) \hat{\sigma}_0 \hat{T}(t) \hat{\psi}_{L,{\bf q}}(t),
\nonum \\
&&\hat{{\bf I}}^s(t) = \frac{i}{2} \sum_{{\bf k},{\bf q}}~\hat{\psi}_{R,{\bf k}}^\dag(t) \hat{\vec{\sigma}}^s \hat{T}(t)\hat{\psi}_{L,{\bf q}}(t)
\eea
where $\hat{\vec{\sigma}}^s = (\hat{\sigma}_1,\, \hat{\sigma}_2^s,\, \hat{\sigma}_3)$
give the three components of the spin current,
the $\hat{\sigma}_i$ ($i=0,\,1,\,2,\,3$) were defined in Eq.~(\ref{PauliSN})
and $\hat{\sigma}_2^s = \hat{\sigma}_2 \hat{\sigma}_0$.
Taking the average of the current yields:
\bse
\label{CSN}
\bea
&&I^c(t) = e\,\Tr \big[ \hat{\sigma}_0 \hat{T}(t) \hat{g}^-_{LR}(t,t) \big],
\label{CCSN} \\
&&{\bf I}^s(t) = -\frac{1}{2}\,\Tr \big[ \hat{\vec{\sigma}}^s \hat{T}(t) \hat{g}^-_{LR}(t,t) \big],
\label{SCSN}
\eea
\ese
where $\hat{g}^-_{LR}(t,t')$ is the lesser Keldysh Green's function defined in Eq.~(\ref{LKGF})
and the currents are, {\it a priori}, time-dependent. 
As in Sec.~\ref{Sec.Tunnel}, the current is computed at a given electrode and the lead index is dropped for simplicity.

The computation of the current in Eqs.~(\ref{CSN}) therefore reduces to the computation of the Keldysh Green's function. 
Treating the transmission amplitude $\hat{T}$ as the perturbation the latter satisfies the following Dyson equation:
\be
\label{Dyson_Keldysh}
\check{g}^-(t,t') = [1 + \check{g}^R \circ \check{T}] \circ \check{g}^{(0)-} \circ [1 + \check{T} \circ \check{g}^A] (t,t'),
\ee
where the free lesser Keldysh Green's function was given in Eq.~(\ref{freeLKGF}).
In Eq.~(\ref{Dyson_Keldysh}) the check denotes the fact that all matrices are $8\times8$ matrices in the left-right$\otimes$spin$\otimes$Nambu space:  
\be
\check{g} = 
\left(
\begin{array}{cc}
\hat{g}_{RR} &  \hat{g}_{RL} \\
\hat{g}_{LR} &  \hat{g}_{LL} 
\end{array} 
\right), \quad
\check{T} = 
\left(
\begin{array}{cc}
0 &  \hat{T} \\
\hat{T} &  0 
\end{array} 
\right),
\label{LRMatrixStructure}
\ee
and symmetric contacts have been assumed. 
The dressed retarded and advanced Green's functions, $\check{g}^{R,A}$, also satisfy their own Dyson equation:
\be
\label{Dyson_RA}
\check{g}^{R(A)}(t,t') = \check{g}^{(0)R(A)}(t,t') + \check{g}^{(0)R(A)} \circ \check{T} \circ \check{g}^{R(A)} (t,t'),
\ee
where $\hat{g}^{(0)R(A)}$ are given by the proper analytic continuation of Eq.~(\ref{QCGF}). 

In the dc limit, that we are interested in, Eq.~(\ref{Dyson_Keldysh}) may be further simplified by assuming that both electrodes have identical occupation functions:
$n_F^\al \equiv n_F$ does not depend on $\al=L,R$, as no bias is applied.
This yields
\bea
\check{g}^-(t,t') = &&- \left( \check{g}^R \circ n_F - n_F \circ \check{g}^A \right) (t,t') 
\label{Dyson_Keldysh2} \\
&&-~ \check{g}^R \circ \left( n_F \circ \check{T} - \check{T} \circ n_F \right) \circ \check{g}^A (t,t'),
\nonum
\eea
which shows that the computation of the Keldysh Green's function further reduces to the computation of
the retarded and advanced functions. 
Notice that, because $\hat{T}(t)$ is locally time-dependent, the second term in Eq.~(\ref{Dyson_Keldysh2}) 
is non-zero for the present problem. 

The coupled Eqs.~(\ref{Dyson_Keldysh2}) and (\ref{Dyson_RA}) are integral equations which
are quite complicated to solve in general. 
Remarkably, the present problem considerably simplifies by going to the rotated frame. 
With the help of the unitary transformation of Sec.~\ref{SubSec.UT}, the Dyson equation for the lesser Keldysh function becomes:
\bea
\label{GFLesserRotated}
\check{\tilde{g}}^-(t,t') &&= - \left( \check{\tilde{g}}^R \circ \check{\tilde{n}}_F - \check{\tilde{n}}_F \circ \check{\tilde{g}}^A \right) (t,t')
\nonum \\
&&- \check{\tilde{g}}^R \circ \left( \check{\tilde{n}}_F \check{\tilde{T}} - \check{\tilde{T}} \check{\tilde{n}}_F \right) \circ \check{\tilde{g}}^A (t,t'),
\eea
where $\check{\tilde{T}}$ is time-independent (and real) which eliminates a time-convolution.
The non-commutativity between the transmission amplitude matrix and the occupation function is preserved because the latter becomes a matrix,
$\check{\tilde{n}}_F$, in the rotated frame. Similarly, the Dyson equations for the retarded and advanced functions, Eq.~(\ref{Dyson_RA}),
become:
\be
\label{Dyson_RARotated}
\check{\tilde{g}}^{R(A)}(t,t') = \check{\tilde{g}}^{(0)R(A)}(t,t') + \check{\tilde{g}}^{(0)R(A)} \check{T} \circ \check{\tilde{g}}^{R(A)} (t,t'),
\ee
and another time convolution has been eliminated. This implies that all Green's functions in the rotated frame depend on the difference of their
time arguments which allows us to go to Fourier space. Extracting the left-right component of Eq.~(\ref{GFLesserRotated}) yields:
\bea
\hat{\tilde{g}}^-_{LR}(\om) =&&- \left( \hat{\tilde{g}}_{LR}^R (\om) \hat{\tilde{n}}_F(\om) - \hat{\tilde{n}}_F (\om) \hat{\tilde{g}}_{LR}^A (\om) \right)
\label{KGFRF} \\
&&- \hat{\tilde{g}}_{LR}^R (\om) \left( \hat{\tilde{n}}_F (\om) \hat{\tilde{T}} - \hat{\tilde{T}} \hat{\tilde{n}}_F(\om) \right) \hat{\tilde{g}}_{LR}^A (\om)
\nonum \\
&&- \hat{\tilde{g}}_{LL}^R (\om) \left( \hat{\tilde{n}}_F (\om) \hat{\tilde{T}} - \hat{\tilde{T}} \hat{\tilde{n}}_F(\om) \right) \hat{\tilde{g}}_{RR}^A (\om),
\nonum
\eea
where $\hat{\tilde{n}}_F$ is given by Eq.~(\ref{nfRotated}) and $\hat{\tilde{T}}$ by Eqs.~(\ref{TunnelAmplitudeRotated}) and (\ref{TASN}).
Notice that the last two terms in Eq.~(\ref{KGFRF}) are non-zero only because the system is out-of-equilibrium, i.e. the precession frequency
is non-zero. Formally, they appear as collision terms where the occupation function is brought away from its equilibrium value.

Similarly, the Dyson equations for the retarded and advanced functions, Eq.~(\ref{Dyson_RA}), reduce to a set of algebraic equations which
are straightforward to solve, yielding:
\bea
&&\hat{\tilde{g}}_{LR}^{R(A)}(\om) =  {D_{L,R}^{(0)R(A)}}^{-1}\,\hat{\tilde{g}}_{L}^{(0)R(A)}(\om)  \hat{\tilde{T}} \hat{\tilde{g}}_{R}^{(0)R(A)} (\om),
\nonum \\
&&\hat{\tilde{g}}_{RR}^{R(A)}(\om) =  {D_{L,R}^{(0)R(A)}}^{-1}\,\hat{\tilde{g}}_{R}^{(0)R(A)}(\om),
\label{RAGFRF} \\
&&\hat{\tilde{g}}_{LL}^{R(A)}(\om) =  {D_{L,R}^{(0)R(A)}}^{-1}\,\hat{\tilde{g}}_{L}^{(0)R(A)}(\om),
\nonum
\eea
where the denominator reads:
\be
D_{\al,\beta}^{(0)R(A)} = {\bf 1}_4 - \hat{\tilde{g}}_{\al}^{(0)R(A)}(\om) \hat{\tilde{T}} \hat{\tilde{g}}_{\beta}^{(0)R(A)}(\om) \hat{\tilde{T}},
\label{denom}
\ee
the indices $\al,\,\beta\, =\,R,\,L$ and ${\bf 1}_4$ is the unit matrix in $4-$dimensional space.
In Eqs.~(\ref{RAGFRF}) and (\ref{denom}), the non-interacting Green's functions
depend on the effective magnetic field, see Eq.~(\ref{MSGFUT}).

Substituting the expressions of Eqs.~(\ref{RAGFRF}) in Eq.~(\ref{KGFRF}) yields the lesser Keldysh Green's function.
The latter has then to be substituted in the expression of the current in order to compute the transport properties
of the junction.
To this end we have to express the current, Eqs.~(\ref{CSN}), in terms of the transformed Green's functions:
\bse
\label{CSN_UTF}
\bea
&&I^c = e\,\int \frac{d \om}{2\pi}\,\Tr \big[ \hat{\sigma}_0 \hat{\tilde{T}} \hat{\tilde{g}}^-_{LR}(\om) \big],
\label{CCSN_UTF} \\
&&{\bf I}^s(t) = -\frac{1}{2}\,\int \frac{d \om}{2\pi}\,\Tr \big[ \hat{\vec{\tilde{\sigma}}}^s(t) \hat{\tilde{T}} \hat{\tilde{g}}^-_{LR}(\om) \big].
\label{SCSN_UTF}
\eea
\ese
Eq.~(\ref{CCSN_UTF}) shows that the charge current is actually time-independent.
On the other hand, Eq.~(\ref{SCSN_UTF}), shows that the spin-current is circularly polarized in the $xy-$plane 
whereas the $z-$component is time-independent. This comes from the fact 
that: $\hat{\vec{\tilde{\sigma}}}^s(t) = (\hat{\tilde{\sigma}}_1(t),\, \hat{\tilde{\sigma}}_2^s(t),\, \hat{\sigma}_3)$,
where:
\be
\hat{\tilde{\sigma}}_1(t) = \hat{\sigma}_1\,e^{-i\hat{\sigma}_3\Omega t},\quad 
\hat{\tilde{\sigma}}_2^s(t) = \hat{\sigma}_2^s\,e^{-i\hat{\sigma}_3\Omega t}.
\ee
These results agree with the results of Sec.~\ref{Sec.Tunnel}.

Formally, Eqs.~(\ref{CSN_UTF}) together with Eqs.~(\ref{KGFRF}), (\ref{RAGFRF}) as well as the self-consistent gap equation,
allow for an exact computation of the charge and spin currents at any temperature, value of the precession frequency and 
transparency of the junction.
Practically, the analytical computation of the current in the transparent limit is extremely tedious in the general case.
As it will be shown in Sec.~\ref{Sec.ResultsCharge}, analytic calculations can easily be performed for a non-tilted/static spin and 
may be significantly simplified in the case where the spin has a large tilt angle $\theta \rightarrow \pi/2$, 
see also Ref.~[\onlinecite{Teber09,Holmqvist09}].
In the general case, however, these equations have to be solved numerically. 
It turns out that, within the present reformulation and simplification
of the problem, the numerical implementation is quite straightforward. 
Indeed, the system of equations above has the same form as the one for an equilibrium problem. 
Out-of-equilibrium effects simply appear as effective magnetic field dependencies of the 
non-interacting Green's functions, the matrix occupation function as well as the sigma-matrices
of the spin-current.

%%%%%%%%%%%%%%%%%%%%%%%%%%
\section{Results for the charge current}
\label{Sec.ResultsCharge}

The main quantity of interest in this section is the charge current which has been computed in the tunnel limit in Sec.~\ref{Sec.Tunnel}
and for which a formal exact expression in the transparent was given in Sec.~\ref{SubSec.TranspEq}, see Eq.~(\ref{CCSN_UTF}) 
together with Eqs.~(\ref{KGFRF}) and (\ref{RAGFRF}). We will start from the simplest cases (zero precession frequency, large tilt angle)
before going to the general case where a fully numerical approach is required. Besides the current itself, we will also be interested in the 
current-carrying states, the nature of which will help us interpret the results obtained for the current. 
Information on these current-carrying states, as well as their occupation, is contained in the (charge) current kernel:
\be
I^c(\om) = e\,\Tr \big[ \hat{\sigma}_0 \hat{\tilde{T}} \hat{\tilde{g}}^-_{LR}(\om) \big], \quad
I^c = \int \frac{d\om}{2\pi}\,I^c(\om).
\label{CKG}
\ee
By definition, the poles of the current kernel, or zeros of Eq.~(\ref{denom}) 
(when they do exist), correspond to well-defined bound-states with energies 
below the superconducting gap. 
The current due to these Andreev states may formally be written as:
\be
I^c_{ABS} = \int_{-\Delta(h_z)}^{+\Delta(hz)}\,\frac{d \om}{2\pi}\,I^c(\om),
\ee
where the limits of integration may be affected by the precession of the spin.
On the other hand, the branch cut structure of the current kernel implies that extended states may carry the current as well, 
see Sec.~\ref{Sec.Tunnel} where this fact has already been discussed in the tunnel limit.
The current due to these extended may be formally written as:
\be
I^c_{ext} = \int_{-\infty}^{-\Delta(hz)}\,\frac{d \om}{2\pi}\,I^c(\om) + \int_{+\Delta(h_z)}^{+\infty}\,\frac{d \om}{2\pi}\,I^c(\om),
\ee
such that the total current: $I^c = I^c_{ABS} + I^c_{ext}$. 
In general, localized and extended states compete in order to carry the current~\cite{Note:Beenakker}. 
This will clearly be seen at the level of the current and its dependence on the precession-frequency and/or 
superconducting phase difference. 
The goal of this section is to discuss these facts.

In the following, numerical simulations are performed with a small energy relaxation rate
that takes into account phenomenologically the damping of the quasi-particles due to inelastic processes,
such as e.g. electron-phonon interactions, in the leads.
We will use $\eta = 10^{-3}\Delta$, which agrees with typical estimates~\cite{Kaplan76} for usual superconductors, 
where $\Delta$ corresponds to the $T=0$ amplitude of the superconducting order parameter.
Moreover, we will run the simulations at low temperatures, $T=10^{-3}\Delta$, unless specified.

%%%%%%%%%%%%%%%%%%%%
\subsection{Static spin in the junction ($\Omega = 0$)}
\label{SubSec.Static}

\begin{figure}
\centering
\mbox{\subfigure{\includegraphics[width=4.cm,height=3.cm]{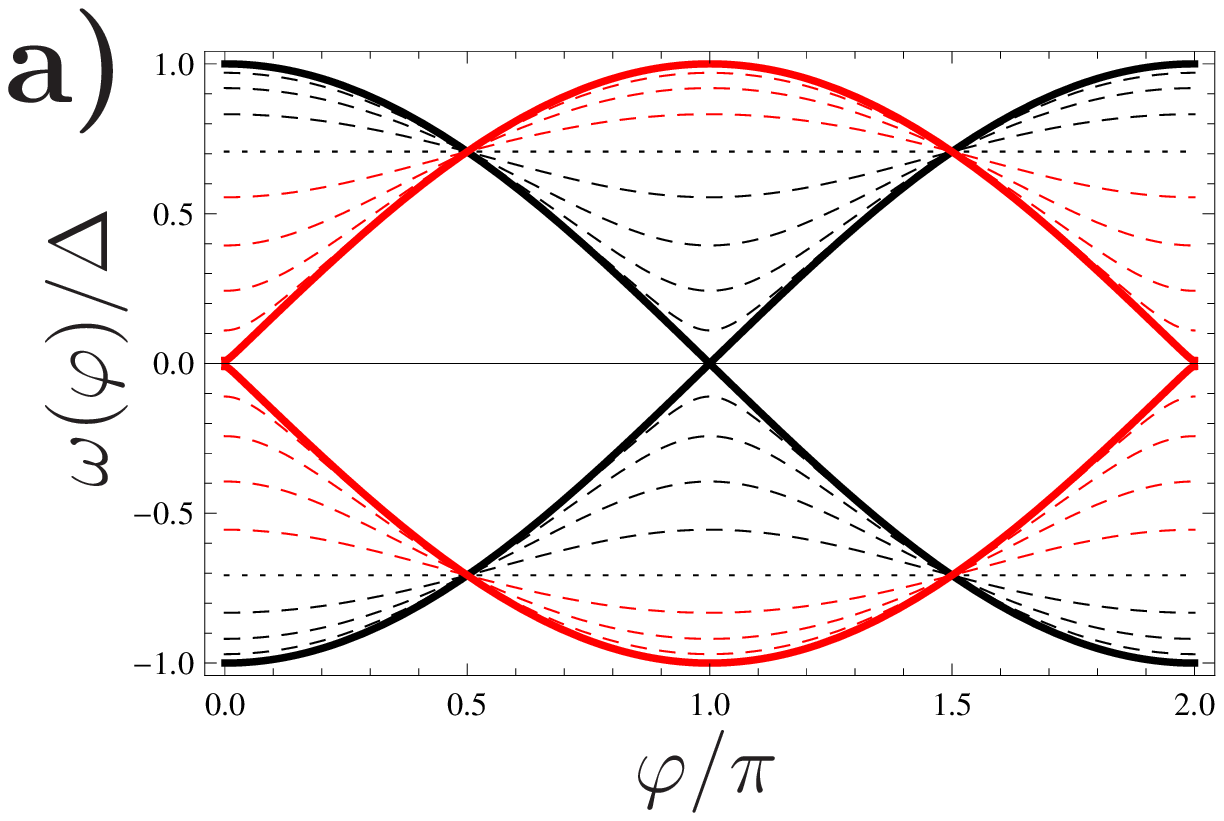}}\quad
\subfigure{\includegraphics[width=4.cm,height=3.cm]{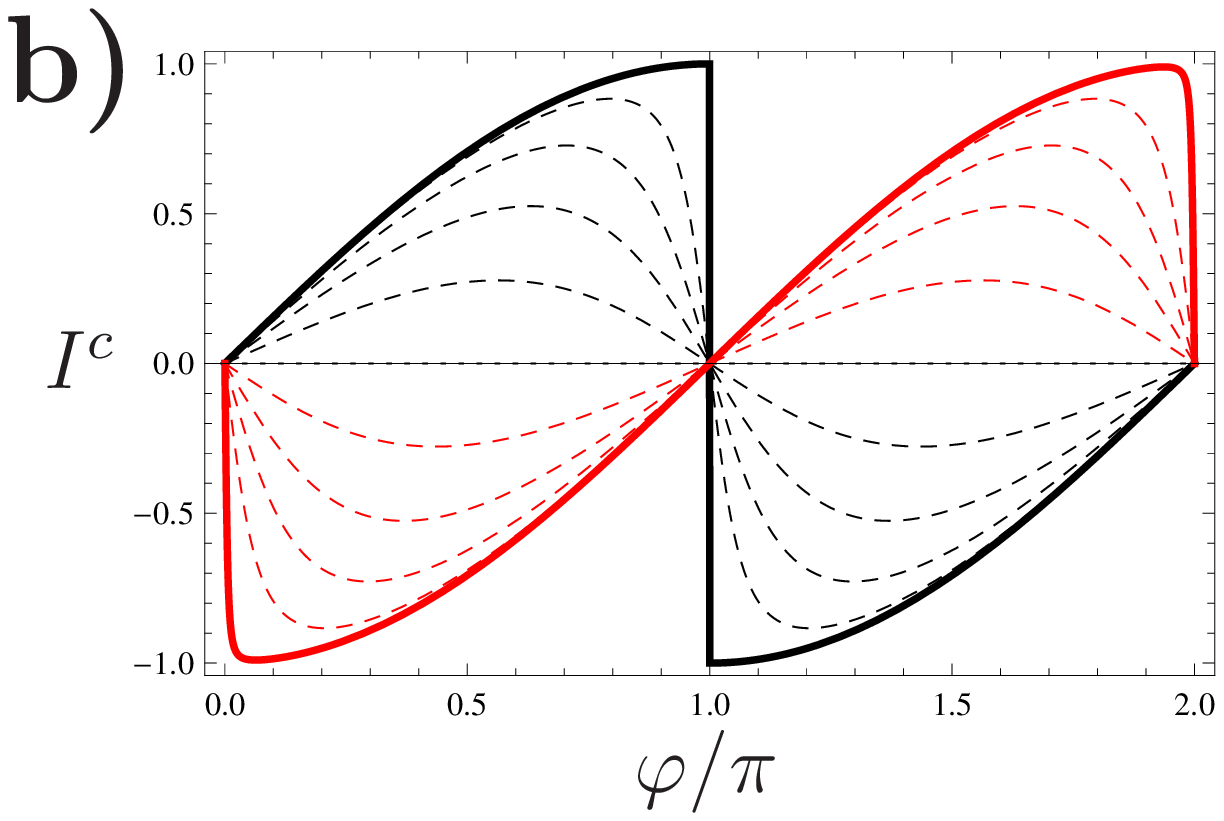}}}
\caption{ \label{Fig:Andreev0Pi} (color online) (a) The spectrum of Andreev bound-states and (b) the charge current
for a static spin with $t_0 = 1 -t_S$ and $t_S$ varies from $0$ (thick solid black line) to $1$ (thick solid red line)
with a step of $0.1$.}
\end{figure}

In the case where the spin in the junction is static
only the first term in the expression of the Keldysh Green's function, 
Eq.~(\ref{KGFRF}), gives a non-zero contribution to the current. 
Some matrix algebra yields a simple analytic expression for the current kernel:
\bea
&&I^c(w) = 
\label{CCKStaticSpin} \\
&&2e\,\left( \mathcal{T}_{0} - \mathcal{T}_{S} \right)\,\Delta^2 \sin \vp\,n_F[\om]\,\mbox{Im}\left \{\frac{1}{(\om+i\eta)^2 - \om^2(\vp)} \right \},
\nonum
\eea
where $\eta = 0^+$, the transparencies of the junctions are defined as:
\be
\mathcal{T}_{0} = \frac{4t_0^2}{(1+t_0^2-t_S^2)^2+4t_S^2},\quad \mathcal{T}_{S} = \frac{4t_S^2}{(1+t_0^2-t_S^2)^2+4t_S^2},
\label{DStaticSpin}
\ee
and, due to the spin-rotational invariance of the static problem, $t_S$ combines both spin-conserving as well as spin-flip tunneling amplitudes:
$t_S^2 = t_\parallel^2 + t_\bot^2$. From Eq.~(\ref{CCKStaticSpin}), we see that the current kernel has two poles at $\om_\pm (\vp) = \pm\, \om(\vp)$
where:
\be
\om (\vp) = \Delta\,\sqrt{1 - \mathcal{T}_{0} \sin^2 \left(\vp / 2\right) - \mathcal{T}_{S} \cos^2 \left(\vp / 2\right) }.
\label{AndreevStaticSpin}
\ee
These poles correspond to bound-states with energies below the superconducting gap, $\om(\vp)\leq \Delta$, 
see Fig.~\ref{Fig:Andreev0Pi}a. 
The integration of the current kernel reduces to an integration over these poles 
which therefore carry all the charge current across the junction. 
As can be seen from Eq.~(\ref{CCKStaticSpin}), the Andreev bound-states carry current in opposite directions:
\be
I_{\pm}^c = \mp\,2e\,\frac{d \om(\vp)}{d \vp}\,n_F\left[\pm \om(\vp) \right],
\label{CCSSCP}
\ee
where, because of the absence of any current-carrying extended state, the total current is given by: $I^c = I^c_+ + I^c_-$. 
The Fermi function then guarantees that, at low enough temperatures, only the lowest sub-gap state is occupied 
yielding a non-zero supercurrent:
\be
I^c = \frac{e\,\left( \mathcal{T}_{0} - \mathcal{T}_{S} \right)\,\Delta \sin \vp}{2\sqrt{1 - \mathcal{T}_{0} \sin^2 \left(\vp / 2\right) - \mathcal{T}_{S} \cos^2 \left(\vp / 2\right) }}\,
\tanh \left(\frac{\om(\vp)}{2T} \right),
\label{CCSS}
\ee
which is plotted on Fig.~\ref{Fig:Andreev0Pi}b.

With these formulas in hand, we now consider the well known case~\cite{Beenakker91} where $t_S=0$ and $t_0$ is arbitrary.
In this case, the spectrum of the Andreev bound-state reads: 
$\om(\vp) = \Delta\,\sqrt{1 - \mathcal{T}_{0} \sin^2 \left(\vp / 2\right) }$ with $\mathcal{T}_{0} = 4t_0^2 / (1+t_0^2)^2$.
Assuming that we are at zero temperature, only the lowest in energy of these Andreev states is occupied. 
We see from the thick black curve in Fig.~\ref{Fig:Andreev0Pi}a that it has a minimum in energy at $\vp=0$, which therefore corresponds
to its ground-state phase difference. 
The current of such a $0-$junction may be straightforwardly obtained from Eq.~(\ref{CCSS}) and it is plotted on the thick black curve
of Fig.~\ref{Fig:Andreev0Pi}b.
On the other hand, for $t_0=0$ and $t_S$ arbitrary, the spectrum of the Andreev bound-state:
$\om(\vp) = \Delta\,\sqrt{1 - \mathcal{T}_{S} \cos^2 \left(\vp / 2\right) }$ with $\mathcal{T}_{S} = 4t_S^2 / (1+t_S^2)^2$,
has a ground-state phase difference of $\vp=\pi$, see the thick red curve in Fig.~\ref{Fig:Andreev0Pi}a. 
The current of such a $\pi-$junction is the opposite of the one of a $0-$junction, see 
Eq.~(\ref{CCSS}) and the thick red curve of Fig.~\ref{Fig:Andreev0Pi}b. As was anticipated in Sec.~\ref{SubSec.ChargeCurr}, when $t_0=t_S$, 
a transition from a $0-$ to a $\pi-$junction takes place. 
This is shown on Fig.~\ref{Fig:Andreev0Pi} where it can be seen that, as soon as $t_S$ 
is non-zero, the bound-state spectrum disconnects from the continuum at $\pm \Delta$. 
Increasing $t_S$, the curvature of the Andreev levels changes sign once $t_S > t_0$ which corresponds to exchanging the 
minimum and maximum bound-state energies and is equivalent to a $\pi-$shift of the phase.

\begin{figure}
\begin{tabular}{cc cc}
        \includegraphics[width=4.cm,height=3.3cm]{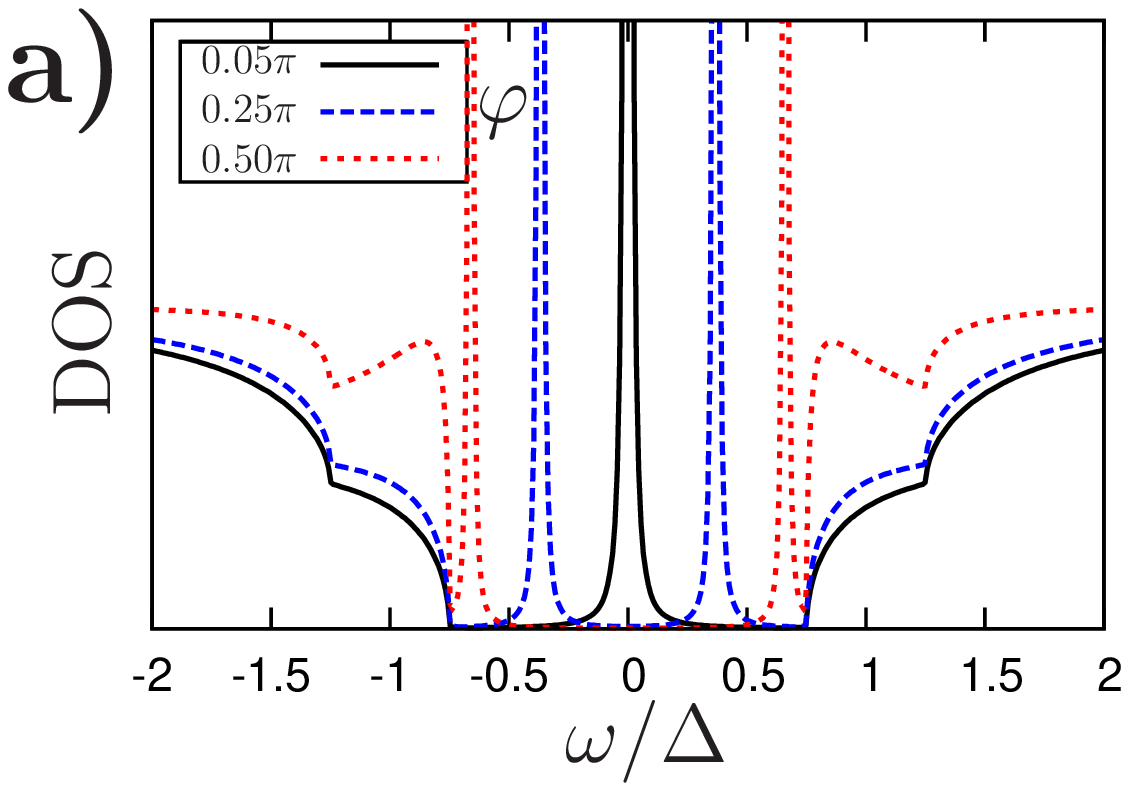}
&
        \includegraphics[width=4.cm,height=3.3cm]{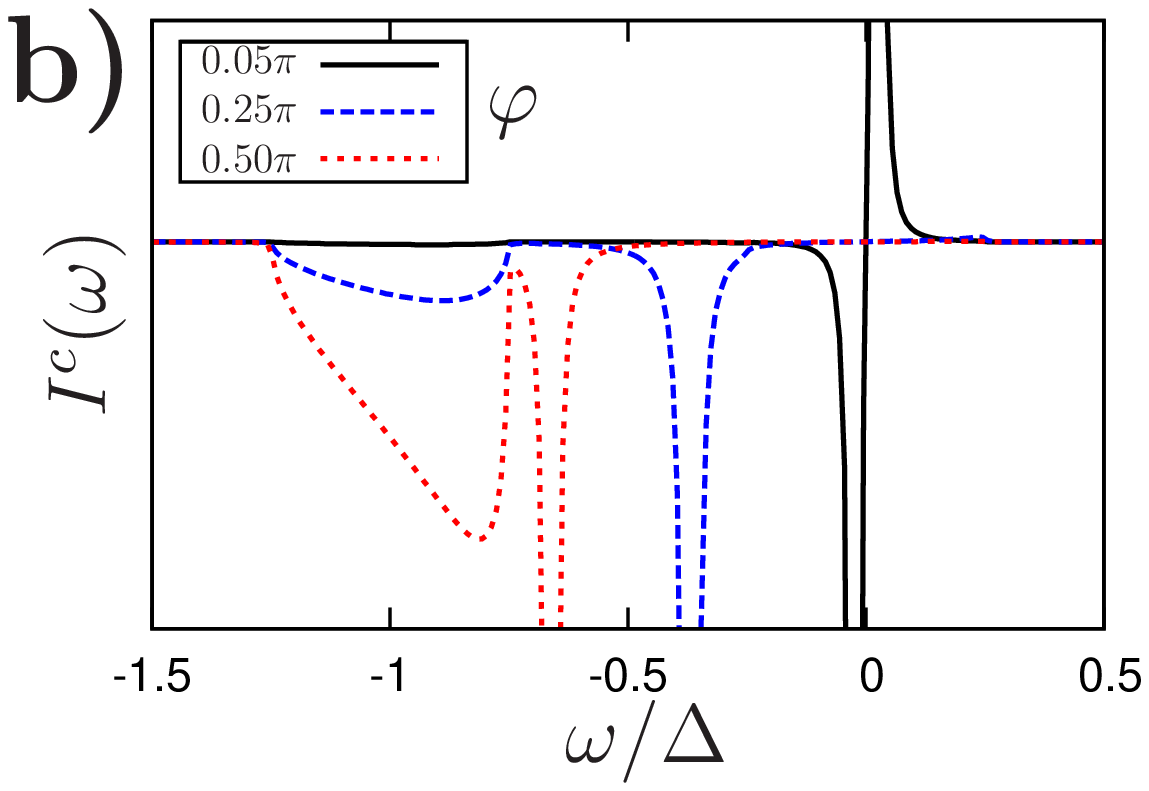}
\\
        \includegraphics[width=4.cm,height=3.3cm]{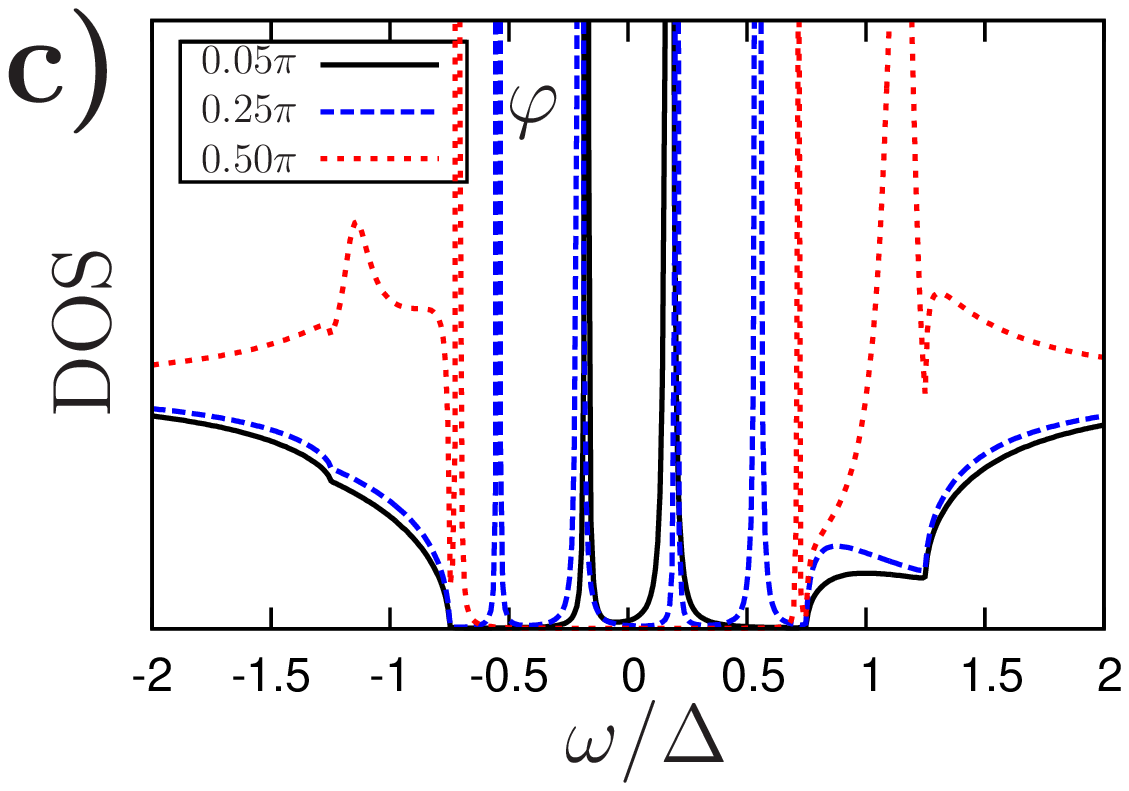}
&
        \includegraphics[width=4.cm,height=3.3cm]{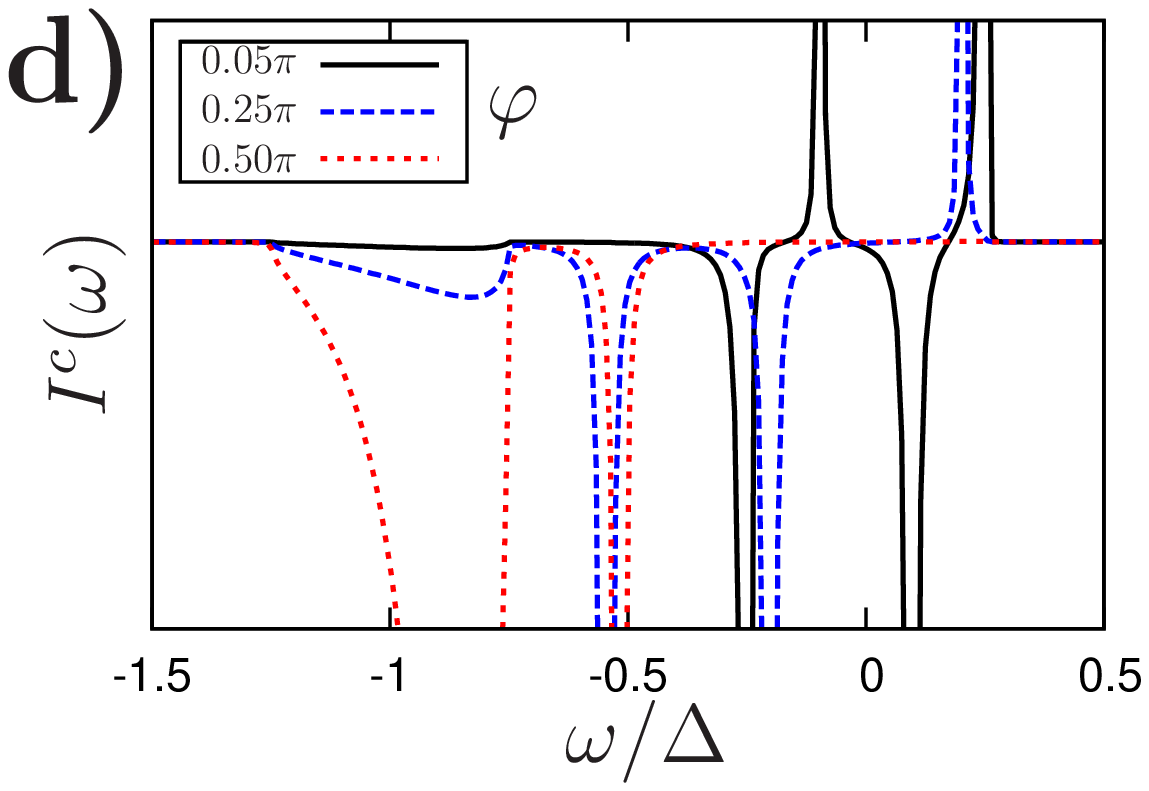}
\end{tabular}
\caption{
(color online)
Density-of-states (DOS) of spin-up particles in the rotated frame and charge current kernel (CCK) for: $t_0=0$, $t_S=1$ and $h_z=0.25\Delta$.
The different line styles (and colors) refer to different values of the superconducting phase difference, $\vp$, see the insets.
The four panels correspond to $\theta=\pi/2$ for (a) and (b) and $\theta=\pi/4$ for (c) and (d).
}
\label{Fig:DOS_CCK} % caption for the whole figure
\end{figure}
%

%%%%%%%%%%%%%%%%%%%%%%%
\subsection{Precessing spin in the junction ($\theta \approx \pi/2$)}
\label{SubSec.Plane}

When the spin in the junction is precessing, significant deviations from the results of Sec.~\ref{SubSec.Static} are obtained.
In this section, we consider the limit where the tilt angle of the spin with respect to the precession axis is large: $\theta \approx \pi/2$. 
From Eq.~(\ref{TA1}) this implies that $t_\parallel \approx 0$. 
For simplicity, we further assume that $t_0=0$. Tunneling is then only possible via spin-flip processes through an in-plane precessing spin.
This case has been reported on in Refs.~[\onlinecite{Teber09,Holmqvist09}]. We will summarize below the basic results related
to the nature of the current-carrying states and compare them with the results of Sec.~\ref{SubSec.Static}.

In the case of a spin precessing in the plane, the simplest quantity which may be extracted analytically is the spectrum of the bound-states.
The latter correspond to the poles of the current kernel of Eq.~(\ref{CKG}) (or equivalently to the zeros of Eq.~(\ref{denom})).
Some algebra yields:
\begin{widetext}
\begin{equation}
\om(\varphi) = \sqrt{ h_z^2 + \frac{\Delta^2}{ 1 - \tilde{\mathcal{T}}^2} \left \{ 1 + \tilde{\mathcal{T}}^2 \cos \varphi  - \sqrt{ \tilde{\mathcal{T}}^2 \left( 1 + \cos \varphi \right)^2 + 4 (1 - \tilde{\mathcal{T}}^2) \left( \frac{h_z}{\Delta} \right)^2 } \right \}},
\label{AndreevPlane}
\end{equation}
\end{widetext}
where the transparency of the junction was redefined as: $\tilde{\mathcal{T}} = 2 t_\bot^2 / (1 + t_\bot^4)$.
Eq.~(\ref{AndreevPlane}) considerably differs from the corresponding expression in the equilibrium case, Eq.~(\ref{AndreevStaticSpin}).
Indeed, Eq.~(\ref{AndreevPlane}) shows that the Andreev spectrum depends now on the precession frequency via $h_z$.
Moreover, the appearance of the square roots suggests that bound-states and extended-states are inter-related. 
This is more easily seen from the asymptotic expressions of the bound-state spectrum:
\be
\om(\vp) =
\left\{
        \begin{array}{ll}
                \Delta\,|1-\tilde{h}_z|,\, & \tilde{\mathcal{T}} \rightarrow 0, \\
                \Delta\,|\tan \left(\vp / 2 \right)| \sqrt{\cos^2\left(\vp/2\right) - \tilde{h}_z^2},\, & \tilde{\mathcal{T}} \rightarrow 1, \\
          \end{array}
\right.
\label{AndreevPlaneLimits}
\ee
where $\tilde{h}_z = h_z / \Delta$ is the reduced effective magnetic field. 
Eq.~(\ref{AndreevPlaneLimits}) shows that, in the tunnel regime where $\tilde{\mathcal{T}} \rightarrow 0$, the Andreev bound-states are dispersionless and cross at $h_z=\Delta$. 
This can be contrasted with the equilibrium behaviour ($h_z=0$) where they touch the
continuum of states (this can also be seen from Eq.~(\ref{AndreevStaticSpin}) with $\mathcal{T}_0,\, \mathcal{T}_1 \rightarrow 0$).
In both cases, however, extended states carry all the current (in the out-of-equilibrium case the currents of both Andreev bound-states
cancel each-other). This agrees with the results of Sec.~\ref{SubSec.ChargeCurr}.
On the other hand, in the transparent limit where $\tilde{\mathcal{T}} \rightarrow 1$, the Andreev spectrum depends on the 
superconducting phase.
As can be seen from Eq.~(\ref{AndreevPlaneLimits}), the bound-states merge with the continuum of states for
high enough values of $h_z$ and/or a phase difference close to $\pi$. 
As has been shown in Refs.~[\onlinecite{Teber09,Holmqvist09}], this merging takes place
at a phase difference $\vp_c$ such that: $d\om(\vp_c)/d\vp=0$. In the transparent limit, 
this leads to the following relation between the critical phase and effective magnetic field:
$h_{z,c}(\vp)= \Delta \cos^2(\vp/2)$ or equivalently $\vp_c(h_z) = 2\arccos(\sqrt{h_z/\Delta})$.
Hence, bound-states exist for phase differences smaller than $\vp_c$ and larger than $2 \pi - \vp_c$. 
For other phase differences the extended states carry all the current. Notice that, from these
arguments, the average critical field at which bound-states and extended states merge 
in the transparent limit is given by:
\be
\bar{h}_{z,c} = \int_0^{2\pi} \frac{d \vp}{2\pi}\,h_{z,c}(\vp) = \frac{\Delta}{2}.
\label{hcrit}
\ee

In order to illustrate these facts on a concrete example,
the density of states in the rotated frame has been plotted on Fig.~\ref{Fig:DOS_CCK}a, for the special 
case where $\tilde{\mathcal{T}} = 1$  and $h_z=0.25\Delta$. 
From the previous arguments, the critical phase associated with this value of the effective magnetic field is given by: $\vp_c \approx 0.67\pi$.
This is confirmed by Fig.~\ref{Fig:DOS_CCK}a which shows that subgap states exist for $\vp < \vp_c$ and that they tend to merge with 
the continuum of states, located at $\pm (\Delta - h_z)$, as $\vp$ approaches $\vp_c$. For $\vp>\vp_c$ (not represented on the figure)
there are no more bound-states. 
The corresponding low$-T$ current kernel has been plotted on Fig.~\ref{Fig:DOS_CCK}b.
In accordance with this figure, the total charge current may be decomposed in the following way ($h_z<\Delta$):
\be
I^c = \left\{ \int_{-\Delta+h_z}^{\Delta-h_z} + \int_{-\Delta-h_z}^{-\Delta+h_z} + \int_{\Delta-h_z}^{\Delta+h_z} \right\}\,\frac{d \om}{2\pi}\,I^c(\om),
\label{PolesAndBC}
\ee
where the first integral is over poles, $\om(\vp)$, corresponding to current-carrying bound-states
whereas the last two integrals are over branch cuts, of width $2 h_z$
around $\pm \Delta$, corresponding to current-carrying extended states.
The integral has to be computed numerically which will be done in the next paragraph.
However, we already see from Fig.~\ref{Fig:DOS_CCK}b that, for $\vp \approx 0$ (black curve),
both Andreev levels are occupied while for intermediate values of the phase, $\vp \approx 0.25\pi$ (blue curve),
only a single Andreev state is occupied and that at larger values of the phase extended states 
become the main current carrying states.
These sharp changes in the occupancy of the Andreev states as a function of the phase difference  
will be seen as a strong suppression of the current in the current-phase relation (CPR) around $\vp \approx 0$ in the next paragraph.

\begin{figure}
\begin{tabular}{cc cc}
        \includegraphics[width=4.2cm,height=3.5cm]{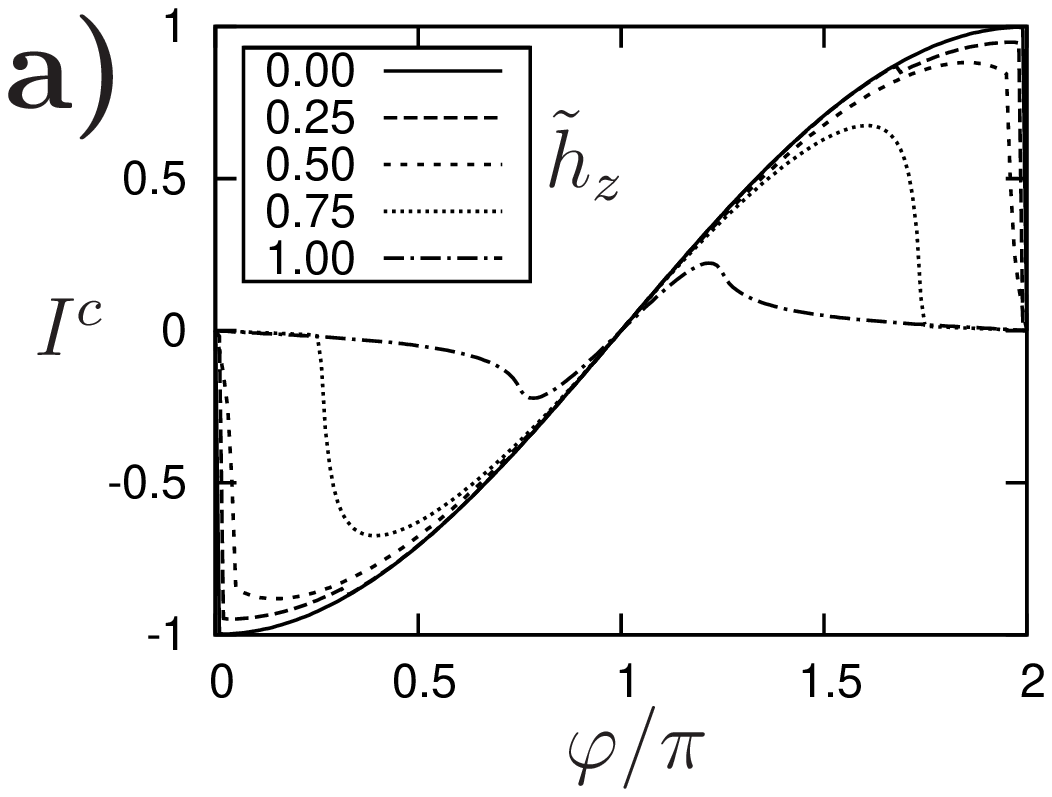}
&
        \includegraphics[width=4.2cm,height=3.5cm]{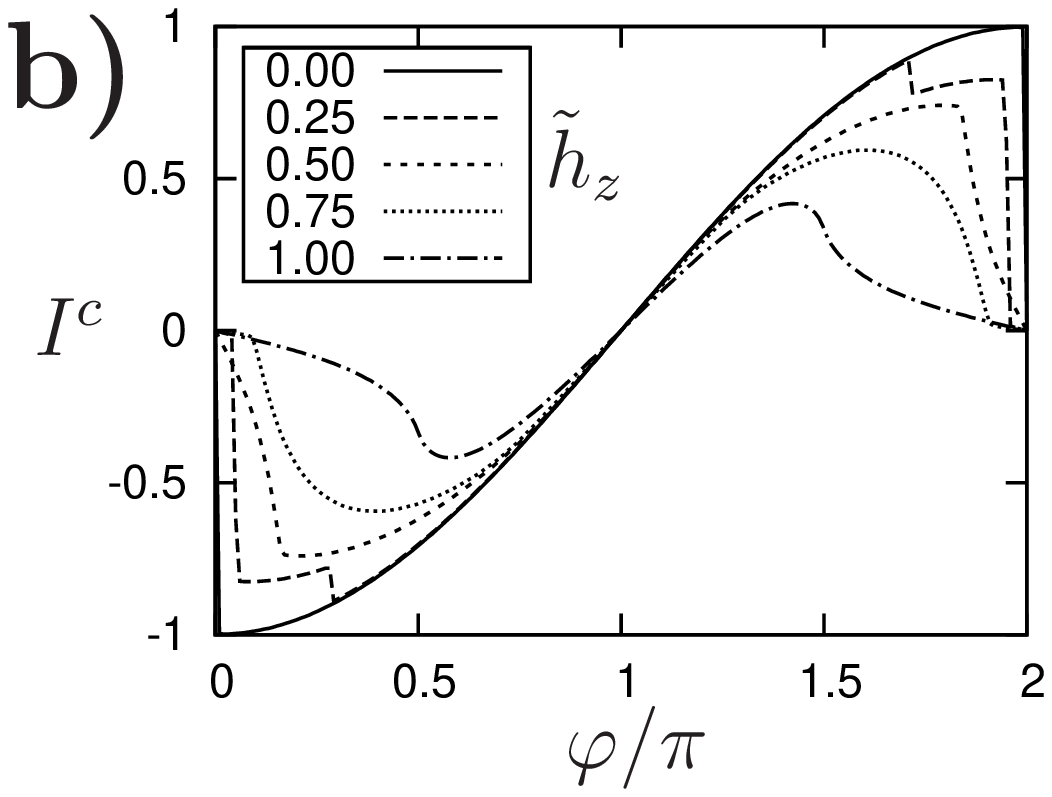}
\\
        \includegraphics[width=4.2cm,height=3.5cm]{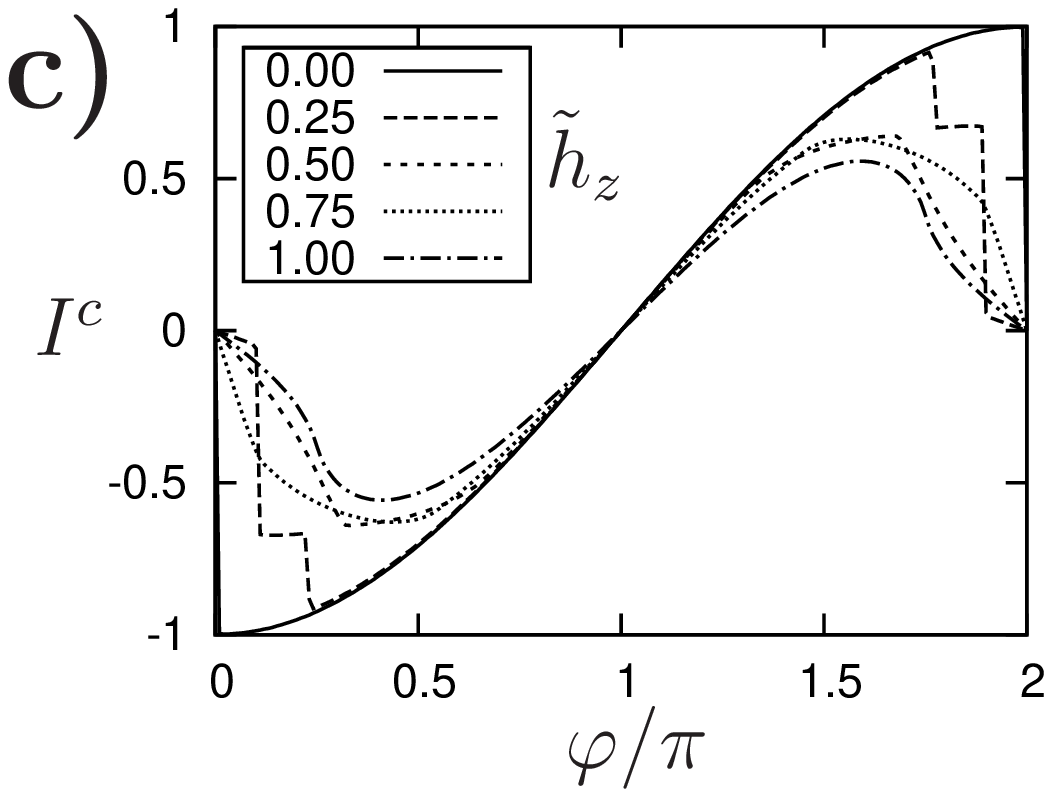}
&
        \includegraphics[width=4.2cm,height=3.5cm]{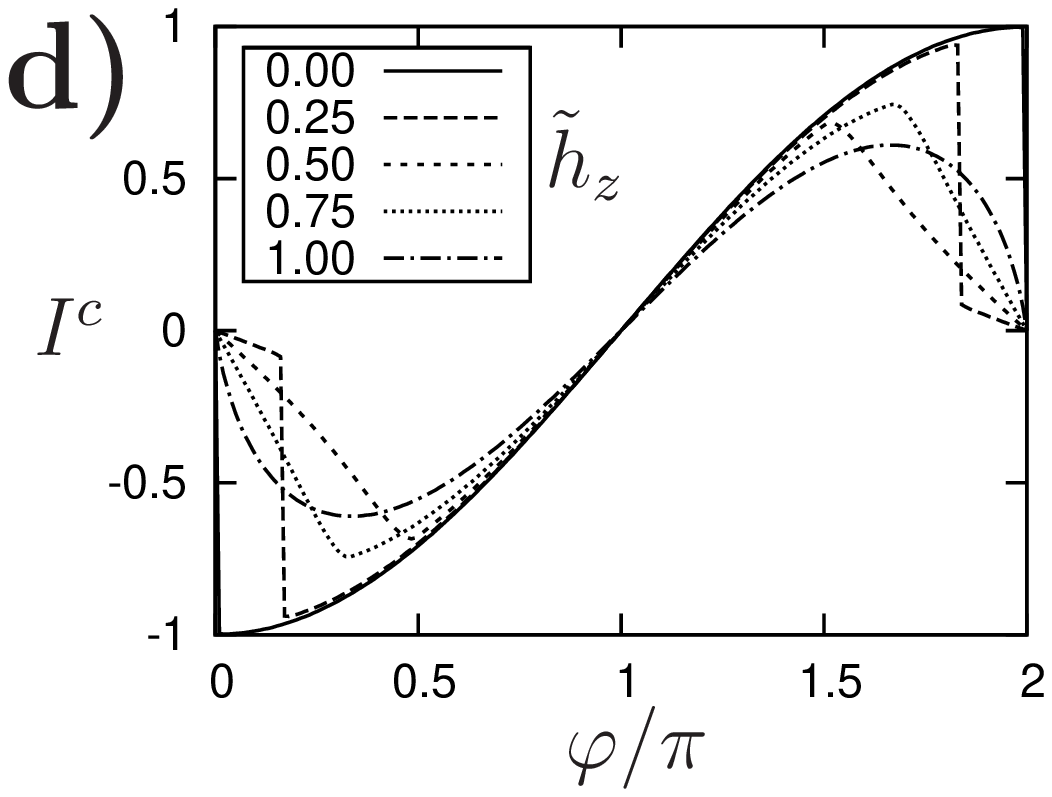}
\end{tabular}
\caption{
Charge CPR for fixed transparencies: $t_0=0$, $t_S=1$ and different reduced
effective magnetic fields, $\tilde{h}_z = h_z / \Delta$, see the insets.
The four panels correspond to:
(a) $\theta = \pi/8$, (b) $\theta = \pi/4$, (c) $\theta = 3\pi/8$ and (d) $\theta = \pi/2$.
The current is in units of $e \Delta / \hbar$.
}
\label{Fig:CCG_hvar} % caption for the whole figure
\end{figure}
\begin{figure}
\begin{tabular}{cc cc}
        \includegraphics[width=4.2cm,height=3.5cm]{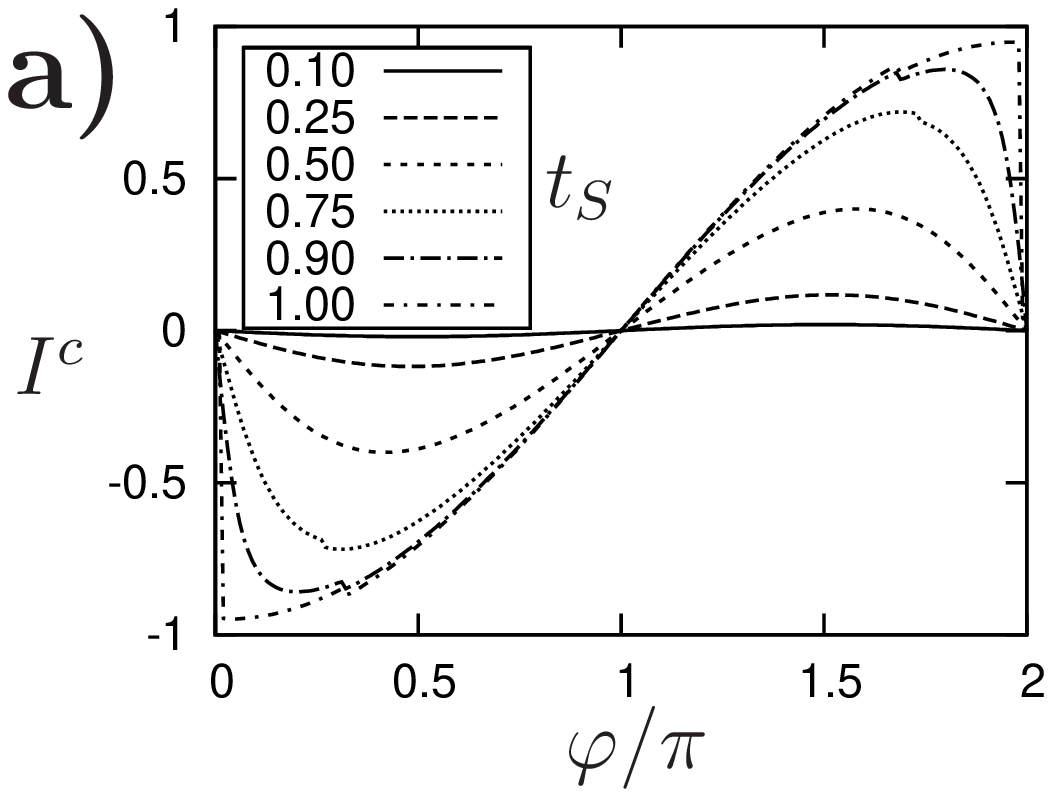}
&
        \includegraphics[width=4.2cm,height=3.5cm]{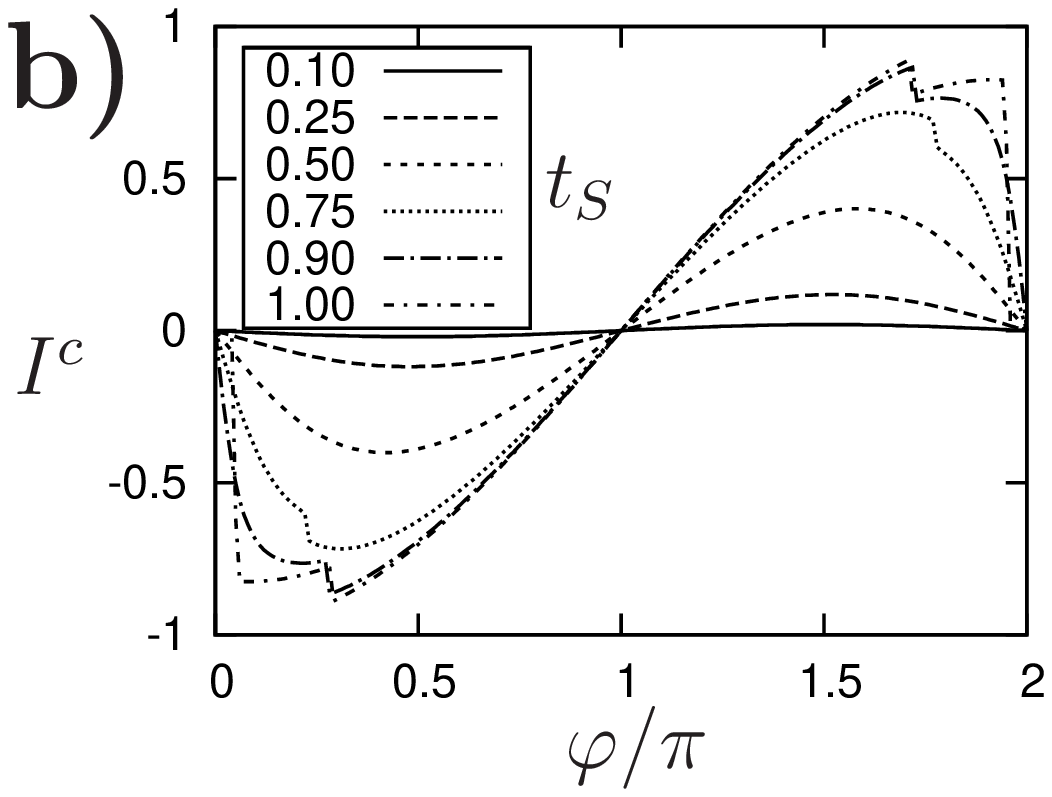}
\\
        \includegraphics[width=4.2cm,height=3.5cm]{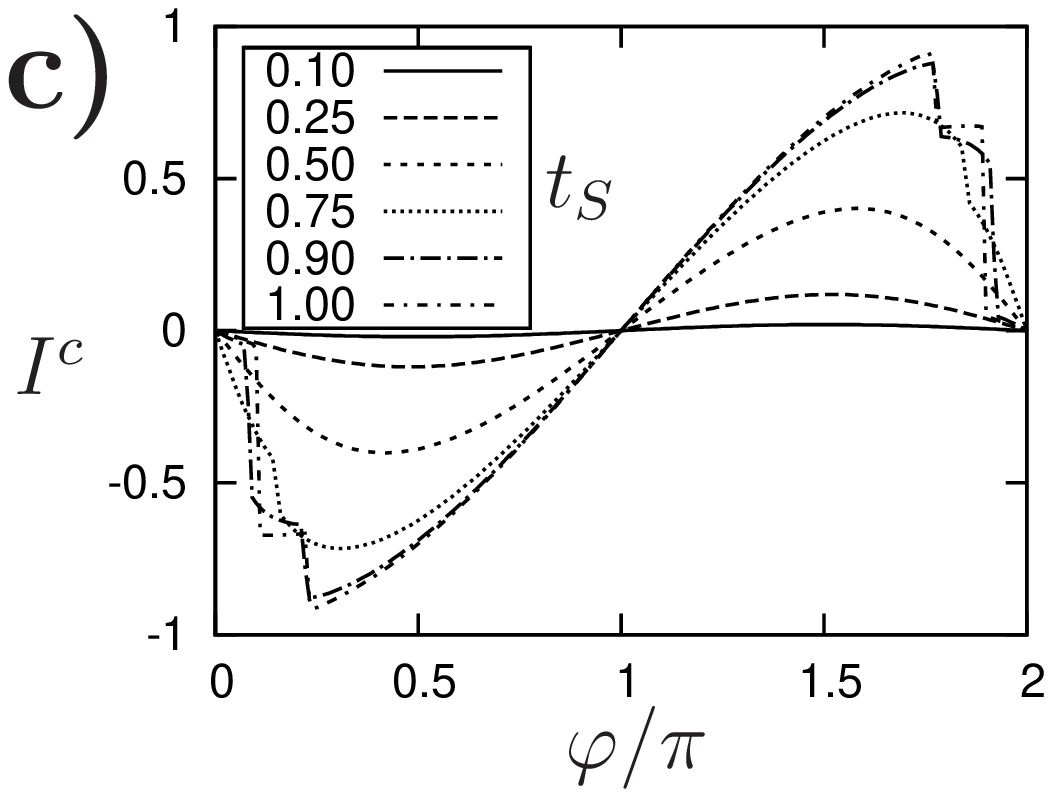}
&
        \includegraphics[width=4.2cm,height=3.5cm]{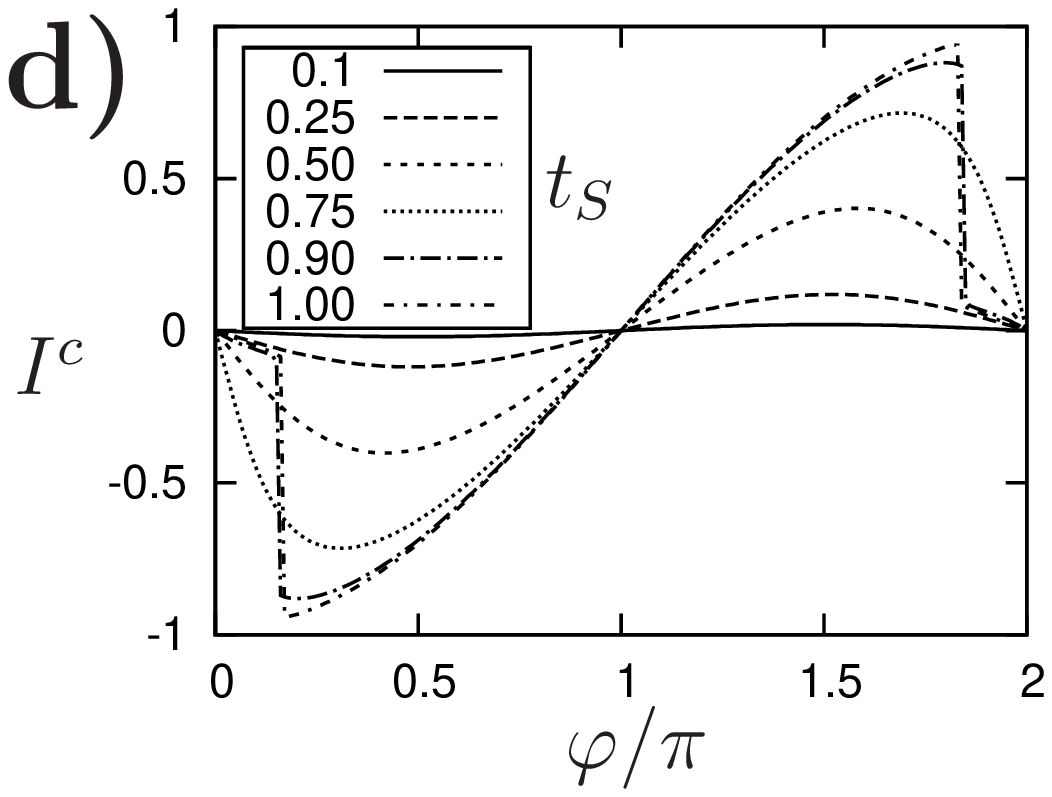}
\end{tabular}
\caption{
Charge CPR for a given: $h_z = 0.25 \Delta$ and $t_0=0$.
The different line styles refer to different transparencies, $t_S$, see the insets.
The four panels correspond to:
(a) $\theta = \pi/8$, (b) $\theta = \pi/4$, (c) $\theta = 3\pi/8$ and (d) $\theta=\pi/2$.
The current is in units of $e \Delta / \hbar$.}
\label{Fig:CCG_ttvar} % caption for the whole figure
\end{figure}
%

%%%%%%%%%%%%%%%%%%%%%%%%%
\subsection{Precessing spin in the junction (arbitrary $\theta$)}
\label{SubSec.AnyAngle}

For arbitrary tilt angles Eq.~(\ref{CCSN_UTF}) of Sec.~\ref{SubSec.TranspEq} 
does not reduce to any simple analytic form and we proceed numerically. 
The results for the CPR are displayed on Fig.~\ref{Fig:CCG_hvar} and
\ref{Fig:CCG_ttvar} in the limit of a $\pi-$junction ($t_S>t_0=0$).
We see from these figures 
that the CPR is characterized by sharp steps
leading to a strong suppression of the current around $\vp \approx 0$
and $\vp \approx 2\pi$.
From Figs.~\ref{Fig:CCG_hvar} and \ref{Fig:CCG_ttvar}, we see that the steps appear clearly 
for high transparencies, low precession frequencies ($\Omega \ll 2 \Delta$) 
and preferably for tilt angles $\theta \approx \pi/4$ or larger. 
For small tilt angles ($\theta \rightarrow 0$)
one needs to go to higher $\Omega$ in order to single out the steps, see Fig.~\ref{Fig:CCG_hvar}a.
Such a structure of the CPR originates from the existence of current-carrying bound-states which 
disappear, in favor of current-carrying extended states, at both higher values of the effective magnetic field and intermediate
values of the phase difference (in particular around $\vp \approx \pi$). 
The strong suppression of the current is due to an abrupt change in the occupation
of the lower and upper Andreev-levels the currents of which cancel each-other
when they are both occupied. 

In the case where $\theta \approx \pi/2$ this agrees with the results of the last paragraph and 
Figs.~\ref{Fig:DOS_CCK} a) and b) (see discussion below Eq.~(\ref{AndreevPlane})). 
This case, which allowed for some analytic estimates, is however rather unphysical.
\begin{figure}
\begin{tabular}{cc cc}
        \includegraphics[width=4.2cm,height=3.5cm]{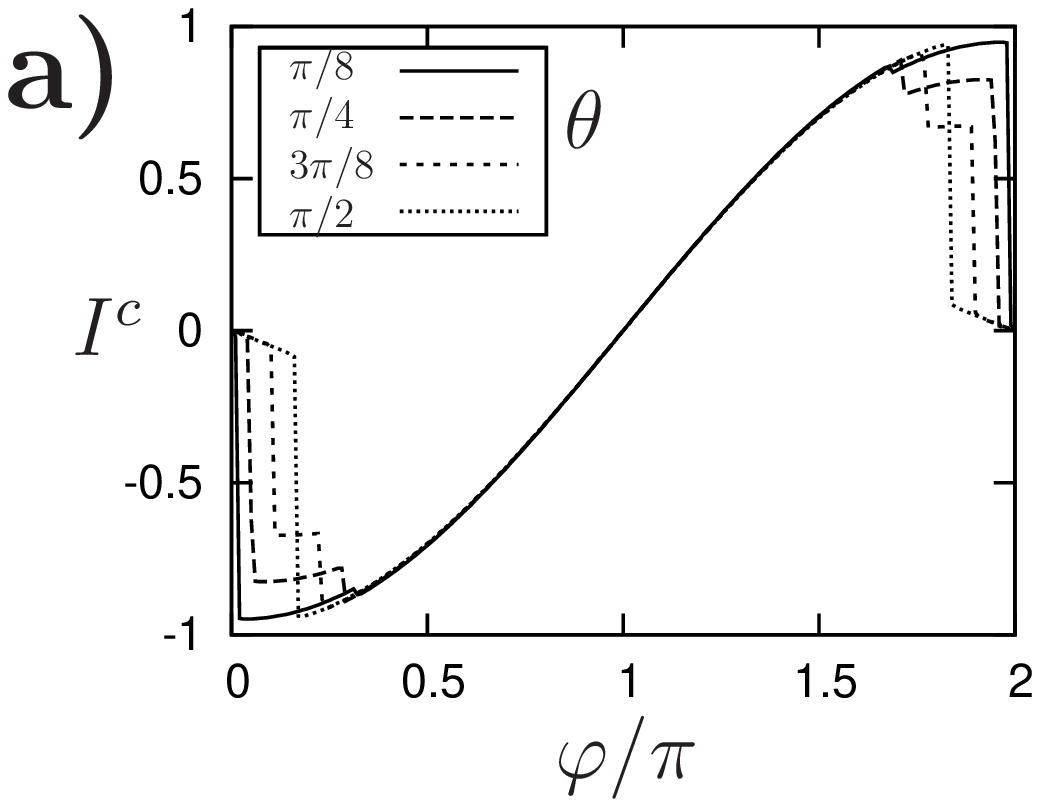}
&
        \includegraphics[width=4.2cm,height=3.5cm]{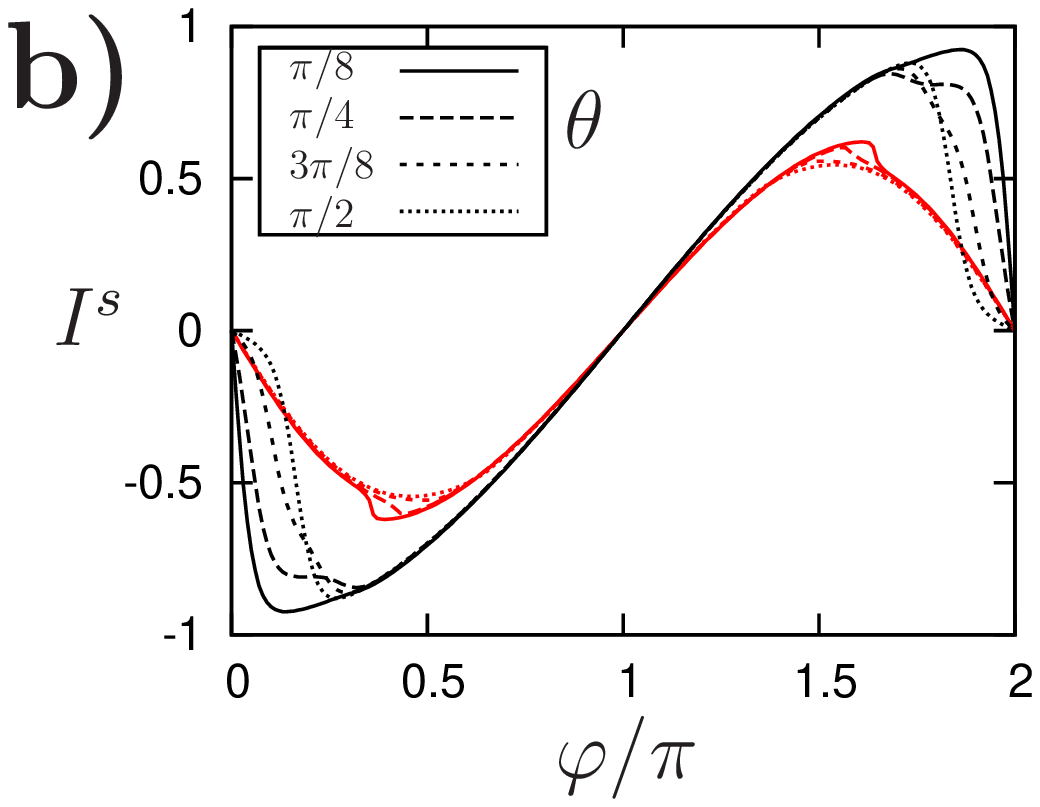}
\\
        \includegraphics[width=4.2cm,height=3.5cm]{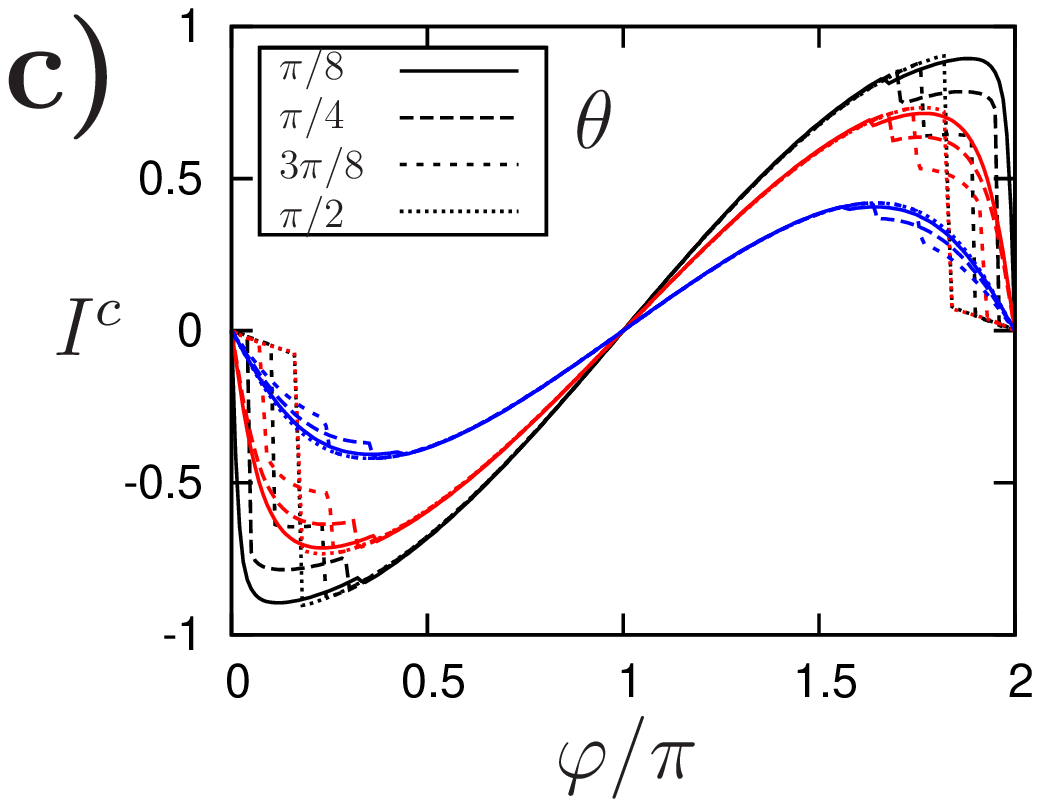}
&
        \includegraphics[width=4.2cm,height=3.5cm]{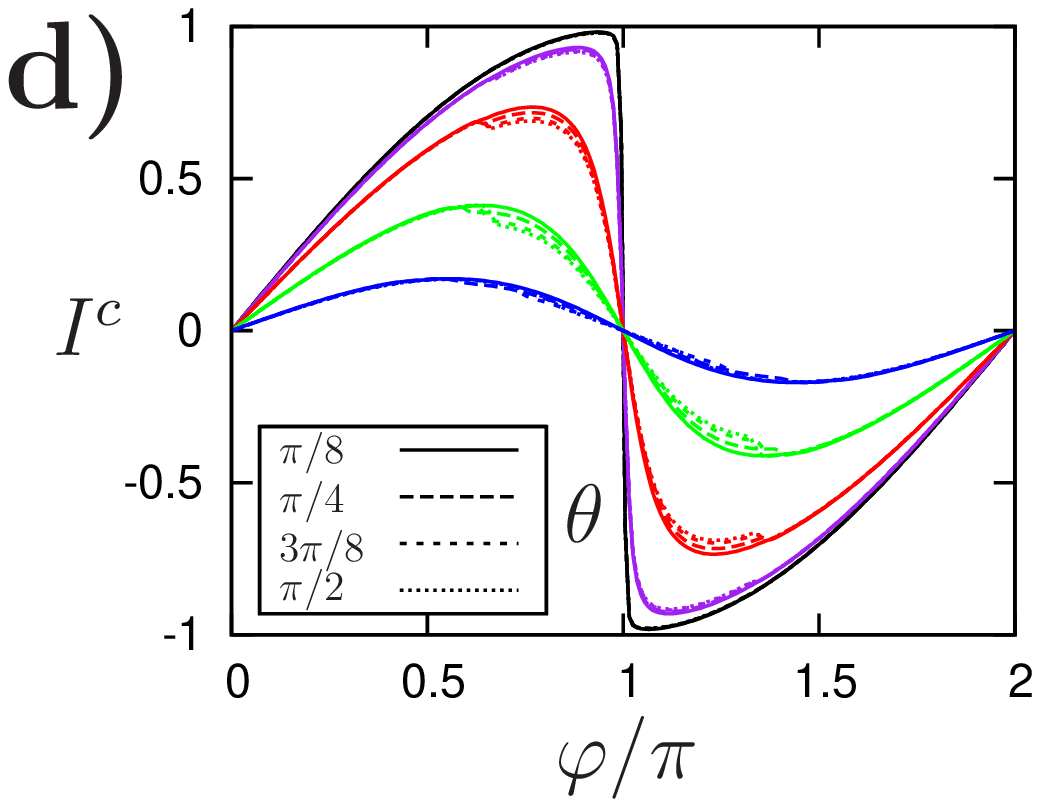}
\end{tabular}
\caption{ \label{Fig:CCG_theta}
(color online)
Charge CPR for $h_z = 0.25 \Delta$.
The different line styles refer to different tilt angles, see the insets.
The four panels correspond to:
(a) $t_S=1$, $t_0=0$, $T=10^{-3}\Delta$;
(b) $t_S=1$, $t_0=0$ and $T=10^{-2}\Delta$ (black) or $T=10^{-1}\Delta$ (red);
(c) $t_S=1$, $T=10^{-3}\Delta$ and $t_0=0.25$ (black) or $t_0=0.5$ (red) or $t_0=0.75$ (blue);
(d) $t_0=1$, $T=10^{-3}\Delta$ and $t_S=0.10$ (black) or $t_S=0.25$ (purple) or $t_S=0.5$ (red) or $t_S=0.75$ (green) $t_S=0.9$ (blue).
The current is in units of $e \Delta / \hbar$.}
\end{figure}
In the more realistic case where $\theta < \pi/2$, we see from Fig.~\ref{Fig:CCG_theta}a 
that the number of current steps doubles with respect to the case where $\theta = \pi/2$. 
In the rotating frame, this feature is due to the Zeeman splitting of the Andreev states by the effective magnetic field,
which is possible only for intermediate tilt angles. 
This was shown on Fig.~\ref{Fig:DOS_CCK}c for the special case $\theta = \pi/4$. 
A doubling of the number of bound-states can be seen on this figure with respect to
Fig.~\ref{Fig:DOS_CCK}a for $\theta = \pi/2$.
The corresponding current kernel is plotted on Fig.~\ref{Fig:DOS_CCK}d. 
The latter shows that for $\vp \approx 0$ (black curve) all bound-states are occupied. 
Because they carry current in opposite direction, the total current is zero.
Upon increasing $\vp$, one bound-state is emptied. 
This leads to a first sharp increase, in absolute value, of the current, e.g. $\vp \approx 0.25\pi$ (blue curve)
on Fig.~\ref{Fig:DOS_CCK}d. Increasing the value of the phase,
another bound-state is emptied which leads to a second step in the CPR.
Upon further increasing $\vp$ bound-states merge with the continuum and the current is again reduced.
This doubling of the bound-state is special to the rotating frame but allows a convenient interpretation of the 
results for the current (which does not depend on the frame one considers).

Notice that all these effects are washed out by temperature, see Fig.~\ref{Fig:CCG_theta}b where
the parameters have the same values as for Fig.~\ref{Fig:CCG_theta}a except for a raise in temperature. 
Indeed, by exciting particles from the lower to the upper bound-state temperature broadens the levels and
the sharp effects seen at very low temperatures disappear. 
The effects are also reduced in magnitude when direct tunneling is increased and the $\pi-$ to $0-$junction transition is approached.
This is seen from Fig.~\ref{Fig:CCG_theta}c which is valid for a $\pi-$junction ($t_S=1>t_0$) but where $t_0$ increases
from $0.25$ (black curves) to $0.75$ (blue curves).
In the limit of a $0-$junction ($t_S<t_0=1$), still with a non-zero tunnel amplitude through the SMM, Fig.~\ref{Fig:CCG_theta}d
shows that steps now appear around $\vp \approx \pi$ but their magnitude is considerably reduced with respect
to the case of a $\pi-$junction. From the point of view of CPR the low$-T$ $\pi-$junction is the most interesting, as expected.

\begin{figure}
\begin{tabular}{cc cc}
        \includegraphics[width=4.2cm,height=3.5cm]{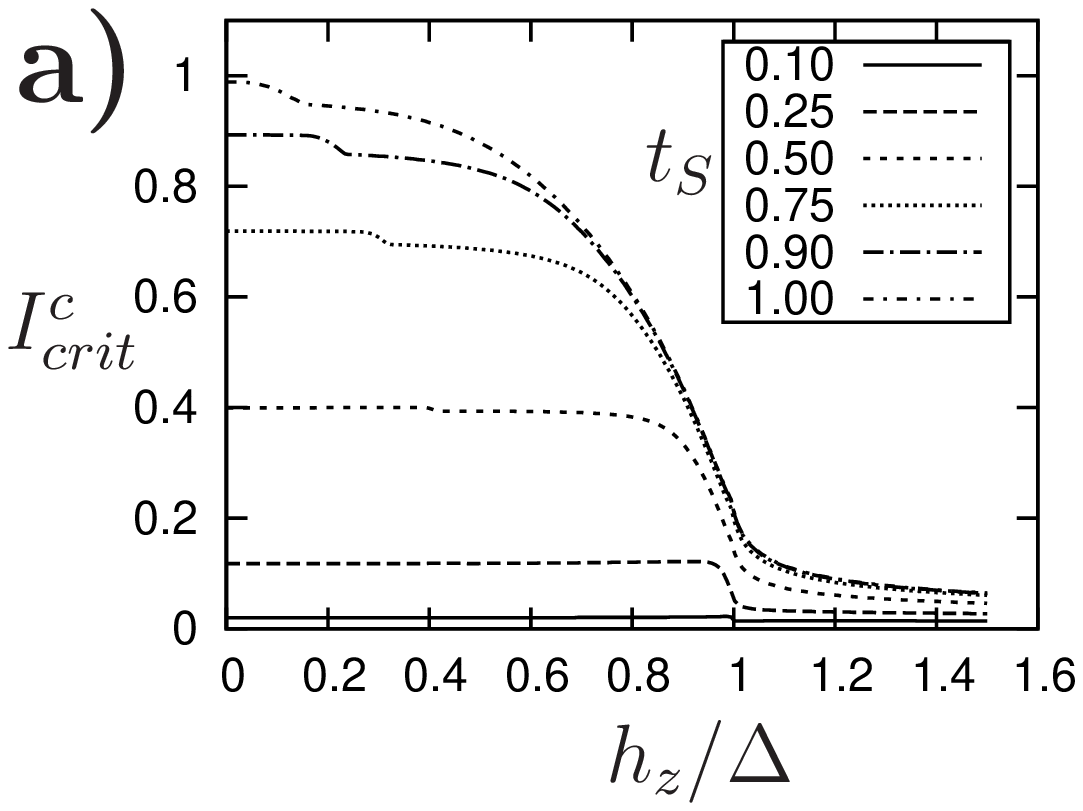}
&
        \includegraphics[width=4.2cm,height=3.5cm]{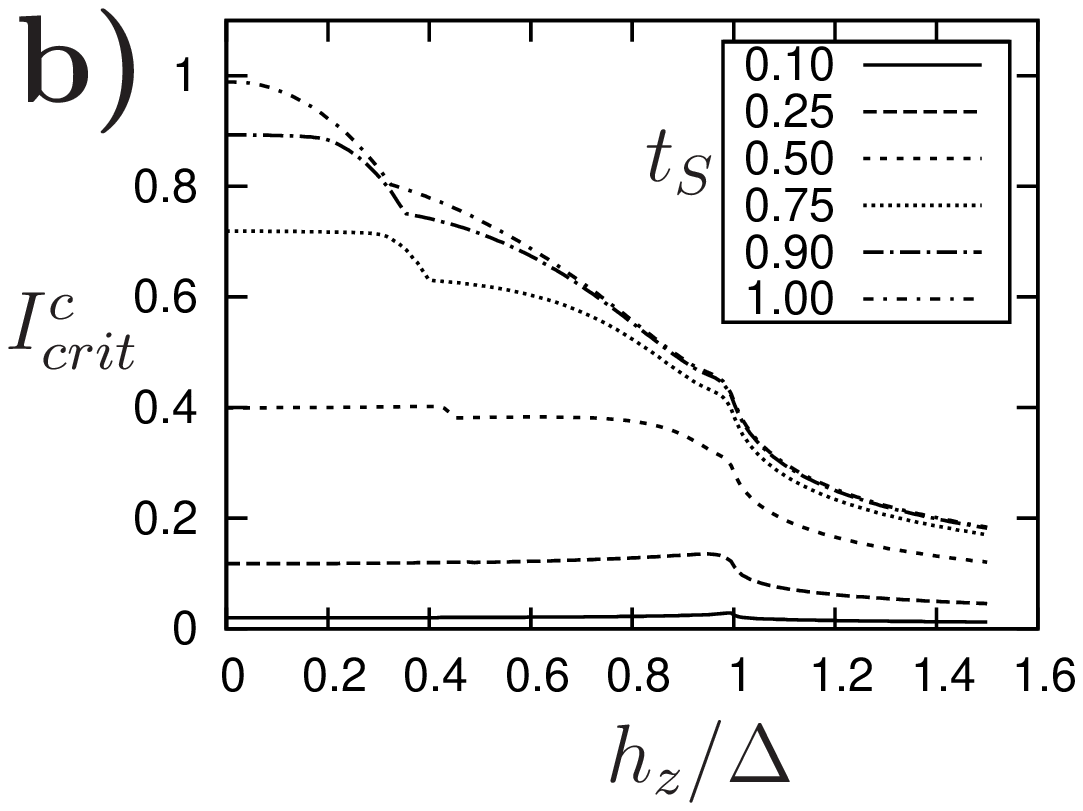}
\\
        \includegraphics[width=4.2cm,height=3.5cm]{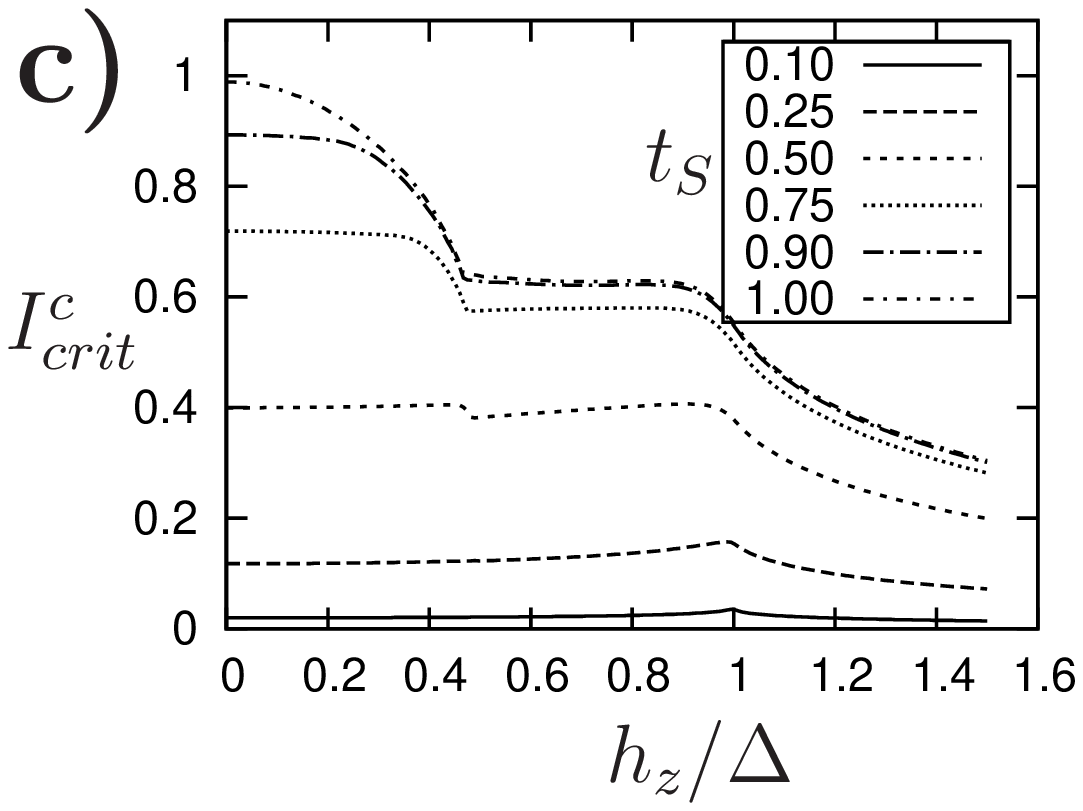}
&
        \includegraphics[width=4.2cm,height=3.5cm]{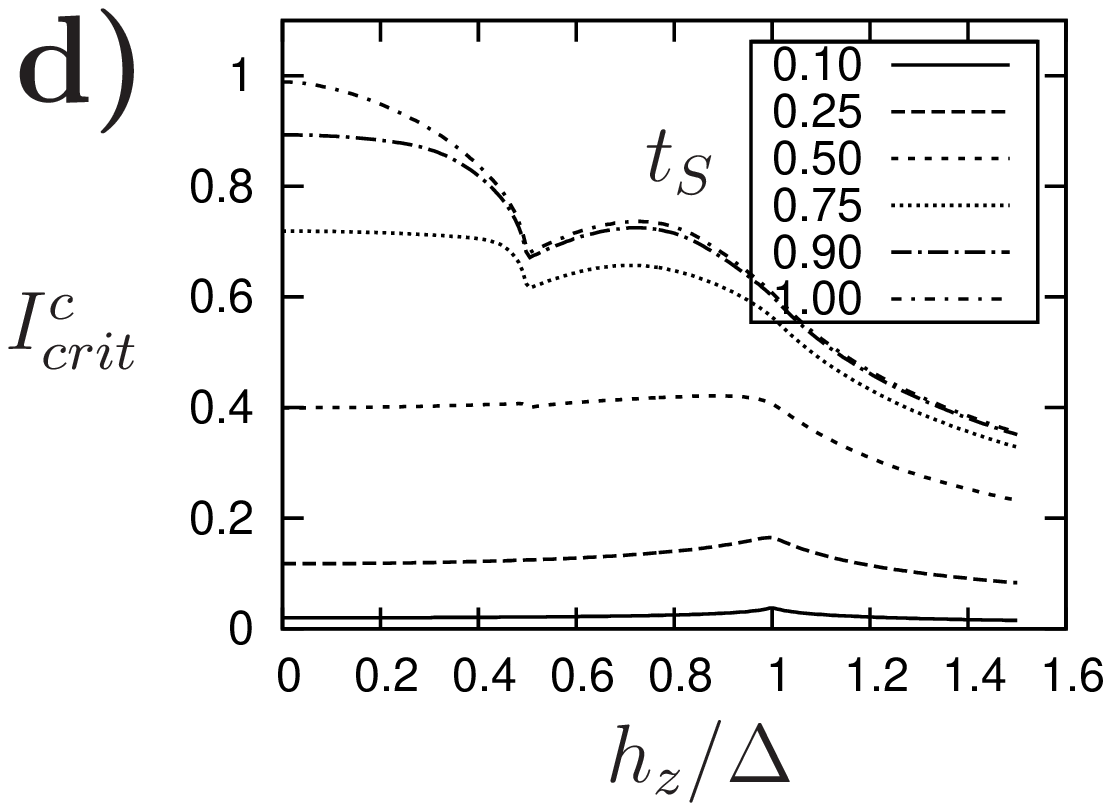}
\end{tabular}
\caption{
Absolute value of the charge critical current, i.e. maximum charge current as a function of the
phase-difference for a given value of $h_z$.
The different line styles refer to different transparencies, $t_S$, see the insets ($t_0=0$).
The four panels correspond to:
(a) $\theta = \pi/8$, (b) $\theta = \pi/4$, (c) $\theta = 3\pi/8$ and (d) $\theta=\pi/2$.
The current is in units of $e \Delta / \hbar$.}
\label{Fig:CCG_cc} % caption for the whole figure
\end{figure}

In Fig.~\ref{Fig:CCG_cc}, the critical current, i.e. the maximum current as a function of the
phase-difference for a given value of $h_z$, is plotted as a function of $h_z$. In the tunnel limit
the logarithmic singularity of Eq.~(\ref{ChargeCurrentTunnelResonance}) is seen at $h_z=1$. 
When the transparency of the junction increases this singularity is broadened and shifted to lower 
frequencies. 
This is quite clear from, e.g. Fig.~\ref{Fig:CCG_cc}d (limit where $\theta \rightarrow \pi/2$ and $t_S \rightarrow 1$) 
where, upon increasing the transparency, the $2\Delta-$resonance shifts to a kink at 
the value of the average critical field of Eq.~(\ref{hcrit}) in accordance with the discussion of the last paragraph. 
When the tilt angle is reduced the corresponding kink shifts to lower values of the precession frequency,
see Figs.~\ref{Fig:CCG_cc}a, \ref{Fig:CCG_cc}b and \ref{Fig:CCG_cc}c.

%

%%%%%%%%%%%%%%%%%%%%%%%%%%%%%%%%%%%%%%%
\section{Results for the spin-current}
\label{Sec.ResultsSpin}

\begin{figure}
\begin{tabular}{cc}
	\includegraphics[width=4.2cm,height=3.5cm]{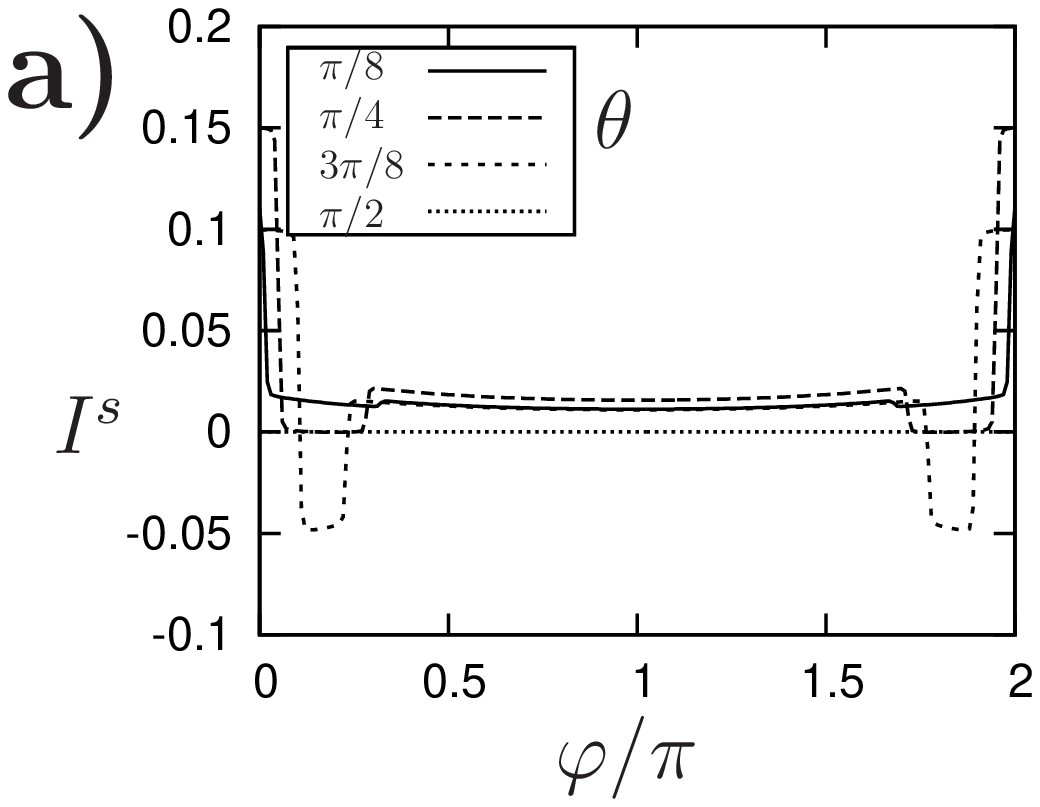}
&
	\includegraphics[width=4.2cm,height=3.5cm]{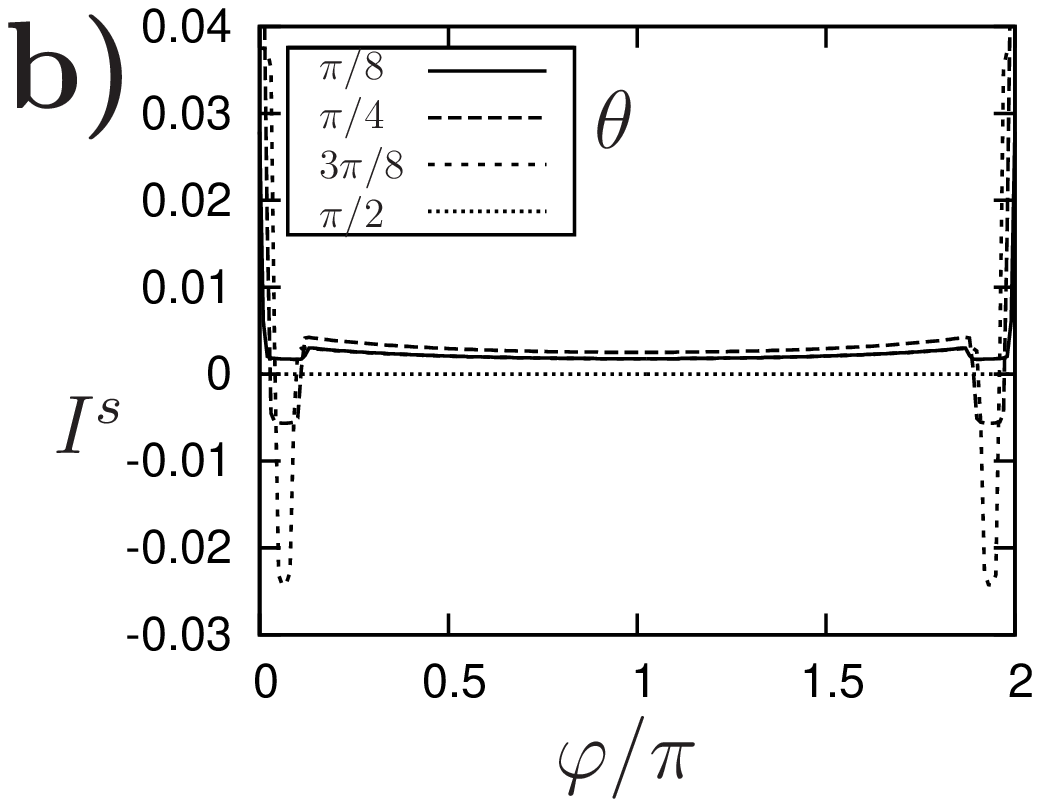}
\end{tabular}
\caption{
Spin CPR as a function of the superconducting phase
for fixed for $t_S=1$ and $t_0=0$.
The different line styles correspond to different tilt angles, see the insets.
a) $h_z=0.25\Delta$ and b) $h_z = 0.1 \Delta$. 
The spin current is in units of $\Delta \hbar /2$.}
\label{Fig:SCG_currphi_theta} % caption for the whole figure
\end{figure}

The main quantity of interest in this section is the spin current which was computed in the tunnel limit in Sec.~\ref{Sec.Tunnel}
and for which a formal exact expression in the transparent limit was given in Sec.~\ref{SubSec.TranspEq}: Eq.~(\ref{SCSN_UTF})
together with Eqs.~(\ref{KGFRF}), (\ref{RAGFRF}) as well as the self-consistent gap equation.
As was shown in Sec.~\ref{Sec.Tunnel}, and contrary to the case of the charge current,
the spin current can be emitted in the leads only when the spin is precessing so it is a purely out-of-equilibrium effect. 
It is circularly polarized in the $xy-$plane at the precession frequency, $\Omega$.
Moreover, as the results of Sec.~\ref{Sec.Tunnel} suggest,
the spin current is generally emitted for an inclined spin. 
A computation at an arbitrary tilt angle is therefore required. 
Nevertheless, the discussion on current-carrying states does not have to be repeated because the spin current kernel:
${\bf I}^s(t,\om) = e\,\Tr \big[ \hat{\vec{\tilde{\sigma}}}^s(t) \hat{\tilde{T}} \hat{\tilde{g}}^-_{LR}(\om) \big]$,
has similar poles and branch-cuts as the charge current kernel of Eq.~(\ref{CKG}).
The spin current will therefore be carried by extended states as well as by bound-states: 
${\bf I}^s = {\bf I}^s_{ABS} + {\bf I}^s_{ext}$, which are the same as the ones carrying the charge current. 
This is confirmed by Fig.~\ref{Fig:SCG_currphi_theta} where it is seen that the amplitude of the spin CPR also has steps
as a function of the superconducting phase difference with strong suppression of the spin current around $\vp \approx 0$
in the case of a $\pi-$junction. For identical values of the parameters the locations of these steps are the same as for
the charge current. 
Similarly to the case of the charge current, sharp suppressions of the spin current therefore originate from an 
abrupt change in the occupation of the Andreev levels. 

The importance of the spin-current is related to its back-action on the motion of the precessing spin in the junction.
For simplicity, we will fix the superconducting phase difference to $\vp=0$ or $\vp=\pi$, which correspond
to the ground-state phase difference of a $0-$junction (where $t_S<t_0$) or a $\pi-$junction (where $t_S>t_0$), respectively.
The fact that the spin-current may be carried by either extended or bound-states allows us then to distinguish
between two different sources for the back-action: one from a normal fluid of quasi-particles (the extended states)
and the other one from the condensate (the Andreev bound-states). Accordingly, the spin transfer torque will be separated into two contributions:
\bse
\label{STT_general}
\bea
&&\vec{\tau}^s = \vec{\tau}_{ext}^s + \vec{\tau}_{ABS}^s
\nonum \\
&&\vec{\tau}_{ext}^s = 2 {\bf I}^s_{ext} = \frac{\al_S(t_S,t_0,T)}{S^2}\,{\bf S} \times \partial_t {\bf S},
\label{STT_general_ext} \\
&&\vec{\tau}_{ABS}^s = 2 {\bf I}^s_{ABS} = \frac{\zeta_S(t_S,t_0,T,\vp)}{S}\,\gamma \,{\bf S} \times {\bf H}_{eff},
\label{STT_general_ABS}
\eea
\ese
where the last equation holds for $\vp=0$ or $\vp=\pi$.
Eqs.~(\ref{STT_general_ext}) and (\ref{STT_general_ABS}) generalize equations 
(\ref{SCTSC_vec}) and (\ref{SCTM_vec}) of the $T=0$ tunnel limit, respectively.
In the tunnel limit, the dimensionless coefficients $\al_S(t_S,t_0,T)$ and $\zeta_S(t_S,t_0,T,\vp)$
reduce to the expressions found in Eqs.~(\ref{GibertC_TM}) and (\ref{FreqShift_TSC}), respectively.
In the general case, they depend on both the transparency of the junction as well as the temperature. 
Notice that $\zeta_S$ also depends on the tilt angle $\theta$, see the cosine term in Eq.~(\ref{FreqShift_TSC}).
In the following, we will fix $\theta=\pi/4$, for simplicity. 
As already mentioned in Sec.~\ref{SubSec.SpinCurr}, $\al_S$ is the Gilbert damping constant
due to the normal fluid whereas $\zeta_S$ is the precession-frequency shift due to the condensate. 

\begin{figure}
\begin{tabular}{cc}
        \includegraphics[width=4.2cm,height=3.5cm]{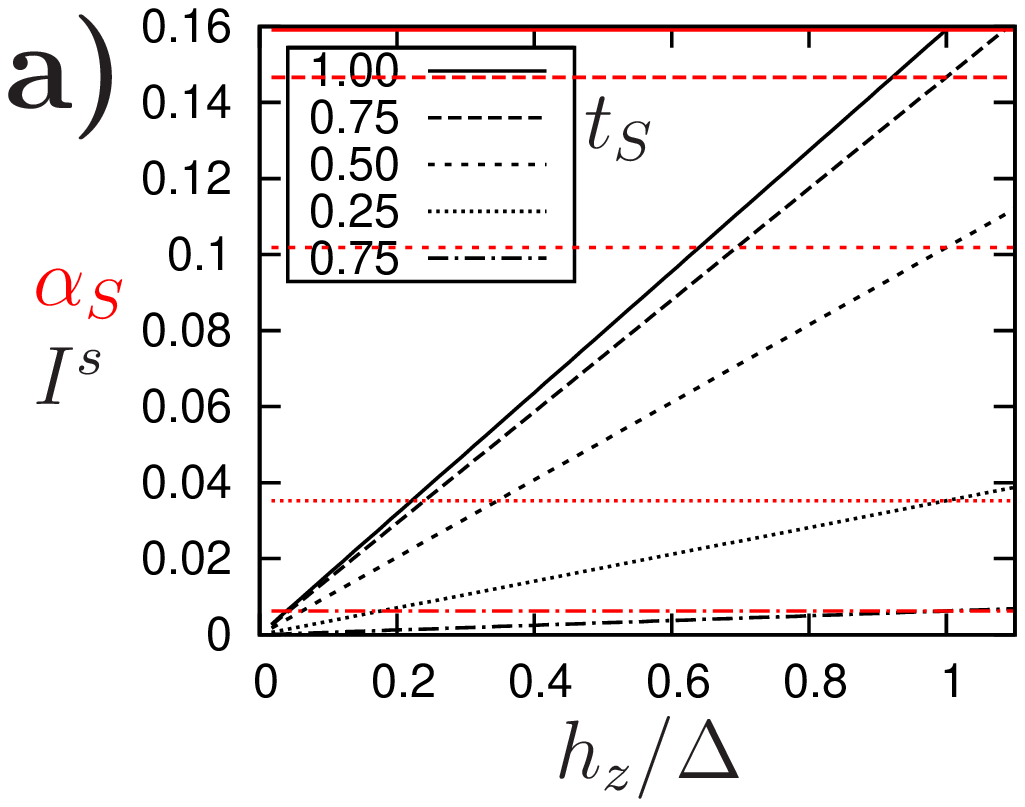}
&
        \includegraphics[width=4.2cm,height=3.5cm]{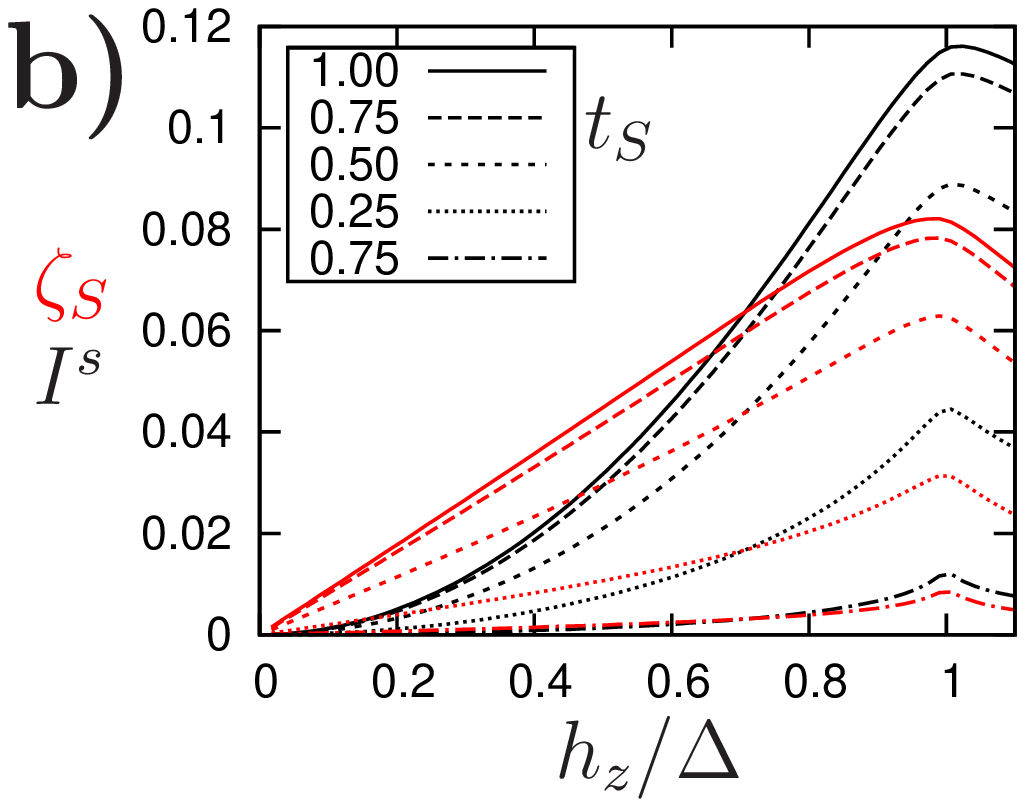}
\end{tabular}
\caption{
(color online)
(a) Case of normal leads (limit $\Delta \rightarrow 0$): absolute value of the quasi-particle spin-current (black) and the dimensionless Gilbert damping constant (red)
as a function of $\tilde{h}_z = h_z/\Delta$.
(b) Case of superconducting leads: absolute value of the superconducting spin-current (black) and the dimensionless shift of the precession frequency (red)
as a function of $\tilde{h}_z = h_z/\Delta$.
The different line styles refer to different transparencies, $t_S$, see the insets.
All figures are plotted for fixed: $t_0=0$, $\theta = \pi/4$, $\vp = \pi$ and $T=10^{-3}\Delta$.
The spin current is in units of $\Delta \hbar /2$.}
\label{Fig:SCG} % caption for the whole figure
\end{figure}

For normal metallic leads, the spin current and the associated Gilbert constant are plotted
at very low-$T$ as a function of the effective magnetic field on Fig.~\ref{Fig:SCG}a 
for different transparencies ranging from the tunnel to the transparent limit. 
This figure shows that the spin current varies linearly as a function of 
the precession frequency in accordance with the results of the tunnel limit.
This pumping of spins from the leads yields a Gilbert constant which 
depends on the transparency of the junction.
This normal contribution does not depend on the phase difference and is only weakly affected
by direct tunneling, e.g. $t_0 \not= 0$. 
When the leads are superconducting, this low-temperature ($T$) normal contribution vanishes altogether.
The remaining low-$T$ superconducting spin-current is plotted on Fig.~\ref{Fig:SCG}b
for a $\pi-$junction ($\vp=\pi$ and $t_0=0$) as a function of the effective magnetic field
and for different values of $t_S$. 
In the adiabatic regime ($h_z \ll \Delta$), and actually up to $h_z \approx \Delta$, 
the superconducting spin-current varies quadratically with $h_z$. 
In accordance with the tunnel limit results, the corresponding $\zeta_S$ is then seen to vary linearly with $h_z$. 
From the dependence of the spin-current on the precession frequency, we therefore recover
the single-particle (linear in $h_z$) versus two-particle (quadratic in $h_z$) 
nature of the normal versus anomalous currents, respectively.

\begin{figure}
\begin{tabular}{cc cc}
        \includegraphics[width=4.2cm,height=3.5cm]{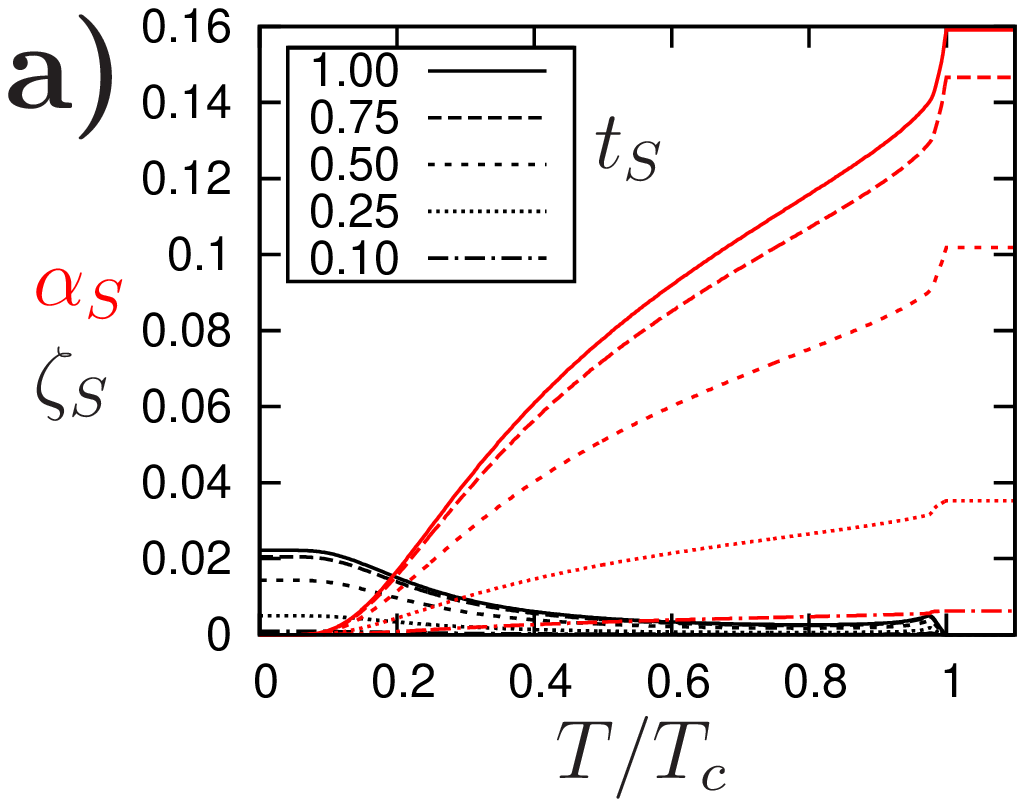}
&
        \includegraphics[width=4.2cm,height=3.5cm]{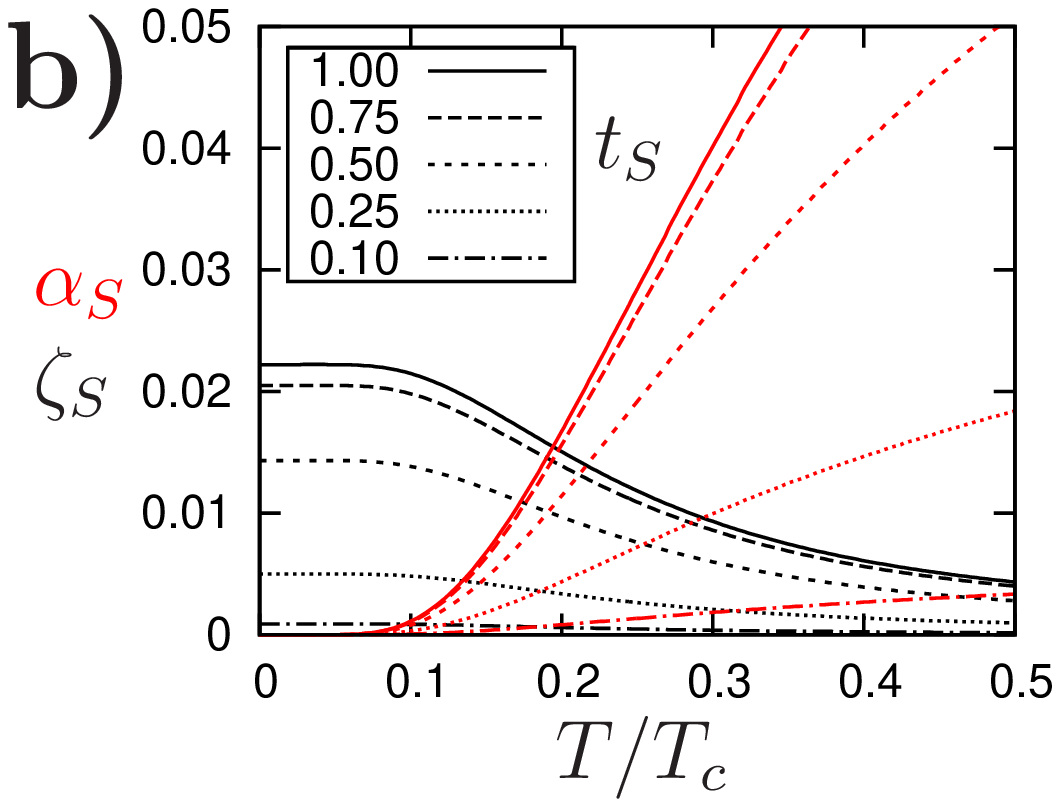}
\\
        \includegraphics[width=4.2cm,height=3.5cm]{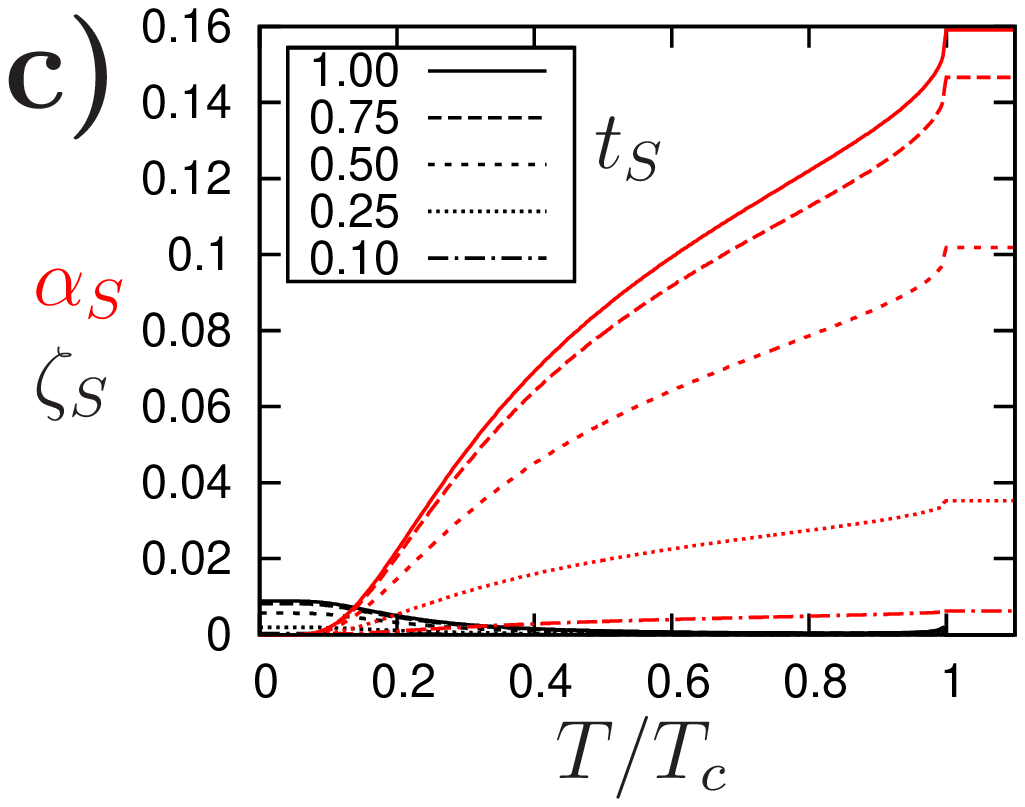}
&
        \includegraphics[width=4.2cm,height=3.5cm]{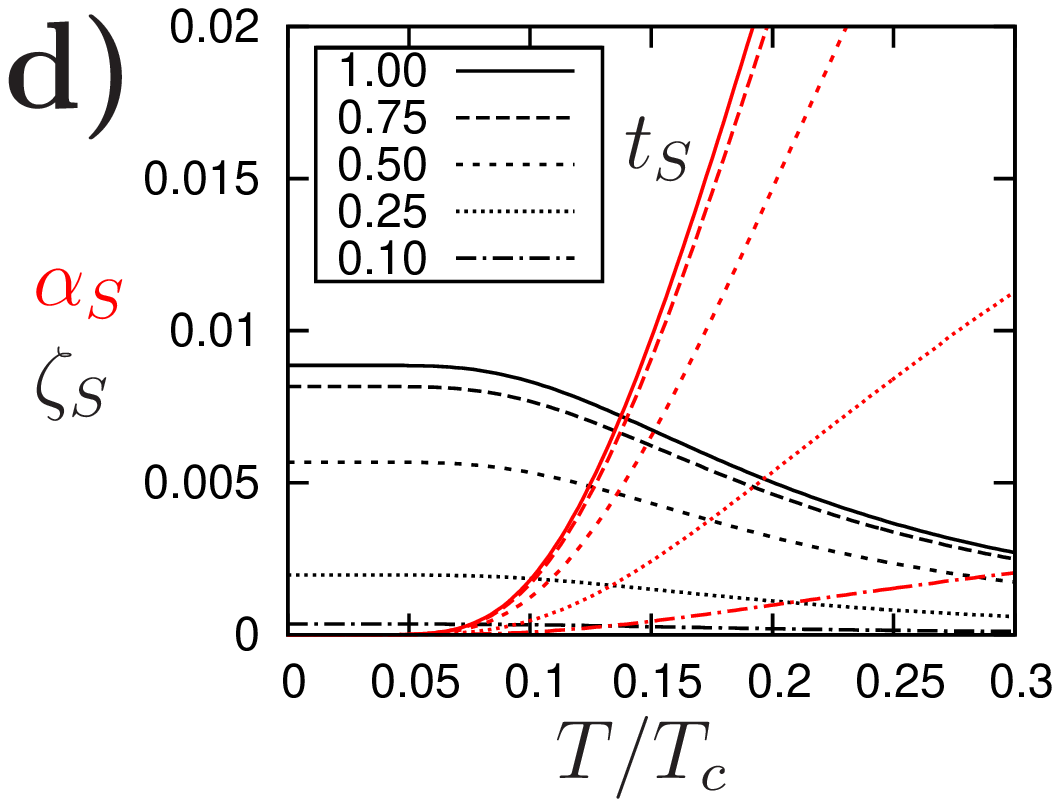}
\end{tabular}
\caption{
(color online)
Case of a $\pi-$junction ($t_S>t_0=0$ and $\vp=\pi$): absolute value of the Gilbert damping constant, $\al_S$, and
the shift of the precession frequency, $\zeta_S$, as a function of temperature $T$
for $\theta=\pi/4$.
The different line styles refer to different transparencies, $t_S$, see the insets.
The four panels correspond to:
(a) $h_z=0.25 \Delta$, (b) zoom on (a), (c) $h_z=0.1\Delta$ and (d) zoom on (c).}
\label{Fig:SCGPS_PiJ} % caption for the whole figure
\end{figure}

Still in the case of a $\pi-$junction, the Gilbert and precession-shift constants are
plotted on Figs.~\ref{Fig:SCGPS_PiJ} as a function of temperature, for different
values of the transparency $t_S$ and two values of the effective magnetic field:
$h_z=0.25\Delta$ (Fig.~\ref{Fig:SCGPS_PiJ}a and \ref{Fig:SCGPS_PiJ}b) and $h_z=0.1\Delta$ (Fig.~\ref{Fig:SCGPS_PiJ}c and \ref{Fig:SCGPS_PiJ}d).
We see clearly from these two figures that the Gilbert damping is strongly suppressed
below temperatures of the order $T \approx 0.1T_c$ for $h_z=0.25\Delta$
and $T \approx 0.075T_c$ for $h_z=0.1\Delta$, where $T_c$ is the superconducting critical temperature.
As shown by Figs.~\ref{Fig:SCGPS_PiJ}b and \ref{Fig:SCGPS_PiJ}d, which zoom on the low-$T$ region of Figs.~\ref{Fig:SCGPS_PiJ}a and \ref{Fig:SCGPS_PiJ}c, respectively,
the reduction of the damping upon decreasing the temperature goes along with an increasing shift of the precession 
frequency due to the superconducting spin-current. This shift even becomes the dominant effect
at low-enough temperatures: $T < 0.1T_c$ for $h_z=0.25\Delta$ and $T < 0.075T_c$ for $h_z=0.1\Delta$. 

\begin{figure}
\begin{tabular}{cc cc}
        \includegraphics[width=4.2cm,height=3.5cm]{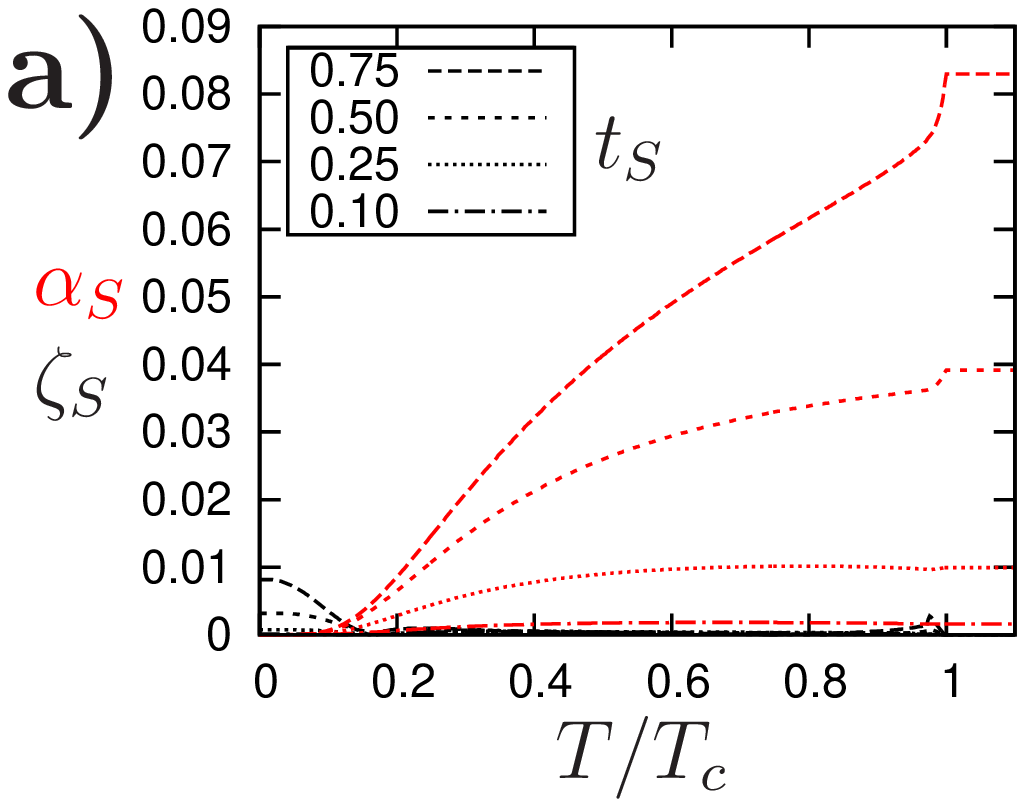}
&
        \includegraphics[width=4.2cm,height=3.5cm]{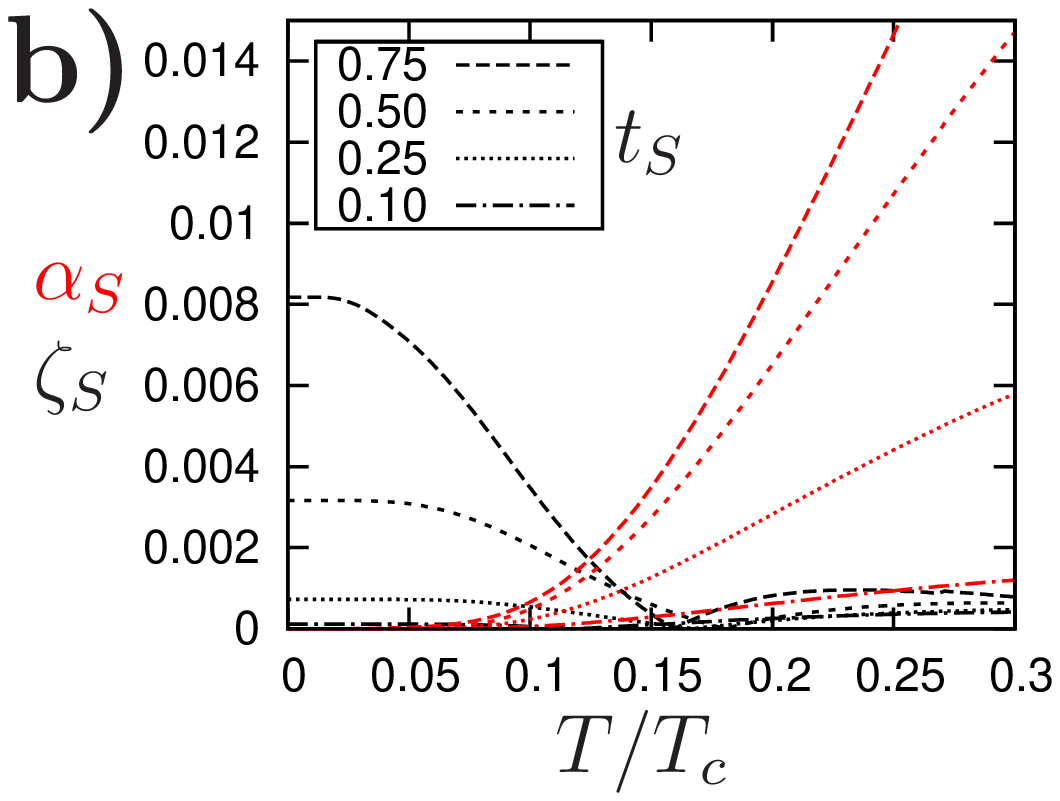}
\\
        \includegraphics[width=4.2cm,height=3.5cm]{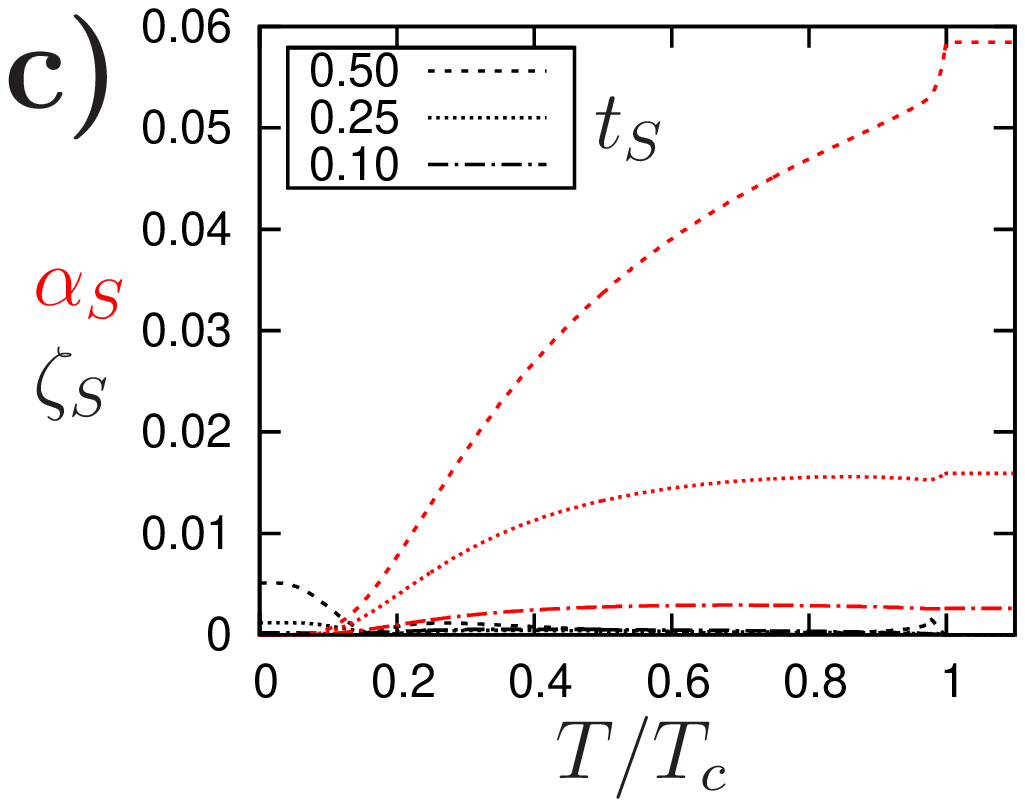}
&
        \includegraphics[width=4.2cm,height=3.5cm]{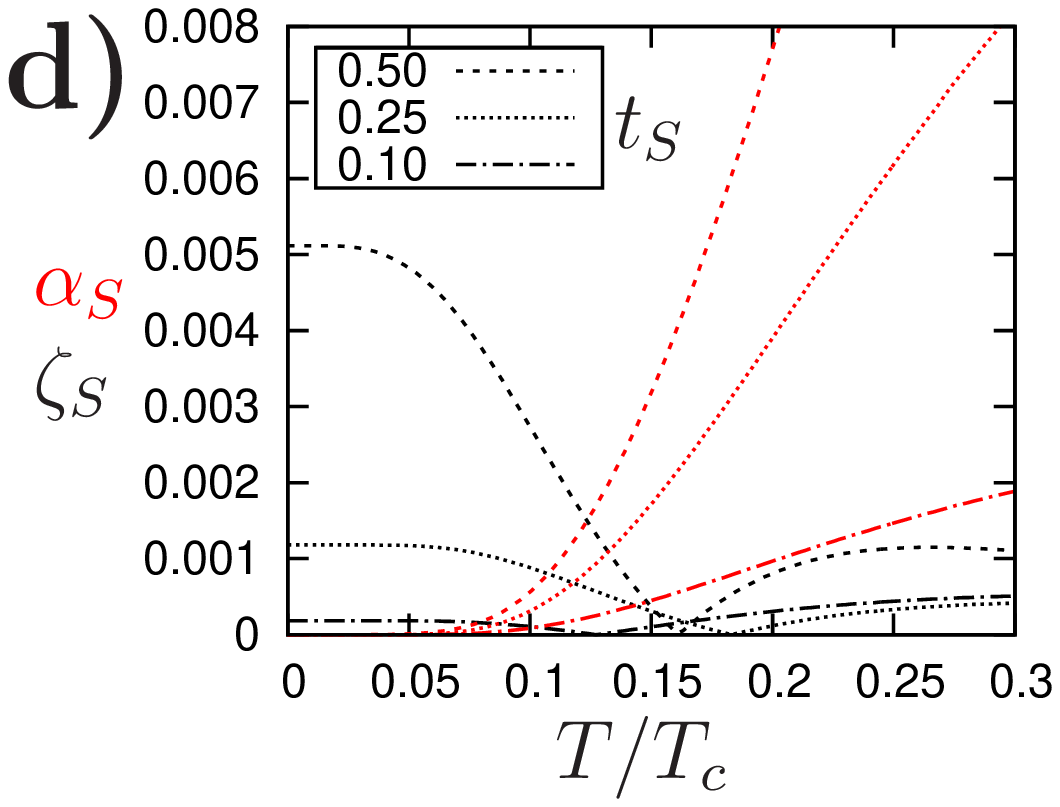}
\end{tabular}
\caption{
(color online)
Case of a $0-$junction ($t_S < t_0$ and $\vp=0$): absolute value of the Gilbert damping constant, $\al_S$, and
the shift of the precession frequency, $\zeta_S$, as a function of temperature $T$
for fixed $\theta=\pi/4$ and $h_z=0.25\Delta$.
The different line styles refer to different transparencies, $t_S$, see the insets.
The four panels correspond to:
(a) $t_S < t_0=1$, (b) zoom on (a), (c) $t_S < t_0=0.75$ and (d) zoom on (c).}
\label{Fig:SCGPS_0J} % caption for the whole figure
\end{figure}

For completeness, a similar plot for a $0-$junction is shown on Figs.~\ref{Fig:SCGPS_0J}. In these figures
$h_z=0.25$ but $t_S<t_0=1$ in Figs.~\ref{Fig:SCGPS_0J}a and \ref{Fig:SCGPS_0J}b, whereas $t_S<t_0=0.75$ in Figs.~\ref{Fig:SCGPS_0J}c and \ref{Fig:SCGPS_0J}d.
Provided that tunneling across the impurity still takes place, i.e. $t_S \not=0$, these figures show the same tendency
as Figs.~\ref{Fig:SCGPS_PiJ} for the damping to be strongly suppressed below $T \approx 0.1T_c$ in favor of a 
shift of the precession frequency due to the superconducting spin-current. 
The magnitudes of the constants are however reduced by an order of magnitude upon reducing $t_0$.

For a junction of arbitrary transparency, we therefore arrive at the conclusion that, deep in the superconducting phase, the precession
of the spin is maintained though the frequency may be slightly lowered, in the adiabatic regime, by the condensate spin-current.
Other non-dissipative effects, 
such as a nutation~\cite{Balatsky04} of the localized spin due to the singlet-nature of the current carriers, may then affect the motion of the
molecular precessing spin but they will not suppress the magnetization dynamics.

%%%%%%%%%%%%%%%%%%%%%%%%%%%%%%%%%%%%%%%%
\section{Conclusion}
\label{Sec.Conclusion}

In conclusion we have presented a theory of charge and spin transport in a S/SMM/S junction
where the single-molecule magnet is precessing at the frequency $\Omega$. 
The theory is based on a usual Green's function approach combined with a unitary transformation
which replaces the time-dependent problem by a static one with non-colinear magnetization.
Starting from the tunnel limit at zero temperature, both analytical and numerical results were presented, in a complementary way,
for a junction of arbitrary transparency and at finite temperatures.
Our goals were to study the out-of-equilibrium effects of the magnetization dynamics
on the current flowing through the junction as well as the back-action of this current on the magnetization dynamics.

For a phase biased junction, the unitary transformation to the rotating frame has revealed that a steady-state superconducting 
charge current flows through the junction and that an out-of-equilibrium circularly polarized spin current, of frequency $\Omega$, is emitted in the leads.
These currents and the corresponding current-phase relations (CPR) were then derived and the nature of the current carrying states was discussed. 
We have shown that the dynamics of the molecular spin in the junction affects both the spectrum of the Andreev states and their occupation.
At high precession frequencies, it eventually leads to their disappearance as current-carrying states in favor of extended (quasi-particle like) states.
At low precession frequencies, in particular in the experimentally reachable adiabatic limit where $\Omega \ll 2\Delta$,
and for high transparencies of the junction, the abrupt change in the occupation of the Andreev states 
is seen as sharp steps in the CPRs with significant variations of the current.
When the transmission amplitude through the magnetic impurity dominates the direct transmission amplitude ($t_S \gg t_0$)
and the junction is in a $\pi-$state, the CPRs (both for charge and spin currents) 
show that the current is strongly suppressed around $\vp \approx 0$ and $2\pi$.
In the opposite case of a $0-$junction, still with a non-zero transmission amplitude through the magnetic channel ($t_0>t_S\not=0$),
a similar suppression takes place around $\vp \approx \pi$, though the magnitude of the effect is much smaller. 
Integrating this contact into a superconducting loop one may hope that such sharp features could be observed experimentally. 
In particular, our results for the charge CPR of a $\pi-$junction with a high transmission amplitude in the magnetic channel 
bear some similarity to experimental results on microwave irradiated SNS $0-$junctions. 
The CPR was recently measured~\cite{Fuechsle09} in such a setup
and the current was shown to be strongly suppressed near $\vp \approx \pi$ upon increasing the rf power.
 
We have also shown that the spin current has a weak back-action on the dynamics of the magnetization
deep in the superconducting phase, i.e. at temperatures lower than the critical superconducting temperature by an order of magnitude or more.
The reason is the strong suppression of quasi-particles well below $T_c$ and in the adiabatic regime where $\Omega \ll 2\Delta$.
This agrees with recent experiments on S/F hybrids~\cite{Bell08} under FMR where the Gilbert damping was shown to be much reduced contrary to 
the case of N/F hybrids~\cite{Mizukami01}. 
We have also shown that the strong suppression of the damping, at low$-T$, goes along with a shift of the precession
frequency due to the back-action of the remaining superconducting part of the spin-current on the molecular magnet.
The fact that the dynamics of the magnetization is only weakly affected by the back-action of the current
provides a self-consistent check for the validity of the above results concerning the CPR.

\acknowledgments

We thank D.~Feinberg and M.~Houzet for useful discussions and input at the early stages of this work.
C.~H. and M.~F. thank T.~Lofwander, J.~Michelsen and V.~Shumeiko for useful discussions.
S.~T. thanks M. Aprili, B.~Dou\c{c}ot, A.~Kamenev and R.~M\'elin for useful discussions at various stages of this work, as well
as R.~Lugumerski for kind help with some figures.
 
%%%%%%%%%%%%%%%%%%%% 
 

\begin{thebibliography}{99}

\bibitem{TedrowMeservey71}
P.~M.~Tedrow and R.~Meservey, \prl {\bf 26} 192 (1971);
\prb {\bf{7}} 318 (1973).

\bibitem{Julliere75}
M.~Julli\`ere, Phys. Lett. A {\bf 54} 225 (1975).

\bibitem{Johnson85}
M.~Johnson and R.~H.~Silsbee, \prl {\bf 55} 1790 (1985).

\bibitem{Baibich88}
M.~N.~Baibich et al., \prl {\bf 61} 2472 (1988).

\bibitem{Binasch89}
G.~Binasch et al., \prb {\bf 39} 4828 (1989).

\bibitem{Wolf01}
S.~A.~Wolf et al., Science {\bf 294} 1488 (2001).

\bibitem{Slonczewski96}
J.~.C.~Slonczewski, J. Magn. Magn. Mater. {\bf 159} L1 (1996).

\bibitem{Berger96}
L.~Berger, \prb {\bf 54} 9353 (1996).

\bibitem{Ralph08}
D.~C.~Ralph and M.~D.~Stiles, J. Magn. Magn. Mater. {\bf 320} 1190 (2008).

\bibitem{TserkovnyakRMP05}
Y.~Tserkovnyak et al., \rmp {\bf 77} 1375 (2005).

\bibitem{Tserkovnyak02}
Y.~Tserkovnyak et al., \prl {\bf 88} 117601 (2002).

\bibitem{Waintal03}
X.~Waintal and P.~W.~Brouwer, \prl {\bf 91} 247201 (2003)

\bibitem{Mizukami01}
S.~Mizukami, Y.~Ando and T.~Miyazaki, J. Magn. Magn. Mater. {\bf 226} 1640 (2001).

\bibitem{Bell08}
C.~Bell et al., \prl {\bf 100} 047002 (2008).

\bibitem{Morten08}
J.~P.~Morten et al., Europhys. Lett. {\bf 84} 57008 (2008).

\bibitem{LeGall09}
C.~LeGall et al., \prl {\bf 102} 127402 (2009);
M.~Goryca et al., \prl {\bf 103} 087401 (2009).

\bibitem{Jo06}
Jo, M.-H. et al., Nano Lett. {\bf 6} 2014 (2006).

\bibitem{Heersche06}
H.~B.~Heersche et al., {\bf 96} 206801 (2006).

\bibitem{Sanvito06}
S.~Sanvito and A.~R.~Rocha, J. Comput. Theor. Nanosci. {\bf 3} 624 (2006).

\bibitem{Wernsdorfer08}
L.~Bogani and W.~Wernsdorfer, Nature Mat. {\bf 7} 179 (2008).

\bibitem{Doh05}
Y.-J.~Doh et al., Science {\bf 309} 272 (2005).

\bibitem{Cleuziou06}
J.~P.~Cleuziou et al., Nature Tech. {\bf 1} 53 (2006).

\bibitem{Jarillo06}
P.~Jarillo-Herrero, J.~A.~van Dam and L.~P.~Kouwenhoven, Nature {\bf 439} 953 (2006).

\bibitem{Jorgensen06}
H.~I.~J{\o}rgensen et al., \prl {\bf 96} 207003 (2006)

\bibitem{Jorgensen07} 
H.~I.~J{\o}rgensen, T.~Novotn\'y et al., Nano Lett. {\bf 7} 2441 (2007).

\bibitem{Winkelmann09}
C.~B.~Winkelmann, N.~Roch, W.~Wernsdorfer, V.~Bouchiat and F.~Balestro, Nature Phys. {\bf 5} 876 (2009)

\bibitem{Kasumov05}
A.~Y.~Kasumov et al., \prb {\bf 72} 033414 (2005).

\bibitem{Appelbaum66}
J.~Appelbaum, \prl {\bf 17} 91 (1966).

\bibitem{Anderson66}
P.~W.~Anderson, \prl {\bf 17} 95 (1966).

\bibitem{Kulik66}
I.~O.~Kulik, Zh. Eksp. Teor. Fiz {\bf 49} 1211 (1965) [Sov. Phys. JETP {\bf 22} 841 (1966)].

\bibitem{Bulaevskii77}
L.~N.~Bulaevskii et al., Zh. Eksp. Tero. Fiz {\bf 25} 314 (1977) [JETP Lett. {\bf 25} 290 (1977)] .

\bibitem{Cuevas01}
J.~C.~Cuevas and M.~Fogelstr\"om, \prb {\bf 64} 104502 (2001). 

\bibitem{Benjamin07}
C.~Benjamin et al., Eur. Phys. J. B {\bf 57} 279 (2007).

\bibitem{Balatsky03}
J.-X.~Zhu and A.~V.~Balatsky, \prb {\bf{67}} 174505 (2003).

\bibitem{Balatsky04}
J.-X.~Zhu, Z.~Nussinov, A.~Shnirman and A.~V.~Balatsky, \prl {\bf{92}} 107001 (2004).

\bibitem{Shumeiko08}
J.~Michelsen, V.~S.~Shumeiko and G.~Wendin, \prb {\bf 77} 184506 (2008).

\bibitem{Teber09}
S.~Teber, C.~Holmqvist, M.~Houzet, D.~Feinberg and M.~Fogelstr\"om, Physica B {\bf 404}, 527 (2009).

\bibitem{Holmqvist09}
C.~Holmqvist, S.~Teber, M.~Houzet, D.~Feinberg, and M.~Fogelstr\"om, J. Phys: Conf. Ser. {\bf 150}, 022027 (2009).

\bibitem{Melin00}
R.~M\'elin, Europhys. Lett. {\bf 51} 202 (2000).

\bibitem{Note:Houzet}
Using a similar approach, the Josephson current through a ferromagnet under FMR was studied
in Ref.~[\onlinecite{Houzet08}].

\bibitem{Houzet08}
M.~Houzet, \prl {\bf 101} 057009 (2008).

\bibitem{Kulik69}
I.~O.~Kulik, Zh.~Eksp.~Teor.~Fiz., {\bf 57} 1745 (1969) [Sov.~Phys.~JETP {\bf 30} 944 (1969)].

\bibitem{Golubov04}
A.~A.~Golubov, M.~Yu.~Kupriyanov and E.~Il'ichev, \rmp {\bf 76} 411 (2004).

\bibitem{Fuechsle09}
M.~Fuechsle et al., \prl {\bf 102} 127001 (2009).

\bibitem{Chiodi09}
F.~Chiodi, M.~Aprili and B.~Reulet, \prl {\bf 103} 177002 (2009).

\bibitem{Virtanen10}
P.~Virtanen, T.~Heikkil\"a, F.~Bergeret and J.~Cuevas, arXiv:1001.5149.

\bibitem{MahanBook}
G.~D.~Mahan, {\it Many-Particle Physics}, Plenum Press (1990).

\bibitem{Note:Analogy1}
Such an integral also appears in the theory of a superconductor in an alternating magnetic field, see Ref.~[\onlinecite{AGD}].

\bibitem{AGD}
A.~A.~Abrikosov, L.~P.~Gorkov and I.~E.~Dzyaloshinski, {\it Methods of Quantum field Theory in Statistical Physics}, Dover (1975).

\bibitem{Note:Analogy2} 
A non-analyticity of similar nature has been found in Ref.~[\onlinecite{Levchenko06}] where the Josephson current is a non-analytic function of the length of the normal disordered region of an SNS junction.

\bibitem{Levchenko06}
A.~Levchenko, A.~Kamenev and L.~Glazman, \prb {\bf 74} 212509 (2006).

\bibitem{Note:SpinCurrent}
The spin current, e.g. at the right lead, is defined as:
${\bf I}^s(t) =  \la {\dot{{\bf S}}}(t)\ra$. In the Heisenberg picture: ${\dot{{\bf S}}} = \frac{i}{\hbar} [H,{\bf S}]$, where:
${\bf S}(t) = \frac{\hbar}{2}\sum_{{\bf k},\sigma,\sigma'}\,c^\dag_{R,{\bf k},\sigma}(t)\,\vec{\sigma}_{\sigma,\sigma'}\,c_{R,{\bf k},\sigma'}(t)$.
Upon computing the average, we go to the interaction representation and perform an expansion to the lowest meaningful order in $H_T$. 
This yields Eqs.~(\ref{SCT}).

\bibitem{Wang03}
B.~Wang, J.~Wang and H.~Guo, \prb {\bf 67} 092408 (2003).

\bibitem{Note:GT}
At the hamiltonian level the unitary transformation is implemented with the help of a time- and spin-dependent unitary operator:
$U_{\al,{\bf k},\sigma}(t) = e^{-i \sigma \frac{\Omega t}{2} c^\dag_{\al,{\bf k},\sigma} c_{\al,{\bf k},\sigma}}$ acting on the fermionic operators of the leads: 
$\tilde{c}^\dag_{\al,{\bf k},\sigma} = e^{-i \sigma \Omega t / 2} c^\dag_{\al,{\bf k},\sigma} = U^\dag_{\al,{\bf k},\sigma}(t) c^\dag_{\al,\sigma} U_{\al,{\bf k},\sigma}(t)$. 
The final hamiltonian then reads: $\tilde{H} = H_R + H_L + H_T(\Omega=0) - \sum_{\al, {\bf k},\sigma}~\sigma h_z~c^\dagger_{\al,{\bf k},\sigma} c_{\al,{\bf k},\sigma}$ 
which is equivalent to the gauge transformed action of Eq.~(\ref{ActionGT}).

\bibitem{KeldyshReviews}
J.~Rammer and H.~Smith, \rmp {\bf 58} 323 (1986); A.~Kamenev and A.~Levchenko, Advances in Physics {\bf 58} 197 (2009). 

\bibitem{Caroli71}
C.~Caroli, R.~Combescot, P.~Nozi\`eres and D.~Saint-James, J. Phys. C: Solid State Phys. {\bf 4} 916 (1971).

\bibitem{Cuevas96}
J.~C.~Cuevas, A.~Mart\'in-Rodero and A.~Levy~Yeyati, \prb {\bf 54} 7366 (1996).

\bibitem{Note:Beenakker}
In SNS junctions, with a normal junction of length $L$, extended and localized states will 
generally both carry the current. A "universal" regime is reached in the limit of a small junction, 
$L \ll \xi$ where $\xi$ is the superconducting coherent length, where only the bound-state carries the current, 
see Ref.~[\onlinecite{Beenakker91}]. In the present out-of-equilibrium problem, though we consider a small junction, 
this statement does not hold.

\bibitem{Beenakker91} 
C.~W.~J.~Beenakker, \prl {\bf 67} 3836 (1991).

\bibitem{Kaplan76}
S.~B.~Kaplan et al., \prb {\bf 14} 4854 (1976). 

\end{thebibliography}
\end{document}